\documentclass[12pt]{article}
\usepackage{amsmath,amssymb}
\usepackage{cancel}
\usepackage{color}
\usepackage{bm}
\usepackage{braket}
\usepackage{slashed}
\usepackage[pdftex]{graphicx}
\usepackage{comment}
\usepackage{ulem}
\usepackage[hang,small,bf]{caption}
\usepackage[subrefformat=parens]{subcaption}

\newcommand{\im}{{\rm Im\,}}
\newcommand{\re}{{\rm Re\,}}
\newcommand{\e}{\varepsilon}

\begin{document}

\title{\vbox{
\baselineskip 14pt
\hfill \hbox{\normalsize }
} \vskip 1.7cm
\bf 
Modular Origin of Mass Hierarchy:
\\
Froggatt-Nielsen like Mechanism
\vskip 0.5cm
}
\author{
Hitomi~Kuranaga$^{1}$, 
Hiroshi~Ohki$^{1}$,
\ and \
Shohei~Uemura$^{2}$
\\*[20pt]
{\it \normalsize 
${}^{1}$Department of Physics, Nara Women's University, 
Nara 630-8506, Japan}
\\
{\it \normalsize 
${}^{2}$CORE of STEM, Nara Women's University, 
Nara 630-8506, Japan}
\\*[50pt]}

\date{
\centerline{\small \bf Abstract}
\begin{minipage}{0.9\linewidth}
\medskip 
\medskip
\small
We study Froggatt-Nielsen (FN) like flavor models with modular symmetry.
The FN mechanism is a convincing solution to the flavor puzzle 
in quark sector.
The FN mechanism requires an extra $U(1)$ gauge symmetry which is broken at high energy.
Alternatively, in the framework of modular symmetry 
the modular weights can play the role of the FN charges of the extra $U(1)$ symmetry.
Based on the FN-like mechanism with modular symmetry
we present new flavor models for quark sector.
Assuming that the three generations have a common representation under modular symmetry, 
our models simply reproduce the FN-like Yukawa matrices.
We also show that the realistic mass hierarchy and mixing angles, 
which are related each other through the modular parameters and a scalar vev, 
can be realized in models with several finite modular groups (and their double covering groups)
without unnatural hierarchical parameters.
\end{minipage}
}

\newpage
\begin{titlepage}
\maketitle
\thispagestyle{empty}
\clearpage
\end{titlepage}

\renewcommand{\thefootnote}{\arabic{footnote}}

\section{Introduction}

The origin of the flavor structure of the Standard Model (SM) 
is one of the most challenging problems in the particle physics.
The observed mass eigenvalues of the matter fields have a large hierarchy, 
which is more than $10^6$ among them.
This hierarchy can not be explained in the framework of the SM, 
since the Yukawa couplings are free parameters in the SM.
Thus we require a beyond the  SM mechanism naturally reproducing it.

The modular symmetry is a recently proposed solution for the flavor puzzle \cite{Feruglio:2017spp, Criado:2018thu}.
In this model, the action is assumed to be invariant under the (inhomogeneous) modular group 
$\Gamma \simeq PSL(2,\mathbb{Z})$, 
which is the quotient group of $SL(2,\mathbb{Z})$ divided by its center $\{I,-I \}$.
In this model, coupling constants are no longer free parameters, 
but modular forms.
The modular forms are specific holomorphic functions of the complex parameter known as the modulus.
They form unitary representations of the quotient group of the modular group: $\Gamma_N = \Gamma / \Gamma(N)$.
$\Gamma(N)$ is known as a principle congruence subgroup of $\Gamma$.
$\Gamma_N$ is isomorphic to a non-Abelian discrete group.
In particular, it is isomorphic to a finite group when the level $N$ is lower than 6 \cite{deAdelhartToorop:2011re}.
Therefore coupling constants, as well as the dynamical fields, transform as unitary representations of the non-Abelian finite group in this model.
This is attractive for the particle phenomenologist since discrete symmetries are well known candidate solutions for the flavor puzzle, especially for the lepton flavor structure, \cite{Harrison:2002er, Harrison:2002kp, Altarelli:2010gt}.%
\footnote{For review, see \cite{Ishimori:2010au, King:2013eh} and other approaches including continuous flavor symmetry and the GUT are shortly reviewed in \cite{Feruglio:2015jfa}.}
Indeed, various modular invariant models which successfully reproduce the SM have been constructed in these years,
e.g., for $\Gamma_2$ \cite{Kobayashi:2018vbk, Kobayashi:2019rzp, Okada:2019xqk, Novichkov:2021evw}, 
$\Gamma_3$ \cite{Feruglio:2017spp, Criado:2018thu, Kobayashi:2018vbk, Novichkov:2021evw, Novichkov:2018yse, Asaka:2019vev, Gui-JunDing:2019wap, Okada:2020ukr, Okada:2019uoy, Okada:2020rjb, Yao:2020qyy},
$\Gamma_4$ \cite{Novichkov:2021evw, Gui-JunDing:2019wap, Penedo:2018nmg, Novichkov:2018ovf, Mishra:2020gxg, Criado:2019tzk},
$\Gamma_5$ \cite{Novichkov:2021evw, Criado:2019tzk, Novichkov:2018nkm}, 
$\Gamma_7$ \cite{Ding:2020msi}, 
and the double covering groups of the modular groups \cite{Novichkov:2021evw, Liu:2019khw, Lu:2019vgm, Novichkov:2020eep, Liu:2020akv, Wang:2020lxk, Yao:2020zml, Wang:2021mkw}. 
A combined symmetry of the modular symmetry with the conventional flavor symmetries or CP-symmetry 
is also considered in \cite{Nilles:2020nnc, Ohki:2020bpo, Nilles:2020kgo, Nilles:2020tdp, Nilles:2020gvu, Baur:2020jwc, Baur:2019iai, Baur:2019kwi}.

The Froggatt-Nielsen (FN) mechanism is another well-known possible solution for the flavor puzzle \cite{Froggatt:1978nt}.
In the original FN mechanism, 
the matter fields such as the left-handed quarks $Q_i$
and the right-handed quarks $u_i^c$ and $d_i^c$ are assumed to be charged under an extra gauge group denoted by $U(1)_{FN}$,
which prohibits the tree level Yukawa coupling except for the top quark.
An extra scalar field $\phi$, which is a trivial singlet under the SM gauge group, 
is introduced to spontaneously break $U(1)_{FN}$. 
Since the effective Yukawa couplings are given by higher order couplings suppressed by 
its vacuum expectation value $(\langle \phi \rangle /\Lambda)^n$,
where $n$ is a difference of $U(1)_{FN}$ charges between generations, 
the mass hierarchy is controlled by the $U(1)_{FN}$ charges of the quark fields.
It is interesting that 
in the FN model the mass ratios are also related to the mixing angles, 
so that it naturally explains a realistic mass hierarchy and the mixing angles of the quark sector simultaneously 
if $\langle \phi \rangle/\Lambda$ is chosen to the Cabibbo angle~\cite{Feruglio:2015jfa}.

We consider FN-like mechanism
in the framework of modular symmetry.
In analogous to the FN mechanism, 
the modular weights of the fermion fields play the role of $U(1)_{FN}$ charges. 
An extra SM singlet $\phi$ with a negative modular weight is introduced, 
which compensates the modular weights of the fermion fields.
The effective couplings are given by higher dimensional operators 
suppressed by powers of $(\phi/\Lambda)^n$, 
where $n$ is a difference of modular weights between generations. 
The Yukawa coupling itself is also controlled by the modular weights of the fermion fields, 
since the modular forms are classified by modular weights.
This FN-like mechanism based on modular symmetry 
has been recently considered to explain the large mass hierarchy 
in \cite{Criado:2019tzk, King:2020qaj, King:2021fhl, Abbas:2020qzc}.\footnote{
Another approach to the mass hierarchy by the residual modular symmetry can be found in \cite{Novichkov:2021evw, Feruglio:2021dte}.}
We note that while in the previous model the mass hierarchy originates from powers of the weighton vev, 
the resulting quark mass matrices do not fully simulate the FN-like structure, 
so that the relationship between two origins of mass hierarchy and the small mixing angles 
becomes more
subtle  than that of the FN model.

In this paper, we present new flavor models for quark sector 
based on the FN-like mechanism with modular symmetry. 
In our models, it is assumed that the three generations of singlet quarks 
have a common representation under the modular symmetry,  
so that we see the same order of suppression factors appearing in each column or each row
of the Yukawa matrices,
where the two different suppression factors (the exponentially suppressed factor in the modular forms
and a power suppression by $\phi/\Lambda$) are incorporated, respectively.
Thus our models can simply reproduce the characteristic structure of the FN-like mass matrices. 
We illustrate this mechanism in models with different finite modular groups of 
$\Gamma_3', \Gamma_4'$ and $\Gamma_5'$ in detail.
Following Ref.~\cite{King:2020qaj}, 
we analyze an approximate expression for the mass ratios and the mixing angles, 
where a hierarchical mass structure which relates to the mixing angles is easily obtained 
with a suitable choice of the modular parameters and the singlet vev. 
We then numerically confirm this mechanism by a fit analysis, 
where we find parameter sets for the realistic mass hierarchy and mixing angles with $\mathcal{O}(1)$ coefficients.
The validity of the approximate estimation and the stability against changes of free parameters  
are also numerically investigated through the parameter dependence of the results.

This paper is organized as follows.
In section 2, we briefly review the modular symmetry.
We also explain possible two origins of the hierarchy for the modular symmetry;
one is the hierarchy among the modular forms, 
and the other is FN mechanism.
In section 3, we consider Froggatt-Nielsen like superpotential with the modular group of level 3.
In this model, hierarchical mass matrix is realized by cooperation between the above two possible origins.
In section 4, we generalize the previous model to the modular group of higher levels.
In section 5, we investigate our models statistically.
Section 6 is devoted to the conclusion.
We also review the modular forms of level 3, 4, and 5 in Appendix.

\section{Modular symmetry and the Froggatt-Nielsen mechanism}

In this section, we briefly review the modular symmetry and the Froggatt-Nielsen mechanism.
We also introduce our notations mostly based on \cite{Feruglio:2017spp}.

The modular symmetry is a recently proposed model building framework.
In this model, the action is assumed to include a complex parameter known as
the complex structure modulus $\tau \in \mathbb{H}$.
The complex structure parameterizes the geometry of torus.
Torus is invariant under the linear fractional transformation,
\begin{align}
\tau \rightarrow \gamma\tau = \frac{a\tau+b}{c\tau+d},
\end{align}
where $\gamma = 
\begin{pmatrix}
a & b \\
c & d
\end{pmatrix}
\in SL(2,\mathbb{Z})$ 
since $\tau$ and $\gamma \tau$ generate the same lattice.
It is obvious that both $\gamma$ and $-\gamma$ equivalently act on $\tau$.
Thus the torus is invariant under
$PSL(2,\mathbb{Z}) \equiv SL(2,\mathbb{Z})/\mathbb{Z}_2$.
This is the (inhomogeneous) modular group $\Gamma$.
The modular group is generated by two generators,
\begin{align}
S = \begin{pmatrix}
0 & 1\\
-1 & 0
\end{pmatrix}
~~{\rm and}~~
T = \begin{pmatrix}
1 & 1\\
0 & 1
\end{pmatrix},
\end{align}
and these generators satisfying the following relations
\begin{align}
S^2 = (ST)^3 = I,
\end{align}
where $I$ is the identity.
In this paper, we abuse elements of $SL(2,\mathbb{Z})$ to denote 
the corresponding elements of the modular group $SL(2,\mathbb{Z})/\mathbb{Z}_2$.
For instance,
$S$ is the element of $SL(2, \mathbb{Z})$ originally,
but the same symbol also denotes the equivalence class in the modular group.
The modular group acts on the effective action.
For instance, coupling constants such as Yukawa couplings depend on the moduli $\tau$.
They transform under the modular group through the modular transformation of $\tau$.
To construct a modular invariant action, 
coupling constants should form representations of the modular group, 
and such functions are known as the modular forms.

Before considering modular forms, 
we introduce the principal congruence subgroup of level $N$ in 
$\Gamma$, which is usually represented by $\Gamma(N)$.
$\Gamma(N)$ is given by 
\begin{align}
\Gamma(N) = \left\{
\begin{pmatrix}
a & b\\
c & d
\end{pmatrix}
\in \Gamma
\;\middle|\;
a=d = 1~{\rm and}~b =c =0~{\rm mod}~ N
\right\}.
\end{align}
The modular forms of level $N$ and weight $k$ are holomorphic functions 
satisfying the following transformation
\begin{align}
f(\gamma\tau) = (c\tau+d)^k f(\tau),
\end{align}
for any $\gamma$ in $\Gamma(N)$.
The prefactor $(c\tau+d)^k$ is so-called automorphy factor.
Since linear combinations of the modular forms of level $N$ and weight $k$
are also modular forms of level $N$ and weight $k$,
they form a linear space, which is denoted by $\mathcal{M}_k(\Gamma(N))$.
$\mathcal{M}_k(\Gamma(N))$ is finite dimensional.
The modular forms form unitary representations of the quotient group $\Gamma_N = \Gamma/\Gamma(N)$ up to the automorphy factor,
\begin{align}
f_i (\gamma\tau) = (c \tau +d)^k \rho(\gamma)_{ij} f_j(\tau),
\end{align}
where $\{ f_i\}$ is a basis of $\mathcal{M}_k(\Gamma_N)$,
and $\rho$ is a unitary representation of $\Gamma_N$.
The generators of $\Gamma_N$ are satisfying the following relations,%
\footnote{We abuse elements of $SL(2,\mathbb{Z})_2$ to denote elements of $\Gamma_N$ too.}
\begin{align}
S^2 = (ST)^3 = T^N = I.
\end{align}
$\Gamma_N$ is isomorphic to the non-Abelian finite group when $N$ is smaller than 6:
$\Gamma_2 \simeq S_3, \Gamma_3 \simeq A_4, \Gamma_4 \simeq S_4, \Gamma_5 \simeq A_5$ \cite{deAdelhartToorop:2011re}.
Note that the definition of the automorphy factor has ambiguity
since the modular group is divided by its center.
Hence the modular forms are well-defined only if the modular weight is even.
If we consider the double covering group of the modular group instead of the usual modular group, 
modular forms of odd modular weights can be defined as well.
The double covering group of the modular group is known as homogeneous modular group $\Gamma'$.
To distinguish $\Gamma$ from $\Gamma'$, 
$\Gamma$ is called inhomogeneous modular group.
$\Gamma'$ is nothing but $SL(2,\mathbb{Z})$ itself,
and there are no ambiguities of sign of $c$ and $d$.
The principal congruence subgroup of level $N$ 
in the homogeneous modular group, and 
its quotient group $\Gamma_N'$ is similarly obtained as 
\begin{align}
\Gamma_N' = \Gamma'/\Gamma'(N),
\end{align}
where $\Gamma'(N)$ is a subgroup of $\Gamma'$ whose element is equivalent to $I$ mod $N$.
The generator of $\Gamma_N'$ are satisfying the similar relations:
\begin{align}
S^2 = R,~~(ST)^3 = I,~~T^N=I,~~TR = RT,~R^2=I.
\end{align}
$\Gamma_N'$ is isomorphic to non-Abelian finite group when $N <6$, too:
$\Gamma_2' = \Gamma_2 \simeq S_3, \Gamma_3' \simeq T', \Gamma_4' \simeq S_4'$ and $\Gamma_5' \simeq A_5'$ \cite{Liu:2019khw}. 
The modular symmetry is originally inspired by string compactifications \cite{Hamidi:1986vh, Dixon:1986qv, Lauer:1989ax, Lerche:1989cs, Ferrara:1989bc, Kobayashi:2016ovu}.
From the stringy perspective,
the coupling constants and dynamical fields
transform under the homogeneous modular group rather than the inhomogeneous modular group \cite{Nilles:2020kgo, Ohki:2020bpo, Kikuchi:2020frp, Kikuchi:2020nxn, Kikuchi:2021ogn, Hoshiya:2020hki, Tatsuta:2021deu} and metaplectic group \cite{Almumin:2021fbk}.
Hence we consider modular symmetric model based on $\Gamma'$ in this paper.

Throughout this paper, we assume global supersymmetry.
The matter fields such as quarks are denoted by chiral superfields.
We assume that they transform as 
the modular forms of level $N$ and weight $k$,
\begin{align}
\gamma~:~\Phi_i \rightarrow (c\tau +d)^{k_{i}} \rho_{\Phi}(\gamma)_{ij} \Phi_j,
\end{align}
where $\Phi_i$ is a matter field and $\rho_{\Phi}(\gamma)$ is a unitary matrix.
The action of chiral superfields $\Phi_i$ is given by two functions, the K\"ahler potential $K$ and the superpotential $W$,
\begin{align}
\mathcal{S} = \int d^4 x d^2\theta d^2\bar \theta K(\Phi^i , \bar \Phi^i, \tau, \bar \tau)
+ \int d^4 x d^2\theta W(\Phi^i, \tau)  + (h.c.).
\nonumber
\end{align}
The typical modular invariant action is given as%
\footnote{This form of K\"ahler potential is given by string compactifications  at tree level.
The general form of the modular invariant K\"ahler potential includes modular forms and other couplings.
Such additional terms may affect results  \cite{Chen:2019ewa}, but we ignore their effects in the present paper for simplicity.
}
\begin{align}
K &= \sum_{i} \frac{\Phi_i \bar \Phi_i}{\im \tau^{-k_i}},
\label{eq:Kahler_tree}
\\
W &= \sum (f_{i_1 i_2 ...i_n}(\tau) \Phi_{i_1} \Phi_{i_2}... \Phi_{i_n})_{\bf 1},
\label{eq:superpotential}
\end{align}
where $f_{i_1 i_2 ...i_n}(\tau)$ is a modular form of weight $k$ satisfying 
\begin{align}
k  + k_{i_1} +  ...+ k_{i_n}=0,
\label{eq:weight}
\end{align}
so that the modular weight of the superpotential is zero.%
\footnote{From local supersymmetry, the modulus $\tau$ is a vacuum expectation value of a dynamical field,
and the K\"ahler potential includes the kinetic term of the moduli field:
\begin{align}
K = -h \ln(\tau + \bar \tau).
\nonumber
\end{align}
This term implies that the modular weight of the superpotential is $h$ rather than zero, and the modular invariant condition \eqref{eq:weight} is changed to $k  + k_{i_1} +  ...+ k_{i_n}=h$.
However it is always possible to cancel it by shifting the modular weight of the chiral superfields.
Therefore we assume \eqref{eq:weight} throughout this paper.
}
The modular weight plays a similar role of the charge of $U(1)$ gauge symmetry.
$(f_{i_1 i_2 ...i_n}(\tau) \Phi_{i_1} \Phi_{i_2}... \Phi_{i_n})_{\bf 1}$ denotes the trivial singlet component of the tensor product of the chiral superfields and the modular form.
We consider canonically normalized couplings rather than the holomorphic couplings.
The canonically normalized field $\tilde \Phi_i$ is given by
\begin{align}
\tilde \Phi_i = \im \tau^{k_i/2} \Phi_i,
\end{align}
and the canonically normalized $n$-point coupling $\tilde f_{i_1 i_2... i_n}$is given by 
\begin{align}
\label{eq:canonicalf}
\tilde f_{i_1 i_2 ... i_n} = \im \tau ^{-\frac{k_1 + k_2 + ... + k_n}{2}}f_{i_1 i_2...i_n}(\tau)= \im \tau^{k/2}f_{i_1 i_2...i_n}(\tau),
\end{align}
where $k$ is the modular weight of $f_{i_1 i_2 ... i_n}$ itself.

In order to be invariant under modular symmetry, 
the allowed Yukawa couplings are also given in term of the modular forms,
which are classified by modular weights.
In the followings, we discuss the properties of the modular forms.

\subsection{Hierarchy in the modular forms}

The modular forms naturally have a hierarchy.
To illustrate this point clearly,
we consider the modular forms of level 3 as a concrete example.
The modular forms of level 3 and weight $k$ form a $k+1$ dimensional linear space $\mathcal{M}_{k}(\Gamma(3))$ \cite{Liu:2019khw, Lu:2019vgm}.
The modular forms of weight 1 are given by
\begin{align}
\hat e_1(\tau) \equiv \frac{\eta^3(3\tau)}{\eta(\tau)},~~~~
\hat e_2(\tau) \equiv \frac{\eta^3(\tau/3)}{\eta(\tau)},
\end{align}
where $\eta(\tau)$ is the Dedekind eta function,
\begin{align}
\eta(\tau) = q^{1/24}\prod_{n=1}^{\infty}(1-q^n),~~~~q\equiv e^{2\pi i\tau}.
\nonumber
\end{align}
Note that since we consider the homogeneous modular group, odd weight modular forms can be defined.
The modular transformations of $\hat e_1$ and $\hat e_2$ are given by
\begin{align}
\begin{cases}
\hat e_1(\tau+1) = e^{i2\pi/3} \hat e_1(\tau),~~
\hat e_2(\tau+1) = 3(1-e^{i2\pi/3})\hat e_1(\tau) +\hat e_2(\tau),
\\
\hat e_1(-1/\tau) = 3^{-3/2} (-i\tau) \hat e_2(\tau),~~
\hat e_2(-1/\tau) = 3^{3/2} (-i\tau) \hat e_1(\tau),
\end{cases}
\end{align}
and we can check that the action of the modular group is closed in this space.  
Since the modular forms form representations of $\Gamma_3'$, they can be decomposed to the irreducible representations of $\Gamma_3' \simeq T'$.
The irreducible representations of $T'$ are as  follows,%
\footnote{Their matrix representations are summarized in Appendix \ref{app:level_3}.}
\begin{align}
{\bf 1, 1', 1'', 2,2',2'', 3}.
\end{align}
$\hat e_1, \hat e_2$ form ${\bf 2}$ of $T'$ by
\begin{align}
Y^{(1)}_{\bf 2}(\tau) =
\begin{pmatrix}
Y_1(\tau)\\
Y_2(\tau)
\end{pmatrix}
=
\begin{pmatrix}
\sqrt{2} e^{i 7 \pi /12} \hat e_1(\tau)\\
\hat e_1(\tau) + \frac 1 3 \hat e_2(\tau)
\end{pmatrix}.
\label{eq:level3_lowest}
\end{align}
The modular forms of higher weights are constructed
by the tensor products of the modular forms of weight 1,
e.g., the modular forms of weight 2 are given by
\begin{align}
Y_{\bf 3}^{(2)} = (Y_{\bf 2}^{(1)} \otimes Y_{\bf 2}^{(1)})_{\bf 3} = (e^{i\pi/6}Y_2^2, \sqrt 2 e^{i 7\pi/12} Y_1 Y_2, Y_1^2)^t,
\end{align}
and they form ${\bf 3}$ of $T'$.
We can obtain the higher weight modular forms in a similar way.

The modular forms have $q$-expansion.
$Y_{\bf2}^{(1)}(\tau)$ is expanded as
\begin{align}
Y_1(\tau) &= \sqrt 2 e^{i7\pi/12} q^{1/3}(1 + q+2q^2 +2q^4 + q^5 +2q^6+\cdots),
\nonumber
\\
Y_2(\tau) &= 1/3 + 2q + 2q^3 +2q^4 +2q^7 +2q^9 +\cdots.
\end{align}
Thus $Y_1 \ll Y_2$ for large $\im \tau$.
This is a general feature for the modular forms which transform as ${\bf 2}$ of $T'$.
The matrix representation of $\rho_{\bf 2}(T)$ are given by
\begin{align}
\rho_{\bf 2}(T) 
= 
\begin{pmatrix}
\omega & 0\\
0 & 1
\end{pmatrix},
~~(\omega = e^{2\pi i /3}).
\label{eq:2matrix_leve3}
\end{align}
It is clear that $q^{1/M}$ transforms under $T$ as
\begin{align}
q^{1/M}\rightarrow e^{2\pi i (\tau+1)/M} = e^{2\pi i /M} q^{1/M}.
\end{align}
Thus the modular forms which transform as ${\bf 2}$ of $T'$
must have the $q$-expansion of the following form,
\begin{align}
Y_{\bf 2}^{(k)} =
\begin{pmatrix}
Y_{{\bf 2}, 1}^{(k)}\\
Y_{{\bf 2}, 2}^{(k)}
\end{pmatrix}
=
\begin{pmatrix}
q^{1/3} \sum_{n\in \mathbb{N}} C_{n}^{(k)}q^n\\
\sum_{n\in \mathbb{N}} D_{n}^{(k)}q^n
\end{pmatrix},
\end{align}
where $C_n^{(k)}$ and $D_n^{(k)}$ are coefficients independent of $\tau$.
Thus we can generally approximate the modular forms of ${\bf 2}$ as
\begin{align}
Y_{\bf 2}^{(k)} \sim (q^{1/3} , 1)^{T},
\end{align}
for large $\im \tau$. 
More precisely, the leading terms of the $q$-expansions are not uniquely determined by the matrix representations of $T$.
They always have ambiguity of integer powers of $q$.
Thus the modular forms are approximated as
\begin{align}
Y_{{\bf 2}, 1}^{(k)} \sim q^{1/3 + M_1}, ~~ Y_{{\bf 2},2}^{(k)} \sim q^{M_2},
\end{align}
where $M_1$ and $M_2$ are appropriate integer numbers.
To determine the correct hierarchy, we must calculate the explicit forms of the modular forms.
Nevertheless we obtain the hierarchical values in either case since the powers of $q$ of the leading terms can not be the same.
The matrix representation of $T$ for other representations are given in Appendix \ref{app:level_3}.
The modular forms of the other representations have $q$-expansion of
\begin{align}
Y_{\bf 1}^{(k)} &= \sum C_n^{(k)} q^n,~~
Y_{\bf 1'}^{(k)} = q^{1/3} \sum C_n^{(k)} q^n,~~
Y_{\bf 1''}^{(k)} = q^{2/3} \sum C_n^{(k)} q^n,
\nonumber
\\
Y_{\bf 2'}^{(k)} &=
\begin{pmatrix}
 q^{2/3} \sum C_n^{(k)} q^n\\
 q^{1/3} \sum D_n^{(k)} q^n
\end{pmatrix},~~
Y_{\bf 2''}^{(k)} =
\begin{pmatrix}
 \sum C_n^{(k)} q^n\\
 q^{2/3} \sum D_n^{(k)} q^n
\end{pmatrix},~~
Y_{\bf 3}^{(k)} =  
\begin{pmatrix}
\sum C_n^{(k)} q^n\\
 q^{1/3} \sum D_n^{(k)} q^n\\
 q^{2/3} \sum E_n^{(k)} q^n
\end{pmatrix},
\label{eq:app_Y3k}
\end{align}
and they are approximated by
\begin{align}
Y_{\bf 1}^{(k)} &\sim 1,~~
Y_{\bf 1'}^{(k)} \sim q^{1/3},~~
Y_{\bf 1''}^{(k)} \sim q^{2/3},
\nonumber
\\
Y_{\bf 2'}^{(k)} &\sim
\begin{pmatrix}
 q^{2/3}\\
 q^{1/3}
\end{pmatrix},~~
Y_{\bf 2''}^{(k)} \sim
\begin{pmatrix}
1\\
 q^{2/3}
\end{pmatrix},~~
Y_{\bf 3}^{(k)}
\sim 
\begin{pmatrix}
1\\
 q^{1/3}\\
 q^{2/3}
\end{pmatrix}.
\label{eq:MF_order_lev3}
\end{align}
Therefore the modular forms naturally have large hierarchy for large \im $\tau$.

This feature is general for the modular group of other levels.
We summarize the result of $\Gamma_4'$ in Appendix \ref{app:level_4}, 
and that of $\Gamma_5'$ in Appendix \ref{app:level_5}.

\subsection{Froggatt-Nielsen mechanism}

The Froggatt-Nielsen (FN) mechanism is a well-known candidate solution for the flavor puzzle.
In FN mechanism, one introduces an extra Abelian gauge group $U(1)_{FN}$,
and assumes that the matter fields such as the left-handed quark field $Q_i$
and the right-handed quark fields $u_i^c$, $d_i^c$ are charged under $U(1)_{FN}$.
We assume that Higgs fields $H_u$ and $H_d$ are neutral under $U(1)_{FN}$.
The Yukawa couplings are prohibited by $U(1)_{FN}$
unless the sum of $U(1)_{FN}$ charges of the corresponding fields are canceled.
We introduce an extra scalar field $\phi$ which has the $U(1)_{FN}$ charge of $-1$.
We also assume that $\phi$ is the trivial singlet under the SM gauge group.
After the $U(1)_{FN}$ is spontaneously broken down at high energy scale $\Lambda$, 
$\phi$ obtains vacuum expectation value $\braket{\phi}$
and the effective superpotential for the quarks would be given by 
\begin{align}
W_u = \sum_{i,j} C^{(u)}_{ij} \tilde \phi^{f_{u_i}+f_{Q_j}} Q_j H_u u_i^c,~~
W_d = \sum_{i,j} C^{(d)}_{ij} \tilde \phi^{f_{d_i}+f_{Q_j}} Q_j H_d d_i^c,
\nonumber
\end{align}
where $\tilde \phi = \braket{\phi}/\Lambda$,
and $f_{Q_i}$ and $f_{u_i,d_i}$ are the charge of the quarks respectively.
$C^{(u)}_{ij}$ and $C^{(d)}_{ij}$ are free parameters supposed to be order 1.
If $\tilde{\phi}$ is sufficiently small (and the $U(1)_{FN}$ charges of the quarks are large enough),
effective Yukawa couplings much less than 1 are obtained.

A typical example for the quark charges reproducing the observed mass eigenvalues and the mixing angles is given by
\begin{align}
f_{Q_1} &=3, f_{Q_2} =2, f_{Q_3} =0,
\nonumber
\\
f_{u_1^c} &= 3, f_{u_2^c} =2, f_{u_3^c} =0,
\nonumber
\\
f_{d_1^c} &= 2, f_{d_2^c} =1, f_{d_3^c} =1.
\nonumber
\end{align}
In this case, Yukawa couplings of order 1 are prohibited except for the top quark,
and the light quark mass terms are given by nonrenormalizable higher order terms.
The quark mass matrix is given by \cite{Kanemura:2007yy}
\begin{align}
M_{u} \sim
\begin{pmatrix}
\tilde{\phi}^6 & \tilde{\phi}^5 & \tilde{\phi}^3\\
\tilde{\phi}^5 & \tilde{\phi}^4 & \tilde{\phi}^2\\
\tilde{\phi}^3 & \tilde{\phi}^2 & \tilde{\phi}^0
\end{pmatrix}\frac{v_u}{\sqrt 2},~~
M_{d} \sim
\begin{pmatrix}
\tilde{\phi}^5 & \tilde{\phi}^4 & \tilde{\phi}^4\\
\tilde{\phi}^4 & \tilde{\phi}^3 & \tilde{\phi}^3\\
\tilde{\phi}^2 & \tilde{\phi}^1 & \tilde{\phi}^1
\end{pmatrix}\frac{v_d}{\sqrt 2}.
\label{eq:mass_FN}
\end{align}
The left-handed quarks $Q_i$ are on the left side of the mass matrix, 
and the right-handed quarks $u_i^c, d_i^c$ are  on the right side in our notation \cite{Zyla:2020zbs}.
The mass matrix is diagonalized by two unitary matrices $V_L^f$ and $V_R^f$ as $V_L^f M_f V_R^{f\dagger} = M_f^{diagonal}$,
and the Cabbibo-Kobayashi-Maskawa (CKM) matrix is given by $V_{CKM} = V_L^u V_L^{d\dagger}$.
The mass eigenvalues of \eqref{eq:mass_FN} are approximated by $m_u \sim \tilde \phi^6 v_u, m_c \sim \tilde \phi^4 v_u$ and $m_t \sim v_u$ for the up sector,
and they are given as 
$m_d \sim \tilde \phi^5 v_d, m_s \sim \tilde \phi^3 v_d$ and $m_b \sim \tilde \phi v_u$ for the down sector.

In terms of the mixing angles,
we consider two-flavor model at first for simplicity.
Suppose that the mass matrix of the up and down quarks are given by
\begin{align}
M_u = 
\begin{pmatrix}
\epsilon_1 \epsilon_2 & \epsilon_2\\
\epsilon_1 & 1 
\end{pmatrix},~~
M_d = 
\begin{pmatrix}
\delta_1 \delta_2 &  \delta_2\\
\delta_1 & 1 
\end{pmatrix},
\nonumber
\end{align}
where 
$\epsilon_i, \delta_j$ are small parameters.
The eigenvalues of $M_u$ are obtained by the following eigenvalue equations,
\begin{align}
\det(M_u M_u^\dagger - \lambda I) =0.
\nonumber
\end{align}
We obtain
\begin{align}
\lambda = 0, (1+\epsilon_1^2) (1 + \epsilon_2^2).
\nonumber
\end{align}
The diagonalizing matrix is given by
\begin{align}
(V_L^u)^\dagger = 
\begin{pmatrix}
1 & \epsilon_2 \\
-\epsilon_2  & 1
\end{pmatrix}.
\nonumber
\end{align}
The calculation for the down sector is completely parallel, and we obtain the CKM matrix
\begin{align}
V_L^u (V_L^d)^\dagger =
\begin{pmatrix}
1- \epsilon_2 \delta_2 & \delta_2 - \epsilon_2\\
\epsilon_2 - \delta_2 & 1 + \epsilon_2 \delta_2
\end{pmatrix}
\label{eq:FN-mix_app}
\end{align}
Thus the mixing angle is approximated by the difference between the $\epsilon_2$ and $\delta_2$.
For three-flavor model, the mixing angles $\theta_{IJ}$ is 
approximately given by the mass matrix focusing on the corresponding two quarks, that is
$
\begin{pmatrix}
M_{II} & M_{IJ}\\
M_{JI} & M_{JJ}
\end{pmatrix}
$.
Hence we obtain
\begin{align}
\theta_{12} \sim \frac{M_{12}}{M_{22}}, ~~ 
\theta_{23} \sim \frac{M_{23}}{M_{33}}, ~~ 
\theta_{13} \sim \frac{M_{13}}{M_{33}}.
\label{eq:mixing_FN_model}
\end{align}
The approximated CKM matrix of \eqref{eq:mass_FN} is written as
\begin{align}
V_{CKM} \sim 
\begin{pmatrix}
1 & \tilde \phi & \tilde \phi^3\\
\tilde \phi & 1 & \tilde \phi^2\\
\tilde \phi^3 & \tilde \phi^2 & 1
\end{pmatrix}.
\end{align}
It is interesting that the observed values of CKM matrix are given by
\begin{align}
\theta_{12} \equiv \theta_{C} \sim 0.2,~~
\theta_{23} \sim 0.03,~~
\theta_{13} \sim 0.003,
\end{align}
and the realistic mixing angles are realized if $\tilde \phi$ is equivalent to the Cabbibo angle $\theta_{C}$.

\section{Froggatt-Nielsen like mechanism with $\Gamma_3'$}

The FN-like mechanism based on the modular symmetry is first proposed in 
\cite{King:2020qaj}, and similar idea is also proposed in \cite{Criado:2019tzk}.
In this model, the extra $U(1)$ symmetry is not required, 
but a similar role is played by the modular group.
In this section, we explain this idea.
We also explain a difference between our model and the previous ones.

We consider a modular symmetric models with $\Gamma_3' = T'$ at first.
It is straightforward to generalize the idea to the modular group of another level.
Suppose that the quark fields $Q_i = (U_L^i, D_L^i)^t, u_i^c, d_i^c$ have modular weight of $k_{Q_i, u_j, d_k}$,
and their representations under the finite modular group are denoted by $\rho_Q, \rho_u, \rho_d$, respectively.
We can set the modular weights of the Higgs fields to zero without loss of generality since 
they can be absorbed by shifting the modular weight of the other fields.
We also assume that the Higgs fields are the trivial singlet of the modular group.
The tree-level superpotential is given by
\begin{align}
W_u = \sum (Y_{ij}^{(-k_{Q_i}-k_{u_j})}(\tau) Q_i u_j^c)_{\bf 1} H_u,~~
W_d =  \sum (Y_{ij}^{(-k_{Q_i} - k_{d_j})}(\tau) Q_i d_j^c)_{\bf 1} H_d,
\end{align}
where the Yukawa couplings $Y_{ij}^{(-k_{Q_i}-k_{u_j})}$ and $Y_{ij}^{(-k_{Q_i} - k_{d_j})}$ are the modular forms of weight 
$-k_{Q_i}-k_{u_j}$ and $-k_{Q_i} - k_{d_j}$.
\begin{table}[t]
\begin{center}
\begin{tabular}{|c|c|} \hline
weight $k$ & representations
\\
\hline \hline
1 & {\bf 2} 
\\ 
2 & {\bf 3} 
\\
3 & ${\bf 2}, {\bf 2''}$ 
\\
4 & ${\bf 1, 1', 3}$ 
\\
5 & ${\bf 2, 2', 2''}$ 
\\
6 & ${\bf 1}, 2\times{\bf  3}$ 
\\
\vdots & \vdots 
\\
\hline
\end{tabular}
\caption{Irreducible decomposition of $\mathcal{M}_{k}(\Gamma(3))$.}
\label{tab:irrepT'}
\end{center}
\end{table}
Suppose $\rho_Q ={\bf 3}$ and $\rho_{u} = \rho_{d} = {\bf 1}$ of $T'$,
and their modular weights are given by
\begin{align}
&k_{Q_1} = k_{Q_2} =k_{Q_3} = k_Q =  0,
\nonumber
\\
&k_{u_1} = +1,~k_{u_2} = -3,~k_{u_3} = -6,
\nonumber
\\
&k_{d_1} = 0,~k_{d_2} = -3,~k_{d_3} = -5,
\nonumber
\end{align}
where odd modular weights to the quark fields are allowed 
if we consider the double covering group of the modular group. 
In this case, the Yukawa couplings must be {\bf 3}.
We show the modular forms of level 3 and weight $k<7$ in Table \ref{tab:irrepT'}.
The tree level couplings are prohibited except for $u_3^c$
because of the absence of the triplet modular forms for the odd weights.
Hence the modular invariant superpotential is obtained as
\begin{align}
W_u = \left[
\gamma_u \left( (Y^{(6)}_{{\bf 3},I} + r_u Y^{(6)}_{{\bf 3},II}) Q \right)_{\bf 1} u_3^c \right]H_u,
~~
W_d = 0,
\end{align}
where $Y^{(6)}_{{\bf 3},I}$ and $Y^{(6)}_{{\bf 3},II}$ denote the two modular forms of weight 6
which transform as the triplet under $T'$.
$\gamma_u$ and $r_u$ are arbitrary coefficients supposed to be order 1.
The explicit form of $Y^{(6)}_{{\bf 3},I}$ and $Y^{(6)}_{{\bf 3},II}$ are given by \eqref{eq:mf_lev3_wei6}.
$\gamma_u$ is real and $r_u$ is complex since the phase of $\gamma_u$ can be absorbed by redefinition of $u_3^c$.
This superpotential would corresponds to the Yukawa couplings of the top quark.
The other couplings are given as non-renormalizable higher order couplings.
We introduce a new chiral superfield $\phi$ whose modular weight is $-1$.
This $\phi$ is called weighton since it carries the unit of modular weight \cite{King:2020qaj}.
We assume that $\phi$ is the trivial singlet under both $T'$ and the SM gauge group.
After breaking the modular symmetry, $\phi$ develops its vev,
and the effective superpotential should be given by
\begin{align}
W_u^{tri} &= 
\left[
\alpha_u \tilde{\phi}^3 (Y^{(2)}_{{\bf 3}} Q)_{\bf 1} u_1^c + 
\beta_u \tilde \phi (Y^{(4)}_{{\bf 3}} Q)_{\bf 1} u_2^c +
\gamma_u ((Y^{(6)}_{{\bf 3},I} + r_u Y^{(6)}_{{\bf 3},II}) Q)_{\bf 1} u_3^c
\right]
H_u,
\nonumber
\\
~~
W_d^{tri} &= \left[
\alpha_d \tilde{\phi}^2 (Y^{(2)}_{{\bf 3}} Q)_{\bf 1} d_1^c + 
\beta_d \tilde \phi (Y^{(4)}_{{\bf 3}} Q)_{\bf 1} d_2^c +
\gamma_d \tilde\phi ((Y^{(6)}_{{\bf 3},I} + r_d Y^{(6)}_{{\bf 3},II}) Q)_{\bf 1} d_3^c
\right]
H_d,
\label{eq:superpotential_A4}
\end{align}
where $\tilde \phi$ denotes $\braket{\phi}/\Lambda$ and $\Lambda$ is the cutoff scale.
$\alpha_{f}, \beta_f, \gamma_f$ and $r_f$, where $f$ denotes $u$ or $d$, are order 1 parameters. 
The superscript $tri$ explicitly indicates that the left-handed quarks form the triplet.
In our basis, the irreducible decomposition of ${\bf 3}\otimes{\bf 3}$ is given by%
\footnote{The Clebsch-Gordon (CG) coefficients of $T'$ in our notation are summarized in Appendix \ref{app:level_3}.}
\begin{align}
\begin{pmatrix}
Y^{(k)}_1\\
Y^{(k)}_2\\
Y^{(k)}_3\\
\end{pmatrix}_{\bf 3}
\otimes
\begin{pmatrix}
Q_1\\
Q_2\\
Q_3\\
\end{pmatrix}_{\bf 3}
= 
\left(
Y^{(k)}_{1} Q_1
+ Y^{(k)}_{2} Q_3
+ Y^{(k)}_{3} Q_2
\right)_{\bf 1} + \cdots.
\end{align}
If \im $\tau$ is large, the modular forms $Y^{(k)}_{1,2,3}$ are approximated by $q^{0,1/3,2/3}$ respectively (see \eqref{eq:app_Y3k}).
Then the superpotential is approximated for large \im $\tau$ as
\begin{align}
W_u^{tri}  \sim
\Big[&
\alpha_u \tilde{\phi}^3 ( Q_1  + q^{1/3} Q_3 + q^{2/3} Q_2) u_1^c + \beta_u \tilde \phi ( Q_1  + q^{1/3} Q_3 + q^{2/3} Q_2) u_2^c 
\nonumber
\\
& + \gamma_u ( Q_1  + q^{1/3} Q_3 + q^{2/3} Q_2) u_3^c
\Big]
H_u,
\nonumber
\\
~~
W_d^{tri} \sim \Big[&
\alpha_d \tilde{\phi}^2 ( Q_1  + q^{1/3} Q_3 + q^{2/3} Q_2) d_1^c + 
\beta_d \tilde \phi ( Q_1  + q^{1/3} Q_3 + q^{2/3} Q_2) d_2^c 
\nonumber
\\
&+\gamma_d \tilde\phi ( Q_1  + q^{1/3} Q_3 + q^{2/3} Q_2) d_3^c
\Big]
H_d,
\end{align}
and the mass matrix is approximated by
\begin{align}
M^{tri}_u \sim&
\begin{pmatrix}
 \tilde{\phi}^3 & \tilde{\phi}^1 & \tilde{\phi}^0\\
 \tilde{\phi}^3 q^{2/3}& \tilde{\phi}^1 q^{2/3} & \tilde{\phi}^0 q^{2/3}\\
 \tilde{\phi}^3 q^{1/3}& \tilde{\phi}^1 q^{1/3} & \tilde{\phi}^0 q^{1/3}
\end{pmatrix}
\frac{v_u}{\sqrt 2},~~
M^{tri}_{d} \sim
\begin{pmatrix}
\tilde{\phi}^2 & \tilde{\phi}^1 & \tilde{\phi}^1\\
\tilde{\phi}^2 q^{2/3} & \tilde{\phi}^1 q^{2/3} & \tilde{\phi}^1 q^{2/3}\\
\tilde{\phi}^2 q^{1/3} & \tilde{\phi}^1 q^{1/3} & \tilde{\phi}^1 q^{1/3}
\end{pmatrix}
\frac{v_d}{\sqrt 2}.
\end{align}
We can always exchange the indices of the quark fields freely. 
We redefine the left-handed quark fields $Q_i$  as
\begin{align}
Q_1 \rightarrow Q_3,~~Q_2 \rightarrow Q_1, ~~Q_3\rightarrow Q_2,
\end{align}
and the mass matrix is rewritten as
\begin{align}
M_u^{tri} \sim
\begin{pmatrix}
\tilde{\phi}^3 q^{2/3}& \tilde{\phi}^1 q^{2/3} & \tilde{\phi}^0 q^{2/3}\\
\tilde{\phi}^3 q^{1/3}& \tilde{\phi}^1 q^{1/3} & \tilde{\phi}^0 q^{1/3}\\
\tilde{\phi}^3 & \tilde{\phi}^1 & \tilde{\phi}^0
\end{pmatrix}\frac{v_u}{\sqrt 2},~~
M_{d}^{tri} \sim
\begin{pmatrix}
\tilde{\phi}^2 q^{2/3} & \tilde{\phi}^1 q^{2/3} & \tilde{\phi}^1 q^{2/3}\\
\tilde{\phi}^2 q^{1/3} & \tilde{\phi}^1 q^{1/3} & \tilde{\phi}^1 q^{1/3}\\
\tilde{\phi}^2 & \tilde{\phi}^1 & \tilde{\phi}^1
\end{pmatrix}\frac{v_d}{\sqrt 2}.
\label{eq:mass_mod_FN}
\end{align}
The determinants of these two matrices are proportional to $\tilde \phi^4 q$, 
and the largest eigenvalue is of order of $\tilde \phi^0 q^0$.
Thus we have a natural hierarchy.
It is the same as the FN mechanism \eqref{eq:mass_FN}.
If $q^{1/3} \sim \tilde \phi^{3/2}$, we obtain the FN-like mass matrix.
Thus we expect that these mass matrices reproduce the mass hierarchies and the mixing angles of the quarks.

Note that the right-handed quarks $u_i^c$ (and $d_i^c$) must be the same representation to realize FN-like mass matrix.
For example, suppose $u_2^c$ is assigned to ${\bf 1''}$, and $u_3^c$ to ${\bf 1'}$,
and the other right-handed quarks are the trivial singlet {\bf 1}.
In this case, the effective superpotential of the up sector is changed to
\begin{align}
W_u' &= 
\left[
\alpha_u \tilde{\phi}^3 (Y^{(2)}_{{\bf 3}} Q)_{{\bf 1}} u_1^c + 
\beta_u \tilde \phi (Y^{(4)}_{{\bf 3}} Q)_{\bf 1'} u_2^c +
\gamma_u ((Y^{(6)}_{{\bf 3},I} + r_u Y^{(6)}_{{\bf 3},II}) Q)_{\bf 1''} u_3^c
\right]
H_u.
\end{align}
The CG coefficients in terms of $({\bf 3}\times{\bf 3})_{\bf 1''}$,
and  $({\bf 3}\times{\bf 3})_{\bf 1'}$
are summarized in \eqref{eq:CG_level_3}.
The mass matrix is approximated by
\begin{align}
M_u' \sim
\begin{pmatrix}
\tilde{\phi}^3 & \tilde{\phi}^1 q^{1/3} & \tilde{\phi}^0 q^{2/3} \\
\tilde{\phi}^3 q^{2/3} & \tilde{\phi}^1 & \tilde{\phi}^0 q^{1/3} \\
\tilde{\phi}^3 q^{1/3}  & \tilde{\phi}^1 q^{2/3} & \tilde{\phi}^0
\end{pmatrix}\frac{v_u}{\sqrt 2}.
\label{eq:mass_OT_m}
\end{align}
The eigenvalues of \eqref{eq:mass_OT_m} are approximated by 
$\tilde \phi^3 v_u, \tilde \phi^1 v_u$, and $\tilde \phi^0 v_u$ 
in the limit of $\im \tau \rightarrow +\infty$.
It may reproduce the small values of the mixing angles since it is close to diagonal matrix.
Indeed, in the previous modular symmetric models, 
the mass matrices are the same form as \eqref{eq:mass_OT_m}.
In this case, however, the source of the mass hierarchy and that of the Cabbibo angles 
are independent each other~\cite{King:2020qaj}.
On the other hand in the case of the ``FN-like'' mass matrices in \eqref{eq:mass_mod_FN}
they are related each other through the moduli parameters and the singlet vev. 
As it will be discussed later, the empirical relations between mass ratios and the mixing angles in \eqref{eq:mixing_FN_model} 
can also be realized with $\mathcal{O}(1)$ coefficients.
For our purpose it is important to assume that the right-handed quarks are the same representation 
under modular symmetry.

To construct a mass matrix similar to \eqref{eq:mass_mod_FN}, 
we have a restriction on the modular weights too.
Since the components of the modular forms are aligned
in the same order at each row in our model,
the weight of the Yukawa couplings of each generation must be different to realize the full rank matrix.
The modular forms of weight higher than 5 are required at least.
The modular forms of modular weight higher than 5 is also required for the CP-phase.
The complex phases of the modular forms do not affect the CP-phase for large $\im \tau$, 
since the mass matrix is approximated by \eqref{eq:mass_mod_FN}, and the phase of $q$, i.e., $\re \tau$,
can be absorbed by field redefinition of $Q_i$.
The phase of coefficients $\alpha_f, \beta_f$ and $\gamma_f$, as well as that of $\phi$ are also absorbed by $u_i$ and $d_j$.
Hence 
$r_u$ and $r_d$ in \eqref{eq:superpotential_A4} 
are the only source of CP-violation in our model.%
\footnote{More generally, $\im \tau = + \infty$ is a point invariant under $\tau \rightarrow -\tau^*$, 
which is the generalized CP-transformation for the modular symmetry \cite{Novichkov:2019sqv, Kobayashi:2019uyt}.
Thus the superpotential must include explicit breaking term for large $\im \tau$.}
Such a CP-phase appears if $\mathcal{M}_k(\Gamma(3))$ has multiple triplets.
It is satisfied when the modular weight is higher than 5.

We also comment on the sub-leading terms.
Suppose that the leading Yukawa term is given by 
\begin{align}
W \supset \tilde \phi^{k} (Y^{(\ell)}_{\bf 3} Q)_{\bf 1} f^c_j H_f,~~~~f = u~{\rm or}~d,
\label{eq:not_leading}
\end{align}
then we have sub-leading couplings,
\begin{align}
\Delta W = \alpha_f' \tilde \phi^{k +2 } (Y_{\bf 3}^{(\ell+2)} Q)_{\bf 1} f^c_j H_f.
\end{align}
The sub-leading term is suppressed by $\tilde \phi^2$ compared to the leading term
because the weight of the modular forms which is {\bf 3} of $T'$ must be even and positive.
As shown in the following analysis, $\tilde \phi$ is of order $10^{-2}$ for the realistic models,
and we can omit the sub-leading terms.
This also implies that the smallest value of $k$ (the power of $\tilde \phi$ in the leading Yukawa term) 
should be lower than 2 in general. 
The only exception is the 
Yukawa term in which the modular weight of the Yukawa coupling is the lowest, 
that is, $\ell=2 $ in the case of $Y_{\bf 3}^{(\ell)}$.
In this case it is possible to assign an arbitrary positive power of $\tilde \phi$ at the leading order,  
since there is no modular form with lower modular weight $\ell < 2$ for $Y_{\bf 3}^{(\ell)}$ 
(See Table~\ref{tab:irrepT'}).

Finally, we obtain the general superpotential of our FN-like model with $\Gamma_3'$:
\begin{align}
W_u^{tri} &= 
\left[
\alpha_u \tilde \phi^{I} (Q Y^{(k_1)}_{\bf 3})_{\bf 1} u_1^c
+ \beta_u \tilde \phi^{J} (Q Y^{(k_2)}_{\bf 3})_{\bf 1} u_2^c
+\gamma_u \tilde \phi^{K} (Q Y^{(k_3)}_{\bf 3})_{\bf 1} u_3^c
\right]H_u,
\nonumber
\\
W_d^{tri} &= 
\left[
\alpha_d \tilde \phi^{L} (Q Y^{(\ell_1)}_{\bf 3})_{\bf 1} d_1^c 
+ \beta_d \tilde \phi^{M} (Q Y^{(\ell_2)}_{\bf 3})_{\bf 1} d_2^c
+\gamma_d \tilde \phi^{N} (Q Y^{(\ell_3)}_{\bf 3})_{\bf 1} d_3^c
\right]H_d,
\label{eq:general_sup_A4L3R1}
\end{align}
where $I, J , K, L, M, N$ and $k_i, \ell_i$ are integer numbers satisfying the following conditions:
\begin{align}
-I+ k_Q + k_1+ k_{u_1} &= 0,
\nonumber
\\
-J+ k_Q + k_2+ k_{u_2} &= 0,
\nonumber
\\
-K+ k_Q + k_3+ k_{u_3} &= 0,
\nonumber
\\
-L+ k_Q + \ell_1+ k_{d_1} &= 0,
\nonumber
\\
-M+ k_Q + \ell_2+ k_{d_2} &= 0,
\nonumber
\\
-N+ k_Q + \ell_3+ k_{d_3} &= 0.
\label{eq:index_G3_L3R1}
\end{align}
As mentioned above, if $k_i$ or $\ell_i$ is equal to 2, the corresponding capital index can be arbitrary integer number, otherwise the capital indices must be 0 or 1.
$(k_1,k_2,k_3) = (\ell_1,\ell_2,\ell_3) = (2,4,6)$ is the smallest weight full rank model, which has the smallest number of free parameters
because the dimension of $\mathcal{M}_{k}(\Gamma(3))$ monotonically increases as the weight $k$ increases.

\subsection{Models with the singlet left-handed quarks}

We construct a similar model by exchanging the representations of $Q_i$ and $(u_i^c, d_i^c)$.
Suppose that $Q_i$ are the trivial singlet of $T'$, and $u_i^c$ and $d_i^c$ form the triplet of $T'$,
then we obtain the following superpotential:
\begin{align}
W_u^{sing} &= 
\left[
\alpha_u \tilde \phi^{I} Q_1 (Y^{(k_1)}_{\bf 3}u^c)_{\bf 1}
+ \beta_u \tilde \phi^{J} Q_2(Y^{(k_2)}_{\bf 3}u^c)_{\bf 1}
+\gamma_u \tilde \phi^{K} Q_3 (Y^{(k_3)}_{\bf 3}u^c)_{\bf 1}
\right]H_u,
\nonumber
\\
W_d^{sing} &= 
\left[
\alpha_d \tilde \phi^{L} Q_1(Y^{(\ell_1)}_{\bf 3} d^c)_{\bf 1} 
+ \beta_d \tilde \phi^{M} Q_2(Y^{(\ell_2)}_{\bf 3}d^c)_{\bf 1}
+\gamma_d \tilde \phi^{N} Q_3(Y^{(\ell_3)}_{\bf 3}d^c)_{\bf 1}
\right]H_d,
\label{eq:general_sup_A4L1R3}
\end{align}
where $u^c = (u_1^c,u_2^c,u_3^c)^t$ and $d^c = (d_1^c,d_2^c,d_3^c)^t$, which form {\bf 3} of $T'$.
In this case, the powers of $\tilde \phi$ and the modular weights of the fields satisfy the following conditions,
\begin{align}
-I+ k_{Q_1} + k_1+ k_{u} &= 0,
\nonumber
\\
-J+ k_{Q_2} + k_2+ k_{u} &= 0,
\nonumber
\\
-K+ k_{Q_3} + k_3+ k_{u} &= 0,
\nonumber
\\
-L+ k_{Q_1} + \ell_1+ k_{d} &= 0,
\nonumber
\\
-M+ k_{Q_2} + \ell_2+ k_{d} &= 0,
\nonumber
\\
-N+ k_{Q_2} + \ell_3+ k_{d} &= 0.
\label{eq:constraints_G3}
\end{align}
The mass matrices are approximated as
\begin{align}
M_u^{sing} \sim
\begin{pmatrix}
\tilde{\phi}^I q^{2/3}& \tilde{\phi}^I q^{1/3}& \tilde{\phi}^I \\
\tilde{\phi}^J q^{2/3}& \tilde{\phi}^J q^{1/3}& \tilde{\phi}^J \\
\tilde{\phi}^K q^{2/3}& \tilde{\phi}^K q^{1/3}& \tilde{\phi}^K
\end{pmatrix}\frac{v_u}{\sqrt 2},~~
M_d^{sing} \sim
\begin{pmatrix}
\tilde{\phi}^L q^{2/3}& \tilde{\phi}^L q^{1/3}& \tilde{\phi}^L \\
\tilde{\phi}^M q^{2/3}& \tilde{\phi}^M q^{1/3}& \tilde{\phi}^M \\
\tilde{\phi}^N q^{2/3}& \tilde{\phi}^N q^{1/3}& \tilde{\phi}^N
\end{pmatrix}\frac{v_d}{\sqrt 2}.
\end{align}
We also obtain the hierarchical mass matrix.
In fact, this mass matrix is the transposed matrix of the previous one in \eqref{eq:mass_mod_FN}. 

$(k_1,k_2,k_3) = (\ell_1,\ell_2,\ell_3) = (2,4,6)$ is the model with the lowest weight modular forms.
In this case, the constraints \eqref{eq:constraints_G3} implies
\begin{align}
I - L  = J - M = K - N = -k_d + k_u.
\label{eq:pq_G3_L1R3}
\end{align}

\subsection{Numerical analysis of mass ratios and the mixing angles of $\Gamma_3'$ models}

The origin of the modular symmetry is the geometrical symmetry of the extra dimensions.
Hence we should evaluate the Yukawa couplings at the compactification scale. 
We assume that the compactification scale is the GUT scale
($2 \times 10^{16}$ GeV).
The Yukawa couplings at high energy scale receives quantum corrections, 
and they are given by solving the renormalization group equation.
They depend on the physics beyond the standard model.
In this paper, we assume a minimal SUSY breaking scenario with $\tan \beta =5$ \cite{Antusch:2013jca, Bjorkeroth:2015ora}.
At the GUT scale, the Yukawa couplings are calculated as
\begin{align}
y_u^{obs} &= (2.92\pm 1.81) \times 10^{-6},~~
y_c^{obs} = (1.43\pm 0.100) \times 10^{-3},~~
y_t^{obs} = 0.534 \pm 0.0341.
\nonumber
\\
y_d^{obs} &= (4.81 \pm 1.06) \times 10^{-6},~~
y_s^{obs} = (9.52 \pm 1.03) \times 10^{-5},~~
y_b^{obs} = (6.95 \pm 0.175) \times 10^{-3},
\nonumber
\end{align}
We explicitly show $1 \sigma$ interval for every observable.
In the following analysis, we concentrate on the ratios of the Yukawa couplings rather than the Yukawa couplings themselves,
since the overall factor is irrelevant to our study.
The ratios of the Yukawa couplings are calculated as
\begin{align}
y_u^{obs}/y_t^{obs} &= (5.47\pm3.41)\times10^{-6},~~
y_c^{obs}/y_t^{obs} = (2.68\pm0.254)\times 10^{-3},~~
\nonumber
\\
y_d^{obs}/y_b^{obs} &= (6.92\pm1.54)\times 10^{-4},~~
y_s^{obs}/y_b^{obs} = (1.37\pm0.152)\times 10^{-2},
\nonumber
\\
y_b^{obs}/y_t^{obs} &= (1.30\pm0.0893)\times 10^{-2}.
\label{eq:massratio}
\end{align}
Similarly, the mixing angles and $CP$-phase consistent with the experimental results at the GUT scale are given by
\begin{align}
\theta_{12}^{obs} & 
= 0.22736 \pm 0.00142,
~~
\theta_{23}^{obs} 
= 0.03585 \pm 0.00670,
\nonumber
\\
\theta_{13}^{obs} &
= 0.003145 \pm 0.000490,
~~
\delta_{CP}^{obs} 
= 1.206 \pm 0.108.
\nonumber
\end{align}
Our notation of the mixing angles and the CP-phase is based on the PDG \cite{Zyla:2020zbs}.
The quark sector has 9 observables to fit.

In this section, we analyze the mass hierarchy and the mixing angles of our FN-like model.
The superpotential with $\Gamma_3'$ are summarized in \eqref{eq:general_sup_A4L3R1} and \eqref{eq:general_sup_A4L1R3}.
We consider the superpotential with the lowest weight modular forms. 

\subsubsection*{Model with triplet left-handed quarks}

First we consider the FN-like model based on the superpotential of \eqref{eq:general_sup_A4L3R1}.
Before investigating the numerical analysis, we should study the structure of the mass matrix analytically.
The physical mass matrix is given in terms of the canonically normalized Yukawa couplings in \eqref{eq:canonicalf} as 
\begin{align}
M_u =& 
\begin{pmatrix}
\alpha_u \tilde{\phi}^I Y_3^{(2)} & \beta_u \tilde{\phi}^J Y_3^{(4)} & 
\gamma_u \tilde{\phi}^K (Y_{3,I}^{(6)} + r_u Y_{3,II}^{(6)} )\\
\alpha_u \tilde{\phi}^I Y_2^{(2)} & \beta_u \tilde{\phi}^J Y_2^{(4)} & 
\gamma_u \tilde{\phi}^K (Y_{2,I}^{(6)} + r_u Y_{2,II}^{(6)} )\\
\alpha_u \tilde{\phi}^I Y_1^{(2)} & \beta_u \tilde{\phi}^J Y_1^{(4)} & 
\gamma_u \tilde{\phi}^K (Y_{1,I}^{(6)} + r_u Y_{1,II}^{(6)} )\\
\end{pmatrix}
\begin{pmatrix}
\im \tau^1 & & \\
 & \im \tau^2 & \\
& & \im \tau^3
\end{pmatrix}
\frac{v_u}{\sqrt 2},
\nonumber
\\
M_d =&
\begin{pmatrix}
\alpha_d \tilde{\phi}^L Y_3^{(2)} & 
\beta_d \tilde \phi^M (Y_{3,I}^{(6)} + r_d Y_{3,II}^{(6)} ) & \gamma_d \tilde \phi^N Y_3^{(4)}\\
\alpha_d \tilde{\phi}^L Y_2^{(2)}  & 
\beta_d \tilde \phi^M (Y_{2,I}^{(6)} + r_d Y_{2,II}^{(6)} ) & \gamma_d \tilde \phi^N Y_2^{(4)}\\
\alpha_d \tilde{\phi}^L Y_1^{(2)} &
\beta_d \tilde \phi^M (Y_{1,I}^{(6)} + r_d Y_{1,II}^{(6)} ) & \gamma_d  \tilde \phi^N Y_1^{(4)}\\
\end{pmatrix}
\begin{pmatrix}
\im \tau^1 & & \\
 & \im \tau^3 & \\
& & \im \tau^2
\end{pmatrix}
\frac{v_d}{\sqrt 2},
\label{eq:mass_G3_3_1}
\end{align}
where we note that in order to clearly see a FN-like hierarchy we redefine the indices of the quark fields as 
\begin{align}
Q_1 \rightarrow Q_3,~~Q_2 \rightarrow Q_1, ~~Q_3\rightarrow Q_2,~~ 
d_2 \leftrightarrow d_3, 
\end{align}
where we implicitly assume the following conditions\footnote{We can consider other possibilities with different powers. 
We find the model in \eqref{eq:mass_G3_3_1} with the conditions \eqref{eq:A4conditinos} the best in our numerical analysis.}
\begin{align}
I \geq J \geq K, \quad & \quad L \geq M \geq N.
\label{eq:A4conditinos}
\end{align}
Thus the largest Yukawa coupling for the down sector would be $\gamma_d \tilde \phi^N Y_1^{(4)}$,
while that for the up sector $\gamma_u \tilde \phi^K (Y^{(6)}_{1,I} + r_u Y^{(6)}_{1,II})$.
Using explicit $q$-expansions in Appendix \ref{app:level_3}
we obtain an approximate estimation of the mixing angles for large $\im \tau$ as
\begin{align}
\theta_{12} &
\sim 
	\left|
	\frac{Y^{(4)}_3}{Y^{(4)}_2} - \frac{Y^{(6)}_{3, I} + r_d Y^{(6)}_{3, II} }{Y^{(6)}_{2, I} + r_d Y^{(6)}_{2, II} }
	\right|
\sim 
	15 
	\left|
	\frac{r_d + \frac 1 5 }{r_d + \frac 1 2} 
	q^{1/3}
	\right|,
\nonumber
\\
\theta_{23} &
\sim 
	\left|
	\frac{Y^{(6)}_{2, I} + r_u Y^{(6)}_{2, II} }{Y^{(6)}_{1, I} + r_u Y^{(6)}_{1, II}}  -\frac{Y^{(4)}_2}{Y^{(4)}_1}
	\right|
\sim 
	12 \left|
	(1 + r_u)q^{1/3}
	\right| 
,~~
\nonumber
\\
\theta_{13} &
\sim 
	\left|
	\frac{Y^{(6)}_{3, I} + r_u Y^{(6)}_{3, II} }{Y^{(6)}_{1, I} + r_u Y^{(6)}_{1, II} }
	- \frac{Y^{(4)}_{3}}{Y^{(4)}_{1}}
	\right| 
\sim 72 \left|
	(1-r_u) q^{2/3}
	\right|.
\label{eq:mixing_A4_31}
\end{align}
We see that the approximate mixing angles do not depend on $\tilde \phi$, $\alpha_{u,d}, \beta_{u,d}$, and $\gamma_{u,d}$.
The first relation implies $15 |q^{1/3}| \sim \theta_{C} \sim 10^{-1}$, 
which is satisfied when $\im \tau \sim 2.4$.
The other two conditions are rewritten as
\begin{align}
\theta_{23} & \sim
\frac{4 |1+ r_u |}{5} \theta_C \sim  8 |1+ r_u| \times 10^{-2},
\nonumber
\\
\theta_{13} & \sim 
\frac{8 |1- r_u|}{25} \theta_C^2 \sim 3 |1 - r_u| \times 10^{-3}.
\nonumber
\end{align}
Then we can realize the observed values for $r_u=\mathcal{O}(1)$. 
We also obtain a natural hierarchical structure for the mass ratios
which are suppressed by powers of $\theta_{C}$ and $\tilde \phi$ as 
\begin{align}
y_u/y_t &\sim
\frac{162}{25} \im \tau^{-2}  \frac{\alpha_u}{\gamma_u} \tilde \phi^{I-K} \theta_{C}^{2}
\sim 1 \times10^{-2} \frac{\alpha_u}{\gamma_u} \tilde \phi^{I-K} 
,
\nonumber
\\
y_c/y_t &\sim
\frac{18}{5} \im \tau^{-1}  \frac{\beta_u}{\gamma_u} \tilde \phi^{J-K} \theta_{C}
\sim 2 \times 10^{-1} \frac{\beta_u}{\gamma_u} \tilde \phi^{J-K} 
,
\label{eq:cond_up_3}
\end{align}
for the up sector, and 
\begin{align}
y_d/y_b  
& \sim \frac{162}{225} \im \tau^{-1} \frac{\alpha_d}{\gamma_d} \tilde \phi^{L - N} \theta_{C}^{2} 
\sim 3 \times 10^{-3} \frac{\alpha_d}{\gamma_d} \tilde \phi^{L - N}  
,
\nonumber
\\
y_s/y_b & \sim \frac89 \frac{|r_d +\frac12| \beta_d}{\gamma_d} \tilde \phi^{M - N} \theta_{C}
\sim 2 \times 10^{-2} \frac{|r_d +\frac12| \beta_d}{\gamma_d} \tilde \phi^{M - N} 
,
~~
\nonumber
\\
y_b/y_t &\sim 
9 \im \tau^{-1} 
\frac{\gamma_d}{\gamma_u}
\tilde \phi^{N - K}
\sim 
4 \frac{\gamma_d}{\gamma_u}
\tilde \phi^{N - K}
,
\label{eq:app_cod_G3_real}
\end{align}
for the down sector, where we use $\im \tau \sim 2.4$.
We solve these equations under the conditions \eqref{eq:A4conditinos}.
For simplicity we assume $\tilde \phi = 10^{-2}$ for the later estimation. 
We then find a solution of $I = 2, J=1, K =0, M=1, N = 1$, and $L=1$ or 2,  
which can reproduce the observed mass ratios in \eqref{eq:massratio} 
with $\mathcal{O}(1)$ coefficients of $\frac{\alpha_{u,d}}{\gamma_{u,d}}, 
\frac{\beta_{u,d}}{\gamma_{u,d}}, 
\frac{\gamma_d}{\gamma_u}$, and $r_{u,d}$.

To confirm our analysis we construct an explicit model which satisfies the above conditions.
The representations and modular weights of the quark fields are summarized in Table \ref{tab:model1_G3}.
The mass matrix is given by \eqref{eq:mass_G3_3_1} with $I=2, J =1, K=0,$ and $L=2, M=N=1$.
\begin{table}[t]
\begin{center}
\begin{tabular}{|c|cccccccccc|} \hline
 & $Q_{1,2,3}$ 
 & $u_1^c$ & $u_2^c$ & $u_2^c$ 
 & $d_1^c$ & $d_2^c$ & $d_3^c$ 
 & $H_u$ & $H_d$ & $\phi$ 
\\
\hline \hline
$\Gamma_3' = T'$ & ${\bf 3}$ 
& ${\bf 1}$ & ${\bf 1}$ & ${\bf 1}$ 
& ${\bf 1}$ & ${\bf 1}$ & ${\bf 1}$ 
& ${\bf 1}$ & ${\bf 1}$ & ${\bf 1}$
\\
$k_I$ & $-2$ 
& $2$ & $-1$ & $-4$ 
& $2$ & $-3$ & $-1$
& $0$ & $0$ & $-1$
\\
\hline
\end{tabular}
\caption{The representations and the modular weights of the chiral superfields.}
\label{tab:model1_G3}
\end{center}
\end{table}
We set $\re \tau =0$ since the complex phase factor in the modular forms is negligibly small for large $\im \tau$.
We also assume absolute value of $r_f$ is 1 at first.
Thus we have 8 free parameters:
$\alpha_u/\gamma_u$, $\beta_u/\gamma_u$, $r_u$, 
$\alpha_d/\gamma_d$, $\beta_d/\gamma_d$, $r_d$, $\im \tau$ and $\gamma_d/\gamma_u$.
$\tilde \phi$ is not counted as a d.o.f since it is absorbed by the coefficients.
The best fit parameters in our search are given by
\begin{align}
\alpha_u/\gamma_u =& 0.1904,~~
\beta_u/\gamma_u = 2.427,~~
r_u = e^{-2.764i},
\nonumber
\\
\alpha_d/\gamma_d =& 4.946,~~
\beta_d/\gamma_d = 0.5614,~~
r_d = e^{2.462 i},
\nonumber
\\
\gamma_d/\gamma_u =& 0.3400,~~
\tau = 2.406 i,~~
\tilde \phi = 1 \times 10^{-2}.
\label{eq:bench_G4_model1}
\end{align}
The most hierarchical parameter is $\beta_u/\alpha_u = 12.7$ in this parameter set.
The FN-like mass matrices are successfully obtained as,
\begin{align}
|M_u| =&
\gamma_u
\begin{pmatrix}
3.846 \times 10^{-9} & 3.932 \times 10^{-6} & 4.464 \times 10^{-5}\\
1.979 \times 10^{-7} & 6.743 \times 10^{-5} & 2.193 \times 10^{-3}\\
5.090 \times 10^{-6} & 1.734 \times 10^{-3} & 1.911 \times 10^{-2}\\
\end{pmatrix}
\frac{v_u}{\sqrt{2}},
\nonumber
\\
|M_d| =&
\gamma_d
\begin{pmatrix}
9.991 \times 10^{-8} & 2.661 \times 10^{-7} & 1.620 \times 10^{-6}\\
5.140 \times 10^{-6} & 1.188 \times 10^{-5} & 2.778 \times 10^{-5}\\
1.322 \times 10^{-4} & 1.073 \times 10^{-4} & 7.147 \times 10^{-4}\\
\end{pmatrix}
\frac{v_d}{\sqrt{2}}.
\end{align}
As the result, we obtain the following mass eigenvalues and mixing angles,
\begin{align}
y_u/y_t &=4.19 \times 10^{-6},~~
y_c/y_t = 2.67 \times 10^{-3},~~
y_d/y_b = 7.69 \times10^{-4},~~
y_s/y_b = 1.56 \times 10^{-2}, 
\nonumber
\\
y_b/y_t &= 1.30 \times 10^{-2},~~
\theta_{12} =  0.227,~~
\theta_{23} =  0.0300,~~
\theta_{13} =  0.00074,~~
\delta_{CP} = 1.20.
\end{align}
In this case, though we have only 8 free parameters, all mixing angles and mass eigenvalues 
except for the mixing angles $\theta_{13}$ are reproduced within $2 \sigma$ range of the observed values.
In this case, $\chi^2$, which is given by
\begin{align}
\chi^2 = \sum_{x \in Observables} 
\left(\frac{x - x^{obs}}{\sigma_x}\right)^2,
\end{align}
where $Observables = \{y_u/y_t, y_c/y_t, y_d/y_b, y_s/y_b, y_b/y_t, \sigma_{12}, \sigma_{23}, \sigma_{13}, \delta_{CP} \}$, is estimated as $\chi^2 \sim 27$.

If we relax the restriction on $|r|$, we can realize the observed values more precisely, 
while the number of free parameters is more than observed values.
A benchmark value is obtained as
\begin{align}
\alpha_u/\gamma_u =& 0.3320,~~
\beta_u/\gamma_u = 2.159,~~
r_u = 0.8888 e^{-2.649i},
\nonumber
\\
\alpha_d/\gamma_d =& 2.976,~~
\beta_d/\gamma_d = 3.665,~~
r_d = 0.6941 e^{-3.375 i},
\nonumber
\\
\gamma_d/\gamma_u =& 0.2509,~~
\tau = 2.448 i,~~
\tilde \phi = 1 \times 10^{-2}.
\nonumber
\end{align}
The most hierarchical parameter is $\beta_u/\gamma_d = 8.61$ in this parameter set, 
and all the coefficients can be the same order.
The mass matrix is given by
\begin{align}
|M_u| =&
\gamma_u
\begin{pmatrix}
5.720 \times10^{-9} & 3.035 \times 10^{-6} & 5.688 \times10^{-5}\\
3.214 \times 10^{-7} & 5.685 \times 10^{-5} & 7.260 \times 10^{-4} \\
9.030 \times 10^{-6} & 1.597 \times 10^{-3} &  2.013 \times 10^{-2}\\
\end{pmatrix} \frac{v_u}{\sqrt{2}},
\nonumber
\\
|M_d| =&
\gamma_d
\begin{pmatrix}
5.127 \times 10^{-8} & 1.755 \times 10^{-6} & 1.406 \times 10^{-6}\\
2.881 \times 10^{-6} & 1.248 \times 10^{-5} & 2.634 \times 10^{-5}\\
8.095 \times 10^{-5} & 7.376 \times 10^{-4} & 7.399 \times 10^{-4}\\
\end{pmatrix}
\frac{v_d}{\sqrt{2}}.
\end{align}
We obtain the following mass eigenvalues and mixing angles 
\begin{align}
y_u/y_t &=4.22 \times 10^{-6},~~
y_c/y_t = 2.68 \times 10^{-3},~~
y_d/y_b = 6.96 \times10^{-4},~~
y_s/y_b = 1.38 \times 10^{-2}, 
\nonumber
\\
y_b/y_t &= 1.30 \times 10^{-2},~~
\theta_{12} =  0.227,~~
\theta_{23} =  0.0355,~~
\theta_{13} =  0.00314,~~
\delta_{CP} = 1.21.
\nonumber
\end{align}
In this case, all the observables are within $0.4\sigma$ range, and $\chi^2 \sim 0.1$. 
Hence we can realize the realistic values without hierarchical parameters.

\subsubsection*{Model with singlet left-handed quarks}

Here we consider a model with singlet left-handed quarks based on the superpotential of \eqref{eq:general_sup_A4L1R3}.
Changing the flavor indices of the right-handed quarks for both up and down sectors as
$f_1^c\rightarrow f_3^c, f_3^c \rightarrow f_2^c$,
and $f_2^c\rightarrow f_1^c$, we obtain
\begin{align}
M_u =& 
\begin{pmatrix}
\im \tau^1 & & \\
 & \im \tau^2 & \\
& & \im \tau^3
\end{pmatrix}
\begin{pmatrix}
\alpha_u \tilde{\phi}^I Y_3^{(2)} & \beta_u \tilde{\phi}^J Y_3^{(4)} & 
\gamma_u \tilde{\phi}^K (Y_{3,I}^{(6)} + r_u Y_{3,II}^{(6)} )\\
\alpha_u \tilde{\phi}^I Y_2^{(2)} & \beta_u \tilde{\phi}^J Y_2^{(4)} & 
\gamma_u \tilde{\phi}^K (Y_{2,I}^{(6)} + r_u Y_{2,II}^{(6)} )\\
\alpha_u \tilde{\phi}^I Y_1^{(2)} & \beta_u \tilde{\phi}^J Y_1^{(4)} & 
\gamma_u \tilde{\phi}^K (Y_{1,I}^{(6)} + r_u Y_{1,II}^{(6)} )\\
\end{pmatrix}^t
\frac{v_u}{\sqrt 2},
\nonumber
\\
M_d =&
\begin{pmatrix}
\im \tau^1 & & \\
 & \im \tau^2 & \\
& & \im \tau^3
\end{pmatrix}
\begin{pmatrix}
\alpha_d \tilde{\phi}^L Y_3^{(2)} & \beta_d \tilde \phi^M Y_3^{(4)} & 
\gamma_d \tilde \phi^N (Y_{3,I}^{(6)} + r_d Y_{3,II}^{(6)} )\\
\alpha_d \tilde{\phi}^L Y_2^{(2)} & \beta_d \tilde \phi^M Y_2^{(4)} & 
\gamma_d \tilde \phi^N (Y_{2,I}^{(6)} + r_d Y_{2,II}^{(6)} )\\
\alpha_d \tilde{\phi}^L Y_1^{(2)} & \beta_d  \tilde \phi^M Y_1^{(4)} & 
\gamma_d  \tilde \phi^N (Y_{1,I}^{(6)} + r_d Y_{1,II}^{(6)} )\\
\end{pmatrix}^t
\frac{v_d}{\sqrt 2},
\label{eq:Mass_G3_2}
\end{align}
which is the transposed matrix of \eqref{eq:mass_G3_3_1}.
In the case of the lowest weight modular forms, 
we have additional conditions of \eqref{eq:pq_G3_L1R3}, 
which implies  $I- J = L- M$ and $J - K = M- N$. 
Thus the same powers of $\tilde \phi$ arise in the mass ratios for both the up and down sectors.
In fact, an approximate expression of the mass ratios are given by
\begin{align}
y_u/y_t &\sim 2\left(\frac 1 3 \right)^{-6} \im \tau^{-2}  \frac{\alpha_u}{\gamma_u} \tilde \phi^{I-K} q^{2/3}
,
\nonumber
\\
y_c/y_t &\sim 
2 \left(\frac 1 3 \right)^{-3} \im \tau^{-1}  \frac{\beta_u}{\gamma_u} \tilde \phi^{J-K} q^{1/3}
,
\nonumber
\\
y_d/y_b &\sim 2\left(\frac 1 3 \right)^{-6} \im \tau^{-2}  \frac{\alpha_d}{\gamma_d} \tilde \phi^{I-K} q^{2/3}
,
~~
\nonumber
\\
y_s/y_b &\sim
2 \left(\frac 1 3 \right)^{-3} \im \tau^{-1}  \frac{\beta_d}{\gamma_d} \tilde \phi^{J-K} q^{1/3}
,
~~
\nonumber
\\
y_b/y_t &\sim \frac{\gamma_d}{\gamma_u} 
.
\end{align}
Therefore unnatural hierarchical coefficients are inevitable to obtain the realistic mass hierarchy.

\begin{table}[t]
\begin{center}
\begin{tabular}{|c|cccccccc|} \hline
 & $Q_{1}$  & $Q_{2}$  & $Q_{3}$ 
 & $u_{1,2,3}^c$  
 & $d_{1,2,3}^c$ 
 & $H_u$ & $H_d$ & $\phi$ 
\\
\hline \hline
$\Gamma_3' = T'$ 
& ${\bf 1}$ & ${\bf 1}$ & ${\bf 1}$ 
& ${\bf 3}$ 
& ${\bf 3}$ 
& ${\bf 1}$ & ${\bf 1}$ & ${\bf 1}$
\\
$k_I$ 
& $0$ & $-3$ & $-6$ 
& $0$  
& $0$ 
& $0$ & $0$ & $-1$
\\
\hline
\end{tabular}
\caption{The representation of the quark and the Higgs fields.}
\label{tab:model2_G3}
\end{center}
\end{table}
We show an explicit model.
The modular weights and representations for the best fit model are summarized in Table \ref{tab:model2_G3}.
The mass matrices are given by \eqref{eq:Mass_G3_2} with $I = L = 2, J = M = 1, K = N =0$.
The best fit values are given by
\begin{align}
\alpha_u/\gamma_u = & 7.851 \times 10^{-3},~~
\beta_u/\gamma_u = 0.1294,~~
r_u = 1.203 e^{-2.369i},
\nonumber
\\
\alpha_d/\gamma_d =& 1.967,~~
\beta_d/\gamma_d = 0.4558,~~
r_d = 2.398 e^{-4.534 i},
\nonumber
\\
\gamma_d/\gamma_u =& 9.617 \times 10^{-3}~~
\tau = 1.573 i,~~
\tilde \phi = 1 \times 10^{-2},
\nonumber
\end{align}
where hierarchical coefficients are required.
We obtain the mass matrix,
\begin{align}
|M_u| =&
\gamma_u
\begin{pmatrix}
3.395 \times10^{-9} & 3.053 \times 10^{-8} & 1.373 \times10^{-7}\\
2.935 \times 10^{-6} & 8.824 \times 10^{-6} & 3.936 \times 10^{-5} \\
7.375 \times 10^{-4} & 2.166 \times 10^{-3} &  5.359 \times 10^{-3}\\
\end{pmatrix} \frac{v_u}{\sqrt{2}},
\nonumber
\\
|M_d| =&
\gamma_d
\begin{pmatrix}
8.507 \times 10^{-7} & 7.651 \times 10^{-6} & 3.441 \times 10^{-5} \\
1.034 \times 10^{-5} & 3.110 \times 10^{-5} & 1.387 \times 10^{-4}\\
1.298 \times 10^{-3} & 5.611 \times 10^{-3} & 5.387 \times 10^{-3}\\
\end{pmatrix} \frac{v_d}{\sqrt{2}}.
\end{align}
The mixing angles and the mass ratios are calculated as
\begin{align}
y_u/y_t &=5.74 \times 10^{-6},~~
y_c/y_t = 2.70 \times 10^{-3},~~
y_d/y_b =7.97 \times10^{-4},~~
y_s/y_b = 1.38 \times 10^{-2}, 
\nonumber
\\
y_b/y_t &= 1.30 \times 10^{-2},~~
\theta_{12} =  0.227,~~
\theta_{23} =  0.00612,~~
\theta_{13} =  0.00306,~~
\delta_{CP} = 1.11,
\nonumber
\end{align}
where $\chi^2 \sim 20$. 
Even though we relax the constraints on $|r|$, we can not realize the observed values in this model.

As shown above, 
in the case of the lowest modular weight forms, 
hierarchical parameters should be required 
even if we consider other modular groups of higher levels.  
Thus singlet left-handed quark model is not suitable for our purpose.
Hereafter we only consider the models where the left-handed quarks form a triplet of the modular group.

\section{Froggatt-Nielsen like mechanism with the modular groups of higher levels.}

It is straightforward to generalize this mechanism to the modular group of the other levels.
The only requirement is the existence of triplet modular forms which have hierarchical components.
This condition is satisfied for the modular group of level $N \geq 3$.

\subsection{FN-like mechanism with the modular group of level 4}

The algebra of $\Gamma_4'$ is summarized in Appendix \ref{app:level_4}.
$\Gamma_4'$ is isomorphic to $S_4'\simeq SL(2,\mathbb{Z}_4)$.
It has four triplet representations, ${\bf 3},$ $\hat{\bf 3}$, ${\bf 3'}$ and $\hat {\bf 3}'$.
The matrix representations of $T$ in this algebra are summarized in Table \ref{tab:irrep_S4'} in Appendix \ref{app:level_4}.
The triplet modular forms are approximated as
\begin{align}
Y^{(k)}_{\bf 3} \sim
\begin{pmatrix}
q^{1/2}\\
q^{3/4} \\
q^{1/4}
\end{pmatrix},
~~
Y^{(k)}_{\hat{\bf 3}} \sim
\begin{pmatrix}
q^{1/4}\\
q^{2/4} \\
1
\end{pmatrix},
~~
Y^{(k)}_{{\bf 3}'} \sim
\begin{pmatrix}
1 \\
q^{1/4} \\
q^{3/4}
\end{pmatrix},
~~
Y^{(k)}_{\hat{\bf 3}'} \sim
\begin{pmatrix}
q^{3/4}\\
1 \\
q^{2/4}
\end{pmatrix},
\label{eq:MF_order_lev4}
\end{align}
for large \im $\tau$.
Thus the FN-like hierarchical mass matrix can be realized in the similar way.
We consider four classes of the modular invariant superpotentials:
\begin{align}
W_1 &= 
\left[
\alpha_f \tilde \phi^{I} (Q Y^{(k_1)}_{\bf 3})_{\bf 1} f_1^c
+ \beta_f \tilde \phi^{J} (Q Y^{(k_2)}_{\bf 3})_{\bf 1} f_2^c
+\gamma_f \tilde \phi^{K} (Q Y^{(k_3)}_{\bf 3})_{\bf 1} f_3^c
\right]H_f,
\nonumber
\\
W_{2} &= 
\left[
\alpha_f \tilde \phi^{I} (Q Y^{(k_1)}_{\bf \hat{3}})_{\bf  1} f_1^c
+ \beta_f \tilde \phi^{J} (Q Y^{(k_2)}_{\bf \hat{3}})_{\bf  1} f_2^c
+\gamma_f \tilde \phi^{K} (Q Y^{(k_3)}_{\bf \hat{3}})_{\bf 1} f_3^c
\right]H_f,
\nonumber
\\
W_{3} &= 
\left[
\alpha_f \tilde \phi^{I} (Q Y^{(k_1)}_{\bf 3'})_{\bf 1} f_1^c
+ \beta_f \tilde \phi^{J} (Q Y^{(k_2)}_{\bf 3'})_{\bf 1} f_2^c
+\gamma_f \tilde \phi^{K} (Q Y^{(k_3)}_{\bf 3'})_{\bf 1} f_3^c
\right]H_f,
\nonumber
\\
W_{4} &= 
\left[
\alpha_f \tilde \phi^{I} (Q Y^{(k_1)}_{\bf \hat{3}'})_{\bf 1} f_1^c
+ \beta_f \tilde \phi^{J} (Q Y^{(k_2)}_{\bf \hat{3}'})_{\bf 1} f_2^c
+\gamma_f \tilde \phi^{K} (Q Y^{(k_3)}_{\bf \hat{3}'})_{\bf 1} f_3^c
\right]H_f.
\end{align}
$f$ denotes the flavor $u$ or $d$, and $\phi$ is the trivial singlet carrying modular weight $-1$, i.e., weighton.
We can assign $f_i^c$ to the trivial singlet ${\bf 3}$ of $\Gamma_4'$ without loss of generality.
$Q_i^c$ is assigned to ${\bf 3}$ for $W_i$, ${\bf \hat 3'}$ for $W_2$, 
${\bf 3'}$ for $W_3$,  and
${\bf \hat 3}$ for $W_4$. 
The CG coefficients of the tensor products of the triplets are given by \eqref{eq:CG_level4_1}.
We obtain the mass matrix,
\begin{align}
M_{1,2,3,4} \sim
\begin{pmatrix}
\tilde{\phi}^I Y_1^{(k_1)} & \tilde{\phi}^J Y_1^{(k_2)} & \tilde{\phi}^K Y_1^{(k_3)}\\
\tilde{\phi}^I Y_3^{(k_1)} & \tilde{\phi}^J Y_3^{(k_2)} & \tilde{\phi}^K Y_3^{(k_3)}\\
\tilde{\phi}^I Y_2^{(k_1)} & \tilde{\phi}^J Y_2^{(k_2)} & \tilde{\phi}^K Y_2^{(k_3)}
\end{pmatrix}\frac{v_f}{\sqrt 2},
\label{eq:mass_lev4}
\end{align}
where $Y_i^{(k_j)}$ is the $i$-th component of the corresponding triplet modular form of weight $k_j$.
The mass matrix is approximated as
\begin{align}
M_{1} \sim 
\begin{pmatrix}
\tilde{\phi}^I q^{1/2} & \tilde{\phi}^J q^{1/2} & \tilde{\phi}^K q^{1/2}\\
\tilde{\phi}^I q^{1/4} & \tilde{\phi}^J q^{1/4} & \tilde{\phi}^K q^{1/4}\\
\tilde{\phi}^I q^{3/4} & \tilde{\phi}^J q^{3/4} & \tilde{\phi}^K q^{3/4}
\end{pmatrix}\frac{v_f}{\sqrt 2},
~~
M_{2} \sim
\begin{pmatrix}
\tilde{\phi}^I q^{1/4} & \tilde{\phi}^J q^{1/4} & \tilde{\phi}^K q^{1/4}\\
\tilde{\phi}^I  & \tilde{\phi}^J & \tilde{\phi}^K \\
\tilde{\phi}^I q^{1/2} & \tilde{\phi}^J q^{1/2} & \tilde{\phi}^K q^{1/2}
\end{pmatrix}\frac{v_f}{\sqrt 2},
\nonumber
\\
M_{3} \sim
\begin{pmatrix}
\tilde{\phi}^I  & \tilde{\phi}^J & \tilde{\phi}^K \\
\tilde{\phi}^I q^{3/4} & \tilde{\phi}^J q^{3/4} & \tilde{\phi}^K q^{3/4}\\
\tilde{\phi}^I q^{1/4} & \tilde{\phi}^J q^{1/4} & \tilde{\phi}^K q^{1/4}
\end{pmatrix}\frac{v_f}{\sqrt 2},
~~
M_{4} \sim
\begin{pmatrix}
\tilde{\phi}^I q^{3/4} & \tilde{\phi}^J q^{3/4} & \tilde{\phi}^K q^{3/4} \\
\tilde{\phi}^I  q^{1/2} & \tilde{\phi}^J q^{1/2} & \tilde{\phi}^K q^{1/2} \\
\tilde{\phi}^I  & \tilde{\phi}^J & \tilde{\phi}^K
\end{pmatrix}\frac{v_f}{\sqrt 2},
\label{eq:mass_G4_app}
\end{align}
for large $\im \tau$.
We obtain the FN-like mass matrix and hierarchical mass eigenvalues.

We can choose the superpotential of the up sector and that of the down sector from $W_{1}$, $W_{2}$, $W_{3}$ and $W_{4}$ individually.
Thus we have 16 classes of FN-like models with $\Gamma_4'$ in principal.
However, we find that it is difficult to obtain the observed mixing angles if we use a different type of the superpotential for each sector. 
As shown in \eqref{eq:MF_order_lev4} the position of the largest component is different for each representation.
If we use a different representation for each sector, 
we can not obtain FN-like Yukawa matrices for both sectors simultaneously, 
where the order of the contribution to a mixing angle from each sector could be different, 
and one of the mixing angles may become large.%
\footnote{
More precisely, ${\bf 3}$ and ${\bf \hat 3}$ have the same order 
with different numerical factors in $q$-expansion (See Appendix~\ref{app:level_4}), 
so that we could obtain the FN-like Yukawa matrices for a model using ${\bf 3}$ and ${\bf \hat 3}$.
We find, however, that this model is not realistic.
In fact, one can check that the modular forms of ${\bf \hat 3}$ of weight $k$ and 
the modular forms of of {\bf 3} of weight $k+3$ are 
linearly dependent for $k=1, 3, 5$.
}

We have the same constraints on the power of $\tilde \phi$ as those in $\Gamma_3'$. 
Namely, the powers of $\tilde \phi$ are 0 or 1 in general, 
but they can be arbitrary positive integer 
if the corresponding Yukawa coupling is the lowest weight modular form.
The lowest weight is 1 for ${\bf \hat 3}$, 2 for ${\bf 3'}$, 3 for ${\bf \hat 3'}$ and 4 for ${\bf 3}$.

\subsection*{Lowest weight models}

The superpotential $W_i$ has several free parameters.
They are proportional to the number of the triplet modular forms in the superpotential.
In order to minimize the number of the free parameters 
we consider the model with the lowest weight modular forms.
They are given by
\begin{align}
W_1 &= 
\left[
\alpha_f \tilde \phi^{I} (Q Y^{(4)}_{\bf 3})_{\bf 1} f_1^c
+ \beta_f \tilde \phi^{J} (Q Y^{(6)}_{\bf 3})_{\bf 1} f_2^c
+\gamma_f \tilde \phi^{K} (Q (Y^{(8)}_{{\bf 3},I} + r_f Y^{(8)}_{{\bf 3},II} ))_{\bf 1} f_3^c
\right]H_f,
\nonumber
\\
W_{2} &= 
\left[
\alpha_f \tilde \phi^{I} (Q Y^{(1)}_{\bf \hat{3}})_{\bf 1} f_1^c
+ \beta_f \tilde \phi^{J} (Q Y^{3)}_{\bf \hat{3}})_{\bf 1} f_2^c
+\gamma_f \tilde \phi^{K} (Q (Y^{(5)}_{{\bf \hat{3}},I} + r_f Y^{(5)}_{{\bf \hat{3}},II} ))_{\bf 1} f_3^c
\right]H_f,
\nonumber
\\
W_{3} &= 
\left[
\alpha_f \tilde \phi^{I} (Q Y^{(2)}_{\bf 3'})_{\bf 1} f_1^c
+ \beta_f \tilde \phi^{J} (Q Y^{(4)}_{\bf 3'})_{\bf 1} f_2^c
+\gamma_f \tilde \phi^{K} (Q (Y^{(6)}_{{\bf 3'},I} + r_f Y^{(6)}_{{\bf 3'},II})_{\bf 1} f_3^c
\right]H_f,
\nonumber
\\
W_{4} &= 
\left[
\alpha_f \tilde \phi^{I} (Q Y^{(3)}_{\bf \hat{3}'})_{\bf 1} f_1^c
+ \beta_f \tilde \phi^{J} (Q Y^{(5)}_{\bf \hat{3}'})_{\bf 1} f_2^c
+\gamma_f \tilde \phi^{K} (Q (Y^{(7)}_{{\bf \hat{3}}', I}+ r_f Y^{(7)}_{{\bf \hat{3}}', II}))_{\bf 1} f_3^c
\right]H_f.
\label{eq:W_S4}
\end{align}
The total number of the free parameters are the same as that of the superpotential of $\Gamma_3'$.
$\alpha_f, \beta_f, \gamma_f$ are real numbers, and $r_f$ is a complex number.
Thus we have five real parameters for each sector.  
$I$ can be an arbitrary positive integer number, but $J$ and $K$ are restricted to $0$ or $1$.

\subsection{FN-like mechanism with the modular group of level 5}

The algebra of $\Gamma_5'$ is summarized in Appendix \ref{app:level_5}.
$\Gamma_5'$ is isomorphic to $A_5'\simeq SL(2,\mathbb{Z}_5)$ having two triplets ${\bf 3}$ and ${\bf 3'}$.
Their matrix representations are shown in Table \ref{tab:irrep_A5'}.
 The triplet modular forms are approximated as
 \begin{align}
Y^{(k)}_{\bf 3} \sim
\begin{pmatrix}
1 \\
q^{1/5} \\
q^{4/5}
\end{pmatrix},
~~
Y^{(k)}_{{\bf 3}'} \sim
\begin{pmatrix}
1 \\
q^{2/5} \\
q^{3/5}
\end{pmatrix},
\label{eq:approximation_level5}
\end{align}
for large \im $\tau$.
We can construct the FN-like model in the same way.
We have two classes of the FN-like superpotential,
\begin{align}
W_1 &= 
\left[
\alpha_f \tilde \phi^{I} (Q Y^{(k_1)}_{\bf 3})_{\bf 1} f_1^c
+ \beta_f \tilde \phi^{J} (Q Y^{(k_2)}_{\bf 3})_{\bf 1} f_2^c
+\gamma_f \tilde \phi^{K} (Q Y^{(k_3)}_{\bf 3})_{\bf 1} f_3^c
\right]H_f,
\nonumber
\\
W_{2} &= 
\left[
\alpha_f \tilde \phi^{I} (Q Y^{(k_1)}_{\bf 3'})_{\bf 1} f_1^c
+ \beta_f \tilde \phi^{J} (Q Y^{(k_2)}_{\bf 3'})_{\bf 1} f_2^c
+\gamma_f \tilde \phi^{K} (Q Y^{(k_3)}_{\bf 3'})_{\bf 1} f_3^c
\right]H_f,
\label{eq:W_tri_G5}
\end{align}
with the triplet left-handed quarks.
We assume that $f_i^c$ is the trivial singlet, and $Q_i$ form a triplet of $\Gamma_5'$.
The CG coefficients are summarized in \eqref{eq:CG_lev5}.
The mass matrices are approximated as
\begin{align}
M_{1} \sim  
\begin{pmatrix}
\tilde{\phi}^I  & \tilde{\phi}^J  & \tilde{\phi}^K \\
\tilde{\phi}^I q^{4/5} & \tilde{\phi}^J q^{4/5} & \tilde{\phi}^K q^{4/5}\\
\tilde{\phi}^I q^{1/5} & \tilde{\phi}^J q^{1/5} & \tilde{\phi}^K q^{1/5}
\end{pmatrix}\frac{v_f}{\sqrt 2},
~~
M_{2} \sim 
\begin{pmatrix}
\tilde{\phi}^I  & \tilde{\phi}^J  & \tilde{\phi}^K \\
\tilde{\phi}^I q^{3/5} & \tilde{\phi}^J q^{3/5} & \tilde{\phi}^K q^{3/5}\\
\tilde{\phi}^I q^{2/5} & \tilde{\phi}^J q^{2/5} & \tilde{\phi}^K q^{2/5}
\end{pmatrix}\frac{v_f}{\sqrt 2},
\nonumber
\end{align}
for large $\im \tau$.
Thus we obtain FN-like hierarchical eigenvalues for both cases.
We have 2 possibilities of the representations of the Yukawa couplings both for the up and down sectors.
We note, however, that since the tensor product of ${\bf 3}$ and ${\bf 3}'$ has no singlet, 
the Yukawa couplings and $Q$ should be the same representation. 
Therefore we have 2 possibilities of either ${\bf 3}$ or ${\bf 3'}$ for both up and down sectors.

We have the same constraints on the power of $\tilde \phi$ as in the previous models.
In this case, the lowest weight is $2$ both for ${\bf 3}$ and ${\bf 3'}$.

\subsection*{Lowest weight models}

The superpotential including the lowest weight modular forms are given by
\begin{align}
W_1 &= 
\left[
\alpha_f \tilde \phi^{I} (Q Y^{(2)}_{\bf 3})_{\bf 1} f_1^c
+ \beta_f \tilde \phi^{J} (Q Y^{(4)}_{\bf 3})_{\bf 1} f_2^c
+\gamma_f \tilde \phi^{K} (Q (Y^{(6)}_{{\bf 3},I} + r_f Y^{(6)}_{{\bf 3},II} ))_{\bf 1} f_3^c
\right]H_f,
\nonumber
\\
W_{2} &= 
\left[
\alpha_f \tilde \phi^{I} (Q Y^{(2)}_{\bf 3'})_{\bf 1} f_1^c
+ \beta_f \tilde \phi^{J} (Q Y^{(4)}_{\bf 3'})_{\bf 1} f_2^c
+\gamma_f \tilde \phi^{K} (Q (Y^{(6)}_{{\bf 3'},I} + r_f Y^{(6)}_{{\bf 3'},II})_{\bf 1} f_3^c
\right]H_f.
\label{eq:W_G5}
\end{align}
The number of the free parameters are the same as that of the superpotential of $\Gamma_3'$ and $\Gamma_4'$.
$\alpha_f, \beta_f, \gamma_f$ are real, and $r_f$ is a complex number.
We have five real parameters for each sector.

\subsection{Realistic models without hierarchical parameters with $\Gamma_4'$ and $\Gamma_5'$}

In this subsection, we analyze the model with the modular group of level 4 and 5.
We show some typical models for illustration purpose.

\subsubsection*{Yukawa couplings of {\bf 3} representation in $\Gamma_4'$}

We consider the FN-like mechanism based on the superpotential $W_1$ in \eqref{eq:W_S4}, i.e., the Yukawa couplings are ${\bf 3}$ in $\Gamma_4'$.
We obtain the mass matrix
\begin{align}
M_u =& 
\begin{pmatrix}
\alpha_u \tilde{\phi}^I Y_2^{(4)} & \beta_u \tilde{\phi}^J Y_2^{(6)} & 
\gamma_u \tilde{\phi}^K (Y_{2,I}^{(8)} + r_u Y_{2,II}^{(8)} )\\
\alpha_u \tilde{\phi}^I Y_1^{(4)} & \beta_u \tilde{\phi}^J Y_1^{(6)} & 
\gamma_u \tilde{\phi}^K (Y_{1,I}^{(8)} + r_u Y_{1,II}^{(8)} )\\
\alpha_u \tilde{\phi}^I Y_3^{(4)} & \beta_u \tilde{\phi}^J Y_3^{(6)} & 
\gamma_u \tilde{\phi}^K (Y_{3,I}^{(8)} + r_u Y_{3,II}^{(8)} )\\
\end{pmatrix}
\begin{pmatrix}
\im \tau^2 & & \\
 & \im \tau^3 & \\
& & \im \tau^4
\end{pmatrix}
\frac{v_u}{\sqrt 2},
\nonumber
\\
M_d =&
\begin{pmatrix}
\alpha_d \tilde{\phi}^L Y_2^{(4)} & 
\beta_d \tilde \phi^M (Y_{2,I}^{(8)} + r_d Y_{2,II}^{(8)} ) & \gamma_d \tilde \phi^N Y_2^{(6)}\\
\alpha_d \tilde{\phi}^L Y_1^{(4)}  & 
\beta_d \tilde \phi^M (Y_{1,I}^{(8)} + r_d Y_{1,II}^{(8)} ) & \gamma_d \tilde \phi^N Y_1^{(6)}\\
\alpha_d \tilde{\phi}^L Y_3^{(4)} &
\beta_d \tilde \phi^M (Y_{3,I}^{(8)} + r_d Y_{3,II}^{(8)} ) & \gamma_d  \tilde \phi^N Y_3^{(6)}\\
\end{pmatrix}
\begin{pmatrix}
\im \tau^2 & & \\
 & \im \tau^4 & \\
& & \im \tau^3
\end{pmatrix}
\frac{v_d}{\sqrt 2}.
\label{eq:mass_G4_3_1}
\end{align}
We assume $I \geq J \geq K$ and $L \geq M \geq N$ to obtain a FN-like matrix.
We also study other possibilities, but the above mass matrix is the best one.
The mixing angles are approximated by
\begin{align}
\theta_{12} &\sim 
	\left|
	\frac{Y^{(6)}_{2}}{Y^{(6)}_{1}} - 
	\frac{Y^{(8)}_{2, I} + r_d Y^{(8)}_{2, II} }{Y^{(8)}_{1, I } + r_d Y^{(8)}_{1, II }}
	\right| 
\sim 
	\frac{13}{\sqrt 2} 
	\left| 
	\left(
	1 + \frac{\sqrt{10}}{13 r_d}
	\right)q^{1/4}
	\right|
\sim 
	9
	\left|
	 (1 + 0.2 r_d^{-1} ) q^{1/4}
	 \right|,
\nonumber
\\
\theta_{23} &\sim 
	\left|
	\frac{Y^{(8)}_{1,I} + r_u Y^{(8)}_{1, II}}{Y^{(8)}_{3, I } + r_u Y^{(8)}_{3,II}} - \frac{Y^{(6)}_{1}}{Y^{(6)}_{3}}
	\right| 
\sim 
	4 \sqrt 5 
	\left|
	\left( 
	r_u - \sqrt{\frac 2 5 }
	\right)q^{1/4}
	\right| 
\sim 
	9
	\left| 
	(r_u - 0.6) q^{1/4}
	\right|,
\nonumber
\\
\theta_{13} &\sim 
	\left|
	\frac{Y^{(8)}_{2,I} + r_u Y^{(8)}_{2, II}}{Y^{(8)}_{3, I } + r_u Y^{(8)}_{3,II}}- \frac{Y^{(6)}_{2}}{Y^{(6)}_{3}}
	\right| 
\sim 
	16 \sqrt{10} 
	\left|
	\left(
	r_u +  \frac 1 2 \sqrt{\frac 52} 
	\right) q^{2/4}
	\right| 
\sim 
	51
	\left|
	(r_u +  0.8 )q^{2/4}
	\right|.
\nonumber
\end{align}
The first condition implies $9 |q^{1/4}| \sim \theta_{C} \sim 10^{-1}$,
which implies $\im \tau \sim 2.8$.
The remaining conditions are rewritten as
\begin{align}
\theta_{23} \sim
	\left| r_u -0.6 \right| \times 10^{-1},~~
	\theta_{13} \sim 6 \left|r_u +  0.8 \right| \times 10^{-3} .
\nonumber
\end{align}
and the realistic mixing angles are realized naturally with $|r_u-0.6| \sim 10^{-1}$.
Then the mass ratios are approximated as
\begin{align}
y_u/y_t 
	&\sim 
	\frac{16\sqrt{10}}{3}  \left( \frac{\sqrt 2}{13}\right)^2 
	\im\tau^{-2} 
	\frac{\alpha_u}{\gamma_u} 
	\tilde \phi^{I-K}  \theta_{C}^2
	\sim 
	3 \times 10^{-4} \frac{\alpha_u}{\gamma_u} \tilde \phi^{I-K}, 
\nonumber
\\
y_c/y_t 
	&\sim 
	\frac{16 \sqrt{10}}{39} 
	\im\tau^{-1} \frac{\beta_u}{\gamma_u} \tilde \phi^{J-K}  \theta_{C} 
	\sim 
	5 \times 10^{-2} \frac{\beta_u}{\gamma_u} \tilde \phi^{J-K}, 
\nonumber
\\
y_d/y_b 
	&\sim 
	\frac{16}{(13)^2} 
	\im \tau^{-1} \frac{\alpha_d}{\gamma_d} \tilde \phi^{L-N} \theta_C^2 
	\sim
	3 \times 10^{-4} \frac{\alpha_d}{\gamma_d} \tilde \phi^{L-N},
\nonumber
\\
y_s/y_b 
	&\sim
	\frac{6 |r_d| }{13}
	\im \tau \frac{\beta_d}{\gamma_d}  \tilde \phi^{M-N}  \theta_C 
	\sim
	1 \times 10^{-1} \frac{\beta_d|r_d|}{\gamma_d} \tilde \phi^{M-N}, 
\nonumber
\\
y_b/y_t 
	&\sim 
	\frac{4}{3} \sqrt{\frac 5 2} \im \tau^{-1}
	\frac{\gamma_d}{\gamma_u} \tilde \phi^{N-K} 
	\sim
	7 \times 10^{-1} \frac{\gamma_d}{\gamma_u} \tilde \phi^{N-K}.
\label{eq:cond_G4_1}
\end{align}
These conditions imply $\tilde \phi \sim 2\times 10^{-2}$, $I =J =1, K =0, 
L = M = N=1$, and all the coefficients are $\mathcal{O}(1)$.
Thus we can naturally reproduce the observed values.

To confirm the above analysis, we construct an explicit example.
The representations and the modular weights of the quark fields are summarized in Table \ref{tab:model1_G4}.
The mass matrix is given by \eqref{eq:mass_G4_3_1} with $I=2, J =1, K=0,$ and $L=2, M=N=1$.
\begin{table}[t]
\begin{center}
\begin{tabular}{|c|cccccccccc|} \hline
 & $Q_{1,2,3}$ 
 & $u_1^c$ & $u_2^c$ & $u_2^c$ 
 & $d_1^c$ & $d_2^c$ & $d_3^c$ 
 & $H_u$ & $H_d$ & $\phi$ 
\\
\hline \hline
$\Gamma_4'$ & ${\bf 3}$ 
& ${\bf 1}$ & ${\bf 1}$ & ${\bf 1}$ 
& ${\bf 1}$ & ${\bf 1}$ & ${\bf 1}$ 
& ${\bf 1}$ & ${\bf 1}$ & ${\bf 1}$
\\
$k_I$ & $-1$ 
& $-1$ & $-4$ & $-7$ 
& $-1$ & $-6$ & $-4$
& $0$ & $0$ & $-1$
\\
\hline
\end{tabular}
\caption{The representations and the modular weights of the quarks.
This is an explicit model for superpotential with Yukawa coupling of ${\bf 3}$ in $S_4'$.}
\label{tab:model1_G4}
\end{center}
\end{table}
The best fit parameter in our analysis is given by
\begin{align}
\alpha_u/\gamma_u =& 0.2514,~~
\beta_u/\gamma_u = 1.968,~~
r_u = e^{-0.3176i},
\nonumber
\\
\alpha_d/\gamma_d =& 0.8793,~~
\beta_d/\gamma_d = 0.3907,~~
r_d = e^{0.1469i},
\nonumber
\\
\gamma_d/\gamma_u =& 0.3906,~~
\tau = 2.851 i,~~
\tilde \phi = 4 \times 10^{-2}.
\label{eq:bench_G4_model1}
\end{align}
The largest hierarchy comes from $\beta_u/\beta_d = 12.9$.
We set  $\re \tau =0$ and $|r_f| =1 $ again.
We obtain the following hierarchical mass matrices,
\begin{align}
|M_u| =&
\gamma_u
\begin{pmatrix}
1.423 \times 10^{-7} & 1.984 \times 10^{-5} & 1.191 \times 10^{-3} \\
8.865 \times 10^{-7} & 4.945 \times 10^{-4} & 1.344 \times10^{-2}\\
2.76 \times 10^{-5} & 7.702 \times 10^{-3} & 1.323 \times 10^{-1}\\
\end{pmatrix} \frac{v_u}{\sqrt{2}},
\nonumber
\\
|M_d| =&
\gamma_d
\begin{pmatrix}
4.977 \times 10^{-8} & 1.877 \times 10^{-5} & 1.009 \times 10^{-5}\\
3.101 \times 10^{-6} & 2.100 \times 10^{-4} & 2.513 \times 10^{-4}\\
9.658 \times 10^{-5} & 2.068 \times 10^{-3} & 3.914 \times 10^{-3}
\end{pmatrix}
\frac{v_d}{\sqrt{2}}.
\end{align}
The mass eigenvalues and mixing angles are given as
\begin{align}
y_u/y_t  &= 5.45 \times10^{-6},~~
y_c/y_t = 2.68 \times 10^{-3},~~
y_d/y_b = 6.53 \times 10^{-4},~~
y_s/y_b = 1.69 \times 10^{-2}, 
\nonumber
\\
y_b/y_t &= 1.30 \times 10^{-2},~~
\theta_{12} =  0.227,~~
\theta_{23} =  0.0430,~~
\theta_{13} =  0.00203,~~
\delta_{CP} = 1.31.
\end{align}
The parameter which is the most apart from the observed value 
is $\theta_{13}$, and $(\theta_{13} - \theta_{13}^{obs})/\sigma_{13} \sim 2.3 \sigma$.
We obtain $\chi^2 \sim 12$, and almost all the parameters 
are within $2\sigma$ range.

Relaxing the restriction on $|r_f|$,
we can find parameter set which reproduce observed values more precisely.
A benchmark is given by
\begin{align}
\alpha_u/\gamma_u =& 1.130,~~
\beta_u/\gamma_u = 4.334,~~
r_u = 1.040 e^{-0.2224i},
\nonumber
\\
\alpha_d/\gamma_d =& 2.193,~~
\beta_d/\gamma_d = 0.7258,~~
r_d = 0.8893 e^{0.1228 i},
\nonumber
\\
\gamma_d/\gamma_u =& 0.6361,~~
\tau = 2.901 i,~~
\tilde \phi = 2 \times 10^{-2}.
\nonumber
\end{align}
and the mot hierarchical term comes from $\alpha_u/\beta_d =9.4$.
Thus all the coefficients are the same order.
We obtain the mass matrices
\begin{align}
|M_u| =&
\gamma_u
\begin{pmatrix}
1.287 \times 10^{-8} & 1.791 \times 10^{-5} & 1.027 \times 10^{-3}\\
8.671 \times 10^{-7} & 4.825 \times 10^{-4} & 1.259 \times 10^{-2}\\
2.920 \times 10^{-5} & 8.123 \times 10^{-3} & 1.289 \times 10^{-1}\\
\end{pmatrix} \frac{v_u}{\sqrt{2}},
\nonumber
\\
|M_d| =&
\gamma_d
\begin{pmatrix}
2.499 \times 10^{-8} & 1.339 \times 10^{-5} & 4.132 \times 10^{-6}\\
1.683 \times 10^{-6} & 1.563 \times 10^{-4} & 1.113 \times 10^{-4}\\
5.668 \times 10^{-5} & 1.871 \times 10^{-3} & 1.874 \times 10^{-3}\\
\end{pmatrix} \frac{v_d}{\sqrt{2}}.
\end{align}
The mass eigenvalues and mixing angles are given by
\begin{align}
y_u/y_t  &=5.70 \times10^{-6},~~
y_c/y_t = 2.68 \times 10^{-3},~~
y_d/y_b = 6.92 \times 10^{-4},~~
y_s/y_b = 1.37 \times 10^{-2}, ~~
\nonumber
\\
y_b/y_t &= 1.30 \times 10^{-2},~~
\theta_{12} =  0.227,~~
\theta_{23} =  0.0359,~~
\theta_{13} =  0.00314,~~
\delta_{CP} = 1.21.
\end{align}
Thus all parameters are included within $0.1\sigma$ interval, and $\chi^2<0.01$.

\subsubsection*{Yukawa couplings of ${\bf \hat{3}'}$ in $\Gamma_4'$}

The $q$-expansions of the modular forms of ${\bf \hat{3}'}$ in $\Gamma_4'$ 
are different from the modular forms of ${\bf 3}$.
If the superpotential is given by $W_4$ in \eqref{eq:W_S4}, 
the approximated mass matrix is given by $M_4$ in \eqref{eq:mass_G4_app}.
This matrix is quite interesting because the mixing angles are approximated by
\begin{align}
\theta_{12} \sim |q^{1/4}|,~~
\theta_{23} \sim |q^{2/4}|,~~
\theta_{13} \sim |q^{3/4}|.
\end{align}
These relations are nothing but approximate mixing angles predicted in the FN mechanism.
Thus this model seems to be the most promising candidate.
We consider the model with the following mass matrices,
\begin{align}
M_u = & 
\begin{pmatrix}
\alpha_u \tilde{\phi}^I Y_1^{(3)} & \beta_u \tilde{\phi}^J Y_1^{(5)} & 
\gamma_u \tilde{\phi}^K (Y_{1,I}^{(7)} + r_u Y_{1,II}^{(7)} )\\
\alpha_u \tilde{\phi}^I Y_3^{(3)} & \beta_u \tilde{\phi}^J Y_3^{(5)} & 
\gamma_u \tilde{\phi}^K (Y_{3,I}^{(7)} + r_u Y_{3,II}^{(7)} )\\
\alpha_u \tilde{\phi}^I Y_2^{(3)} & \beta_u \tilde{\phi}^J Y_2^{(5)} & 
\gamma_u \tilde{\phi}^K (Y_{2,I}^{(7)} + r_u Y_{2,II}^{(7)} )\\
\end{pmatrix}
\begin{pmatrix}
\im \tau^{3/2} & & \\
 & \im \tau^{5/2} & \\
& & \im \tau^{7/2}
\end{pmatrix}
\frac{v_u}{\sqrt 2},
\nonumber
\\
M_d =&
\begin{pmatrix}
\alpha_d \tilde{\phi}^L Y_1^{(3)} & 
\beta_d \tilde \phi^M (Y_{1,I}^{(7)} + r_d Y_{1,II}^{(7)} ) & \gamma_d \tilde \phi^N Y_1^{(5)}\\
\alpha_d \tilde{\phi}^L Y_3^{(3)}  & 
\beta_d \tilde \phi^M (Y_{3,I}^{(7)} + r_d Y_{3,II}^{(7)} ) & \gamma_d \tilde \phi^N Y_3^{(5)}\\
\alpha_d \tilde{\phi}^L Y_2^{(3)} &
\beta_d \tilde \phi^M (Y_{2,I}^{(7)} + r_d Y_{2,II}^{(7)} ) & \gamma_d \tilde \phi^N Y_2^{(5)}\\
\end{pmatrix}
\begin{pmatrix}
\im \tau^{3/2} & & \\
 & \im \tau^{7/2} & \\
& & \im \tau^{5/2}
\end{pmatrix}
\frac{v_d}{\sqrt 2}.
\label{eq:mass_G4_3hp}
\end{align}

The precise $q$-expansion of the modular forms are summarized in Appendix \ref{app:level_4}.
The mixing angles are estimated as
\begin{align}
\theta_{12} &\sim 
\left|
\frac{Y^{(5)}_{1}}{Y^{(5)}_{3}} - 
\frac{Y^{(7)}_{1, I} + r_d Y^{(7)}_{1, II} }{Y^{(7)}_{3, I } + r_d Y^{(7)}_{3, II }}
\right| \sim 
\frac{22\sqrt{2}}{3} 
\left| 1 + \frac{14}{22 \sqrt{37}} \frac{1}{r_d}\right| |q^{1/4}|
\sim 10 \left| 1 + 0.1 r_d^{-1} \right|
|q^{1/4}|,
\nonumber
\\
\theta_{23} 
	&\sim
	\left|
	\frac{Y^{(7)}_{3,I} + r_u Y^{(7)}_{3, II}}{Y^{(7)}_{2, I } + r_u Y^{(7)}_{2,II}}
	- \frac{Y^{(5)}_{3}}{Y^{(5)}_{2}}
	\right| 
	\sim 
	12 \sqrt{37}
	\left| r_u -  \frac 1 {\sqrt{37}} \right| 
	|q^{2/4}|
	\sim 
	73 \left| r_u -  0.2 \right| |q^{2/4}|,
\nonumber
\\
\theta_{13} 
	&\sim 
	\left|
	\frac{Y^{(7)}_{1,I} + r_u Y^{(7)}_{1, II}}{Y^{(7)}_{2, I } + r_u Y^{(7)}_{2, II}}
	- \frac{Y^{(5)}_{1}}{Y^{(5)}_{2}}
	\right| 
	\sim 
	24 \sqrt{74}  
	\left|r_u +  \frac{5}{\sqrt{37}} \right| 
	|q^{3/4}|
	\sim 206 \left|r_u + 0.8 \right||q^{3/4}|.
\nonumber
\end{align}
The first relation implies $10|q^{1/4}| \sim \theta_{C} \sim 10^{-1}$, which implies $\im \tau \sim 2.9$.
Substituting this relation, we obtain
\begin{align}
\theta_{23} \sim 0.7 \left| r_u -  0.2 \right| \theta_{C}^2,~~
\theta_{13} \sim 0.2 \left|r_u + 0.8 \right| \theta_{C}^3.
\end{align}
Thus we obtain approximate relation of the mixing angles.
The mass ratios are approximated as
\begin{align}
y_u/y_t 
	& \sim
	\frac{36 \sqrt{74}}{11^3} 
	\im \tau^{-2}
	\frac{\alpha_u}{\gamma_u}
	\tilde \phi^{I-K}\theta_{C}^3
	\sim
	3 \times 10^{-5} \frac{\alpha_u}{\gamma_u} 
	\tilde \phi^{I-K},
\nonumber
\\
y_c/y_t 
	& \sim
	\frac{9\sqrt{37}}{242}  
	\im \tau^{-1}
	\frac{\beta_u}{\gamma_u}
	\tilde \phi^{J-K}\theta_{C}^2
	\sim 
	8 \times 10^{-4} \frac{\beta_u}{\gamma_u}
	\tilde \phi^{J-K},
\nonumber
\\
y_d/y_b 
	&\sim 
	\frac{216}{11^3\sqrt 2} 
	\im \tau^{-1}
	\frac{\alpha_d}{\gamma_d} \tilde \phi^{L-N} \theta_C^3
	\sim
	4 \times 10^{-5} \frac{\alpha_d}{\gamma_d} \tilde \phi^{L-N},
\nonumber
\\
y_s/y_b 
	& \sim 
	\frac{81}{242}  
	\im \tau \frac{\beta_d}{\gamma_d} |r_d| \tilde \phi^{M-N}
	\theta_C^2
	\sim 
	1\times 10^{-2} \frac{\beta_d}{\gamma_d}  |r_d| \tilde \phi^{M-N},
\nonumber
\\
y_b/y_t 
	& \sim 
	\frac{\sqrt{37}}{3} \im \tau^{-1} \frac{\gamma_d}{\gamma_u} \tilde \phi^{N-K} 
	\sim 
	7 \times 10^{-1}\frac{\gamma_d}{\gamma_u} \tilde \phi^{N-K}.
\label{eq:condition_G4_3}
\end{align}
These relations imply 
$\tilde \phi \sim 2\times 10^{-2}$, $I  = 1, J = K = 0$  and $ L= M= N =1.$

To confirm the above estimation,
we construct an explicit examples.
The representations and the modular weights of 
the fields are summarized in Table \ref{tab:mode2_G4}.
This model generates the mass matrix with $I =2, J=1, K=0, L =2, M = N = 1$.
\begin{table}[t]
\begin{center}
\begin{tabular}{|c|cccccccccc|} \hline
 & $Q_{1,2,3}$ 
 & $u_1^c$ & $u_2^c$ & $u_2^c$ 
 & $d_1^c$ & $d_2^c$ & $d_3^c$ 
 & $H_u$ & $H_d$ & $\phi$ 
\\
\hline \hline
$\Gamma_4'$ & ${\bf \hat 3}$ 
& ${\bf 1}$ & ${\bf 1}$ & ${\bf 1}$ 
& ${\bf 1}$ & ${\bf 1}$ & ${\bf 1}$ 
& ${\bf 1}$ & ${\bf 1}$ & ${\bf 1}$
\\
$k_I$ & $-1$ 
& $0$ & $-3$ & $-6$ 
& $0$ & $-5$ & $-3$
& $0$ & $0$ & $-1$
\\
\hline
\end{tabular}
\caption{The representations and the modular weights of the quarks.
This is an explicit model for superpotential with Yukawa coupling of ${\bf \hat 3'}$ in $S_4'$.}
\label{tab:mode2_G4}
\end{center}
\end{table}
In this model, 
we can not find a parameter set whose $\chi^2 < 50$ with $|r_f|=1$.
We relax this constraint.
The best fit parameters in our analysis are given by
\begin{align}
\alpha_u/\gamma_u =& 1.516,~~
\beta_u/\gamma_u = 2.908,~~
r_u = 0.4535 e^{2.051 i},
\nonumber
\\
\alpha_d/\gamma_d =& 3.287,~~
\beta_d/\gamma_d = 0.9015,~~
r_d = 0.5514 e^{0.5518 i},
\nonumber
\\
\gamma_d/\gamma_u =& 0.3402,~~
\tau = 2.250i,~~
\tilde \phi = 3 \times 10^{-2}.
\nonumber
\end{align}
The largest hierarchy is $\beta_u/\beta_d = 8.6$ among them.
The mass matrix is given by
\begin{align}
|M_u| =&
\gamma_u
\begin{pmatrix}
2.588 \times 10^{-6} & 2.632 \times 10^{-4} & 3.347 \times 10^{-3}\\
2.352 \times 10^{-5} & 1.196 \times 10^{-3} & 4.197 \times 10^{-2}\\
2.303 \times 10^{-3} & 1.171 \times 10^{-1} & 1.490 \times 10^{-0}
\end{pmatrix} 
\frac{v_u}{\sqrt{2}},
\nonumber
\\
|M_d| =&
\gamma_d
\begin{pmatrix}
5.610 \times 10^{-6} & 1.861 \times 10^{-4} & 9.051 \times 10^{-5}\\
5.099 \times 10^{-5} & 1.380 \times 10^{-3} & 4.114 \times 10^{-4}\\
4.994 \times 10^{-3} & 4.030 \times 10^{-2} & 4.027 \times 10^{-2}\\
\end{pmatrix}
\frac{v_d}{\sqrt{2}}.
\end{align}
The mixing angles and the mass ratios are calculated as
\begin{align}
y_u/y_t  &=5.70 \times10^{-6},~~
y_c/y_t = 2.68 \times 10^{-3},
y_d/y_b = 6.92. \times 10^{-4},~~
y_s/y_b = 1.37 \times 10^{-2}, ~~
\nonumber
\\
y_b/y_t &= 1.30 \times 10^{-2},~~
\theta_{12} =  0.227,~~
\theta_{23} =  0.359,~~
\theta_{13} =  0.00314,~~
\delta_{CP} = 1.21,
\end{align}
and we can realize all the parameters within $0.1\sigma$ range.
$\chi^2 < 0.01$.

\subsubsection*{Yukawa couplings of ${\bf 3'}$ in $\Gamma_5'$}

We investigate the model based on $W_2$ in \eqref{eq:W_G5}.
The explicit form of the $q$-expansion of the modular forms are summarized in 
Appendix \ref{app:level_5}.
We also assume the lightest up-type quark (up-quark) corresponds to the modular form of weight 4, 
and the second lightest up-type quark (charm quark) corresponds to the modular form of weight 2.
Thus the mass matrix is given by
\begin{align}
M_u =  & 
\begin{pmatrix}
\beta_u \tilde{\phi}^I Y_3^{(4)}& \alpha_u \tilde{\phi}^J Y_3^{(2)} & 
\gamma_u \tilde{\phi}^K (Y_{3,I}^{(6)} + r_u Y_{3,II}^{(6)} )\\
\beta_u \tilde{\phi}^I Y_2^{(4)} & \alpha_u \tilde{\phi}^J Y_2^{(2)} & 
\gamma_u \tilde{\phi}^K (Y_{2,I}^{(6)} + r_u Y_{2,II}^{(6)} )\\
\beta_u \tilde{\phi}^I Y_1^{(4)} & \alpha_u \tilde{\phi}^J Y_1^{(2)} & 
\gamma_u \tilde{\phi}^K (Y_{1,I}^{(6)} + r_u Y_{1,II}^{(6)} )\\
\end{pmatrix}
\begin{pmatrix}
 \im \tau^2& & \\
 & \im \tau^1 & \\
& & \im \tau^3
\end{pmatrix}
\frac{v_u}{\sqrt 2},
\nonumber
\\
M_d =  &
\begin{pmatrix}
\alpha_d \tilde{\phi}^L Y_3^{(2)} & \beta_d \tilde \phi^M Y_3^{(4)} & 
\gamma_d \tilde \phi^N (Y_{3,I}^{(6)} + r_d Y_{3,II}^{(6)} )\\
\alpha_d \tilde{\phi}^L Y_2^{(2)} & \beta_d \tilde \phi^M Y_2^{(4)} & 
\gamma_d \tilde \phi^N (Y_{2,I}^{(6)} + r_d Y_{2,II}^{(6)} )\\
\alpha_d \tilde{\phi}^L Y_1^{(2)} & \beta_d  \tilde \phi^M Y_1^{(4)} & 
\gamma_d  \tilde \phi^N (Y_{1,I}^{(6)} + r_d Y_{1,II}^{(6)} )\\
\end{pmatrix}
\begin{pmatrix}
\im \tau^1 & & \\
 & \im \tau^2 & \\
& & \im \tau^3
\end{pmatrix}
\frac{v_d}{\sqrt 2}.
\label{eq:mass_G5_3_1}
\end{align}
The approximate mixing angles are given by
\begin{align}
\theta_{12} \sim \frac{23}{14}|q^{1/5}|,~~
\theta_{23} \sim 40 \sqrt{2} \left| \left( \frac{1}{r_u} - \frac{1}{r_d} \right) q^{2/5} \right| ,~~
\theta_{13} \sim 120\sqrt{2} \left|\left( \frac{1}{r_u} - \frac{1}{r_d} \right) q^{3/5} \right|.
\label{eq:mixing_G5}
\end{align}
Thus $2|q^{1/5}| \sim \theta_{C} \sim 10^{-1}$, which implies $\im \tau \sim 2.4$.
We obtain
\begin{align}
\theta_{23} &\sim 
	2 \left|  \frac{1}{r_u} - \frac{1}{r_d} \right| \times 10^{-1},~~
\theta_{13} \sim 
	4 \left|  \frac{1}{r_u} - \frac{1}{r_d} \right| \times 10^{-2}.
\nonumber
\end{align}
In this case we can realize the observed values if $|\frac 1{r_u}-\frac 1{r_d}| \sim 10^{-1}$.
On the other hand, conditions for the mass eigenvalues are given by
\begin{align}
y_u/y_t 
	& \sim
	\frac{255}{\sqrt{12} |r_u|} \left(\frac{14}{23}\right)^3 \theta_{C}^3
	\im \tau^{-1} \frac{\alpha_u}{\gamma_u}
	\tilde \phi^{I-K} 
	\sim 
	7 \times 10^{-3} \frac{\alpha_u}{\gamma_u |r_u|}
	\tilde \phi^{I-K},
\nonumber
\\
y_c/y_t 
	&\sim
	\frac{5}{|r_u|}
	\left(\frac{14}{23}\right)^2 \theta_{C}^2
	\im \tau^{-2}
	\frac{\beta_u}{\gamma_u}
	\tilde \phi^{J-K} 
	\sim
	3 \times 10^{-3} \frac{\beta_u}{\gamma_u|r_u|} 
	\tilde \phi^{J-K},
\nonumber
\\
y_d/y_b 
	&\sim 
	\frac{10}{|r_d|}
	\left(\frac{14}{23}\right)^3 \theta_{C}^3
	\im \tau^{-2}
	\frac{\alpha_d}{\gamma_d}
	\tilde \phi^{L-N} 
	\sim
	4 \times 10^{-4} \frac{\alpha_d}{\gamma_d |r_d|} \tilde \phi^{L-N}
	,
\nonumber
\\
y_s/y_b
	&\sim 
	\frac{70}{\sqrt{12}|r_d|}
	\left(\frac{14}{23}\right)^2 \theta_{C}^2
	\im \tau^{-1}
	\frac{\beta_d}{\gamma_d}
	\tilde \phi^{M-N} 
	\sim
	3\times 10^{-2} \frac{\beta_d}{\gamma_d |r_d|}  \tilde \phi^{M-N}
	,
	\nonumber
\\
y_b/y_t 
	&\sim 
	\frac{\gamma_d}{\gamma_u} \tilde \phi^{N-K} 
	.
\label{eq:condition_G5_1}
\end{align}
These conditions imply $\tilde \phi \sim 10^{-2}$, $I=1, J = K=0$ and $L=M=N=1 $.
We can realize the observed values naturally.
An explicit model is given by Table \ref{tab:mode1_G5}, which generates the mass matrix of \eqref{eq:mass_G5_3_1} with $I = 1, J = K =0, L=M=N =1$.
\begin{table}[t]
\begin{center}
\begin{tabular}{|c|cccccccccc|} \hline
 & $Q_{1,2,3}$ 
 & $u_1^c$ & $u_2^c$ & $u_2^c$ 
 & $d_1^c$ & $d_2^c$ & $d_3^c$ 
 & $H_u$ & $H_d$ & $\phi$ 
\\
\hline \hline
$A_5'$ & ${\bf 3}'$ 
& ${\bf 1}$ & ${\bf 1}$ & ${\bf 1}$ 
& ${\bf 1}$ & ${\bf 1}$ & ${\bf 1}$ 
& ${\bf 1}$ & ${\bf 1}$ & ${\bf 1}$
\\
$k_I$ & $-1$ 
& $-2$ & $-1$ & $-5$ 
& $0$ & $-2$ & $-4$
& $0$ & $0$ & $-1$
\\
\hline
\end{tabular}
\caption{The representations and the modular weights of the quarks.
This is an explicit model for superpotential with Yukawa coupling of ${\bf 3'}$ in $A_5'$.}
\label{tab:mode1_G5}
\end{center}
\end{table}
In this model, we can not find a parameter set which 
reproduces the mixing angles and the mass ratios whose $\chi^2 < 30$
with the constraints of $|r_f| =1$.
Thus we relax this constraint.
One benchmark value is given by
\begin{align}
\alpha_u/\gamma_u =& 0.1089,~~
\beta_u/\gamma_u = 0.4899,~~
r_u = 1.344e^{-3.500i},
\nonumber
\\
\alpha_d/\gamma_d =& 1.199,~~
\beta_d/\gamma_d = 2.026,~~
r_d = 1.719e^{-2.979i},
\nonumber
\\
\gamma_d/\gamma_u =& 0.7421~~
\tau = 2.560 i,~~
\tilde \phi = 1 \times 10^{-2}.
\label{eq:coe_best_G5}
\end{align}
We obtain
\begin{align}
|M_u| &=
\gamma_u
\begin{pmatrix}
1.195 \times 10^{-4} & 2.798 \times 10^{-3} & 7.022 \times 10^{-1}\\
8.026 \times 10^{-4} & 3.490 \times 10^{-2} & 4.461 \times 10^{+0}\\
5.046 \times 10^{-2} & 3.072 \times 10^{+0} & 7.811 \times 10^{+1}\\
\end{pmatrix}
\frac{v_u}{\sqrt{2}},
\nonumber
\\
|M_d| &=
\gamma_d
\begin{pmatrix}
6.849 \times 10^{-5} & 2.223 \times 10^{-3} & 7.251 \times 10^{-3}\\
8.542 \times 10^{-4} & 1.493 \times 10^{-2} & 4.166 \times 10^{-2}\\
7.518 \times 10^{-2} & 9.385 \times 10^{-1} & 9.987 \times 10^{-1}\\
\end{pmatrix} 
\frac{v_d}{\sqrt{2}}.
\end{align}
The mixing angles and mass ratios are obtained as
\begin{align}
y_u/y_t  &=4.20 \times10^{-6},~~
y_c/y_t = 2.67 \times 10^{-3},
y_d/y_b = 6.69 \times 10^{-4},~~
y_s/y_b = 1.41 \times 10^{-2}, ~~
\nonumber
\\
y_b/y_t &= 1.30 \times 10^{-2},~~
\theta_{12} =  0.227,~~
\theta_{23} =  0.0367,~~
\theta_{13} =  0.00304,~~
\delta_{CP} = 1.20.
\end{align}
In this case $\chi^2 \sim 0.3$ and realistic observed values are reproduced.

\section{Stability of parameters}

In the previous section we find realistic models which reproduce the observed quark mass ratios and mixing angles with $\mathcal{O}(1)$ parameters.
Our approximate estimations show that the mixing angles depend only on $\im \tau$ and $r_{u,d}$.
We expect that they are rather stable prediction in our model, 
if the overall coefficients are $\mathcal{O}(1)$.  
We study the validity of the approximations used in our models
and the stability of the results against changes of the free parameters.
For these purposes, we investigate the coefficient dependence of the results.
In the followings we use the previous model of Table \ref{tab:mode1_G5} in $\Gamma_5'$ as an example.

\begin{figure}[t]
	\begin{minipage}[b]{0.45\linewidth}
		\centering
		\includegraphics[keepaspectratio, scale=0.4]{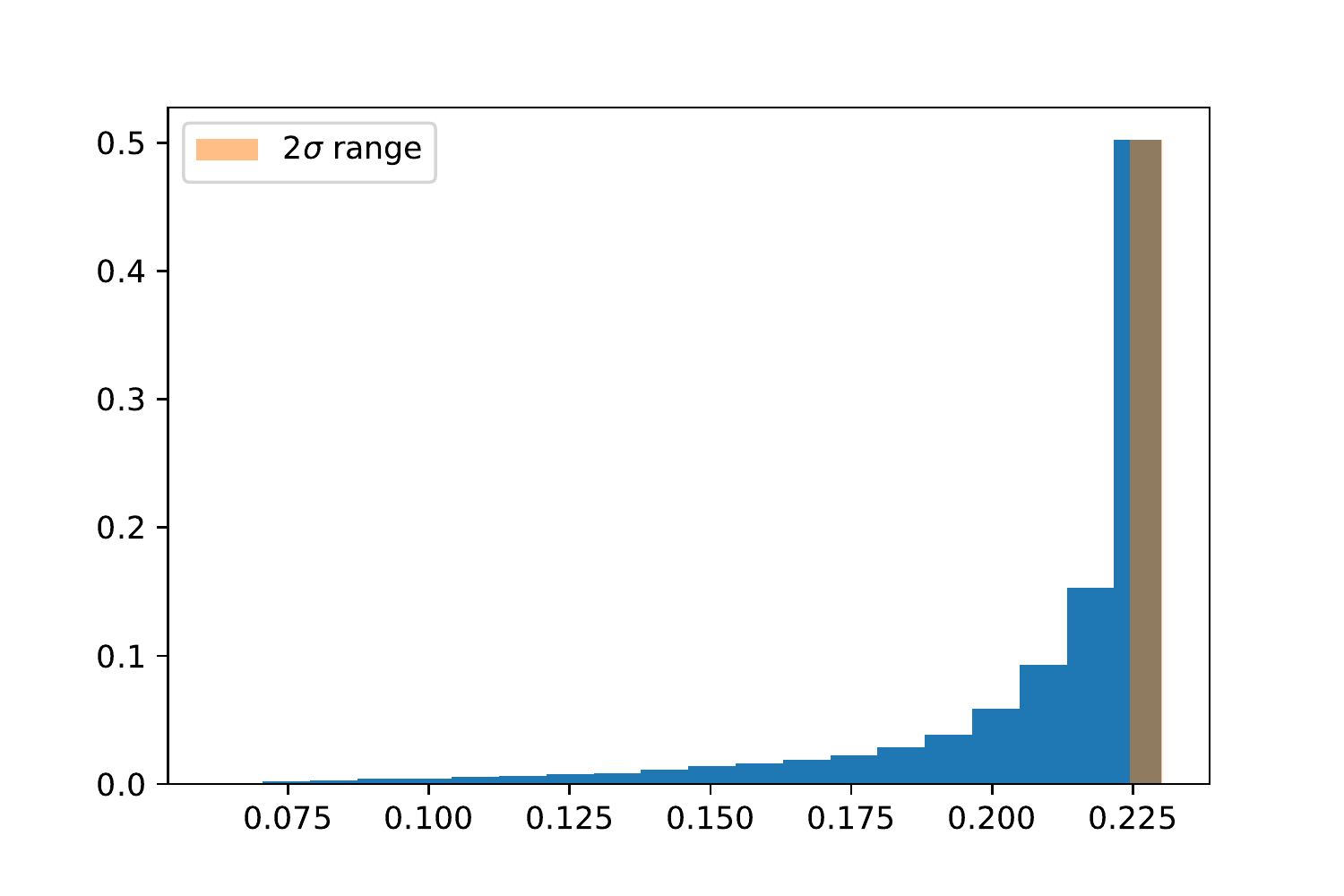}
		\subcaption{Distribution of $\theta_{12}$.}
	\end{minipage}
	\begin{minipage}[b]{0.45\linewidth}
		\centering
		\includegraphics[keepaspectratio, scale = 0.4]{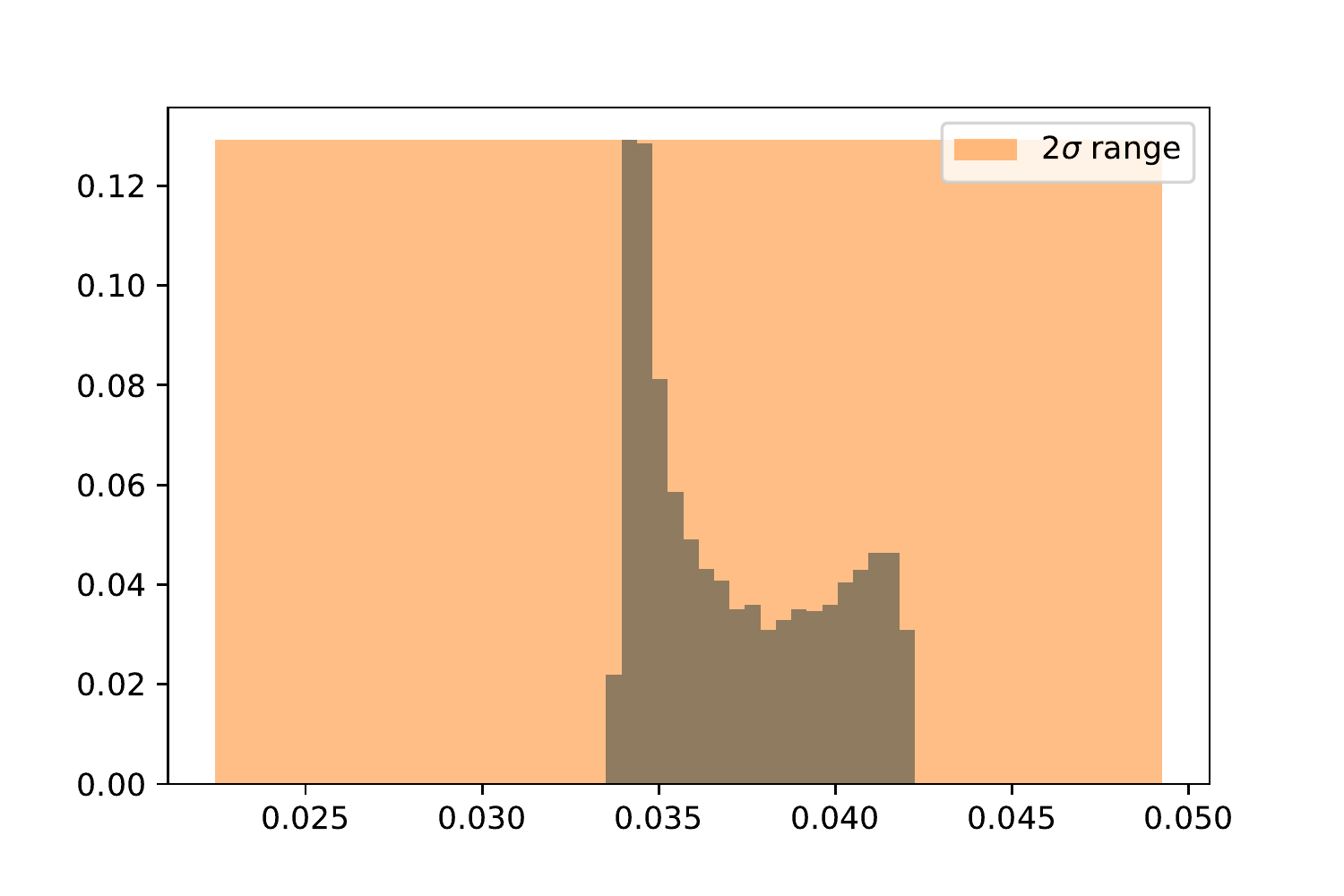}
		\subcaption{Distribution of $\theta_{23}$.}
	\end{minipage}
	\\
\begin{minipage}[b]{0.45\linewidth}
		\centering
		\includegraphics[keepaspectratio, scale=0.4]{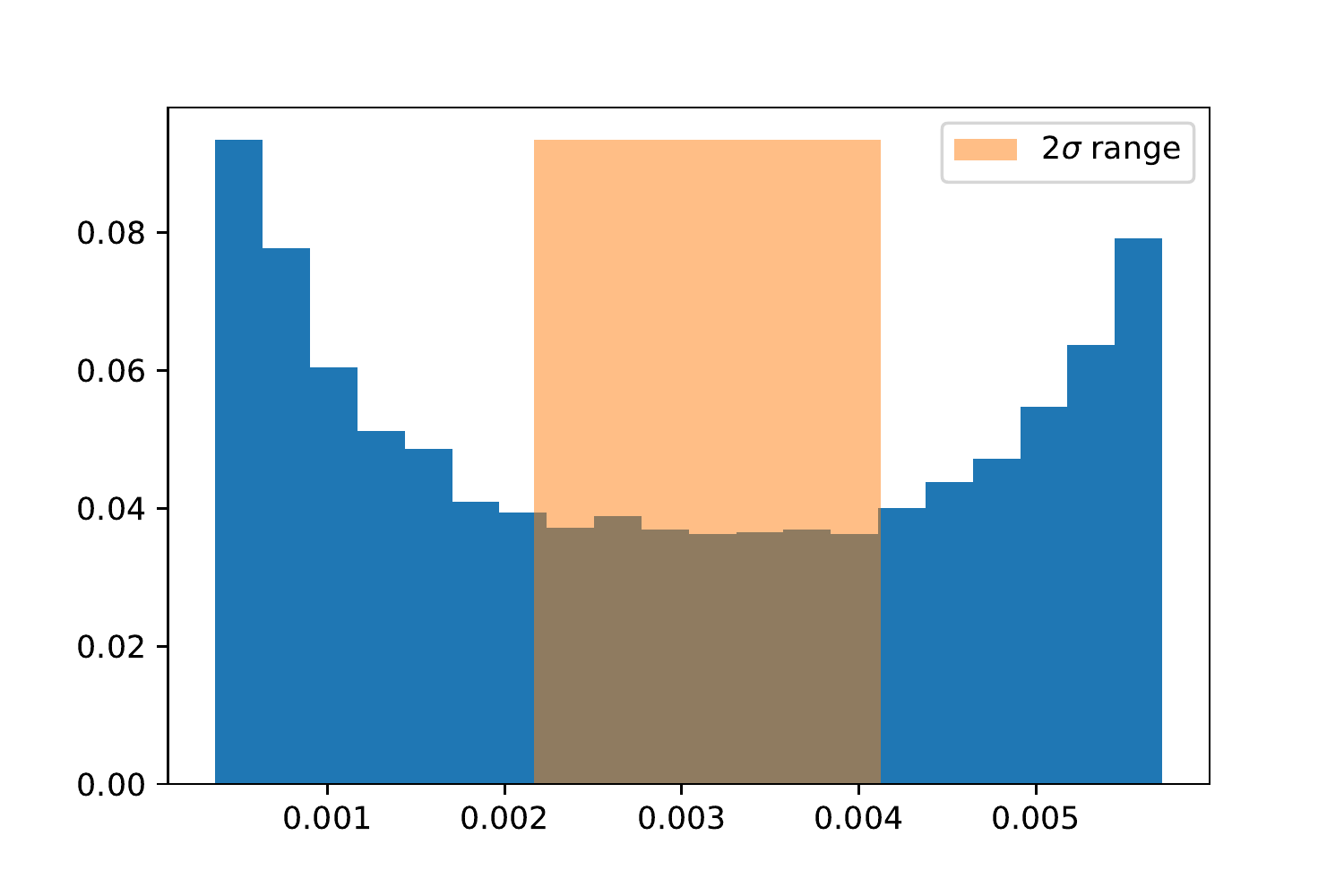}
		\subcaption{Distribution of $\theta_{13}$.}
	\end{minipage}
	\begin{minipage}[b]{0.45\linewidth}
		\centering
		\includegraphics[keepaspectratio, scale = 0.4]{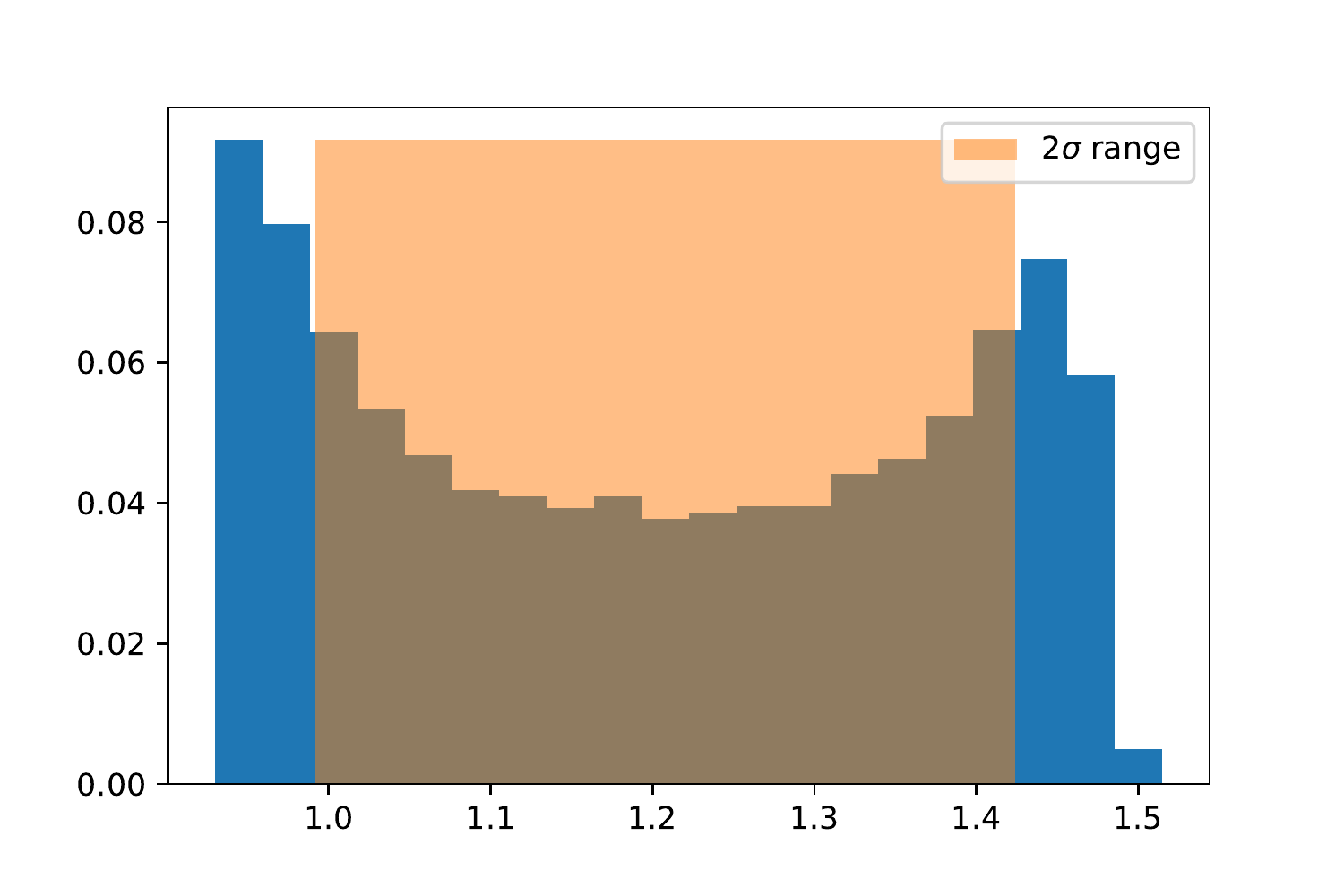}
		\subcaption{Distribution of $\delta_{CP}$.}
	\end{minipage}
\caption{Distribution of the mixing angles and CP-phase.
The blue bars denote the distributions of the mixing angles and the CP-phase. 
The distributions have been normalized already.
The orange regions denote the $2\sigma $ ranges of the corresponding value.}
\label{fig:mixing}
\end{figure}

\begin{figure}[ht]
	\begin{minipage}[b]{0.45\linewidth}
		\centering
		\includegraphics[keepaspectratio, scale=0.4]{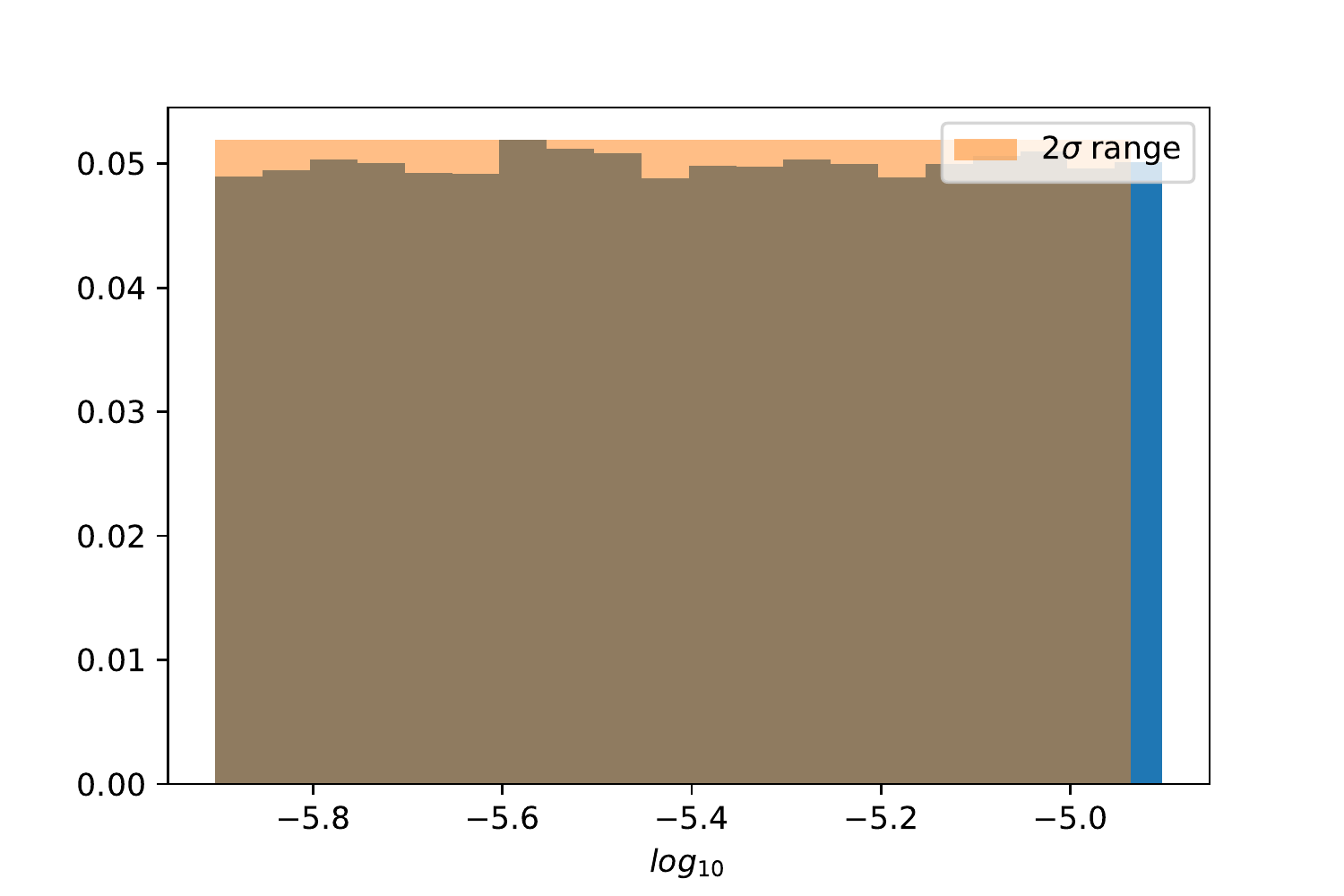}
		\subcaption{Distribution of $y_u/y_t$.}
		\label{Fig:mass_yuyt}
	\end{minipage}
	\begin{minipage}[b]{0.45\linewidth}
		\centering
		\includegraphics[keepaspectratio, scale = 0.4]{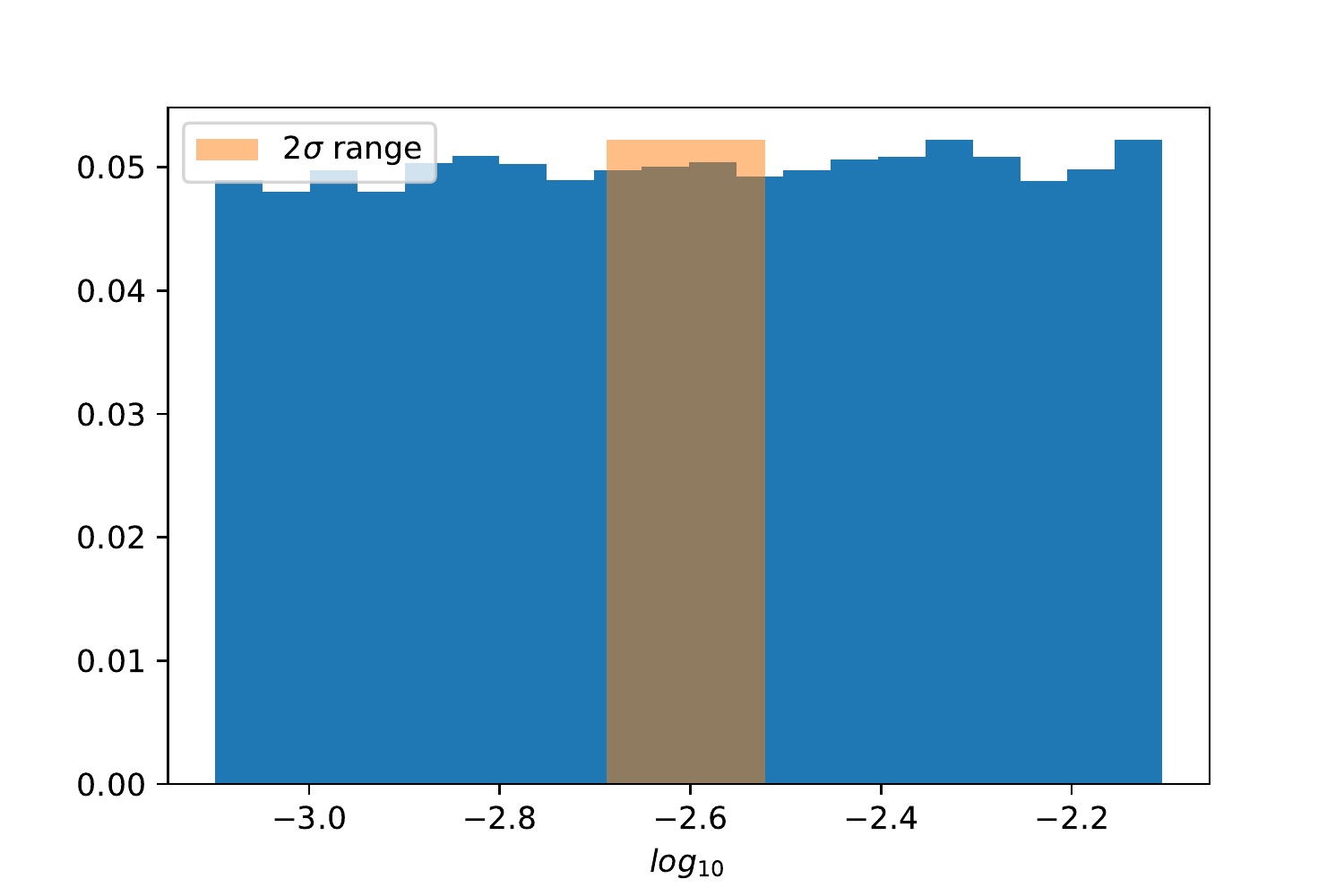}
		\subcaption{Distribution of $y_c/y_t$.}
		\label{Fig:mass_ycyt}
	\end{minipage}
	\\
\begin{minipage}[b]{0.45\linewidth}
		\centering
		\includegraphics[keepaspectratio, scale=0.4]{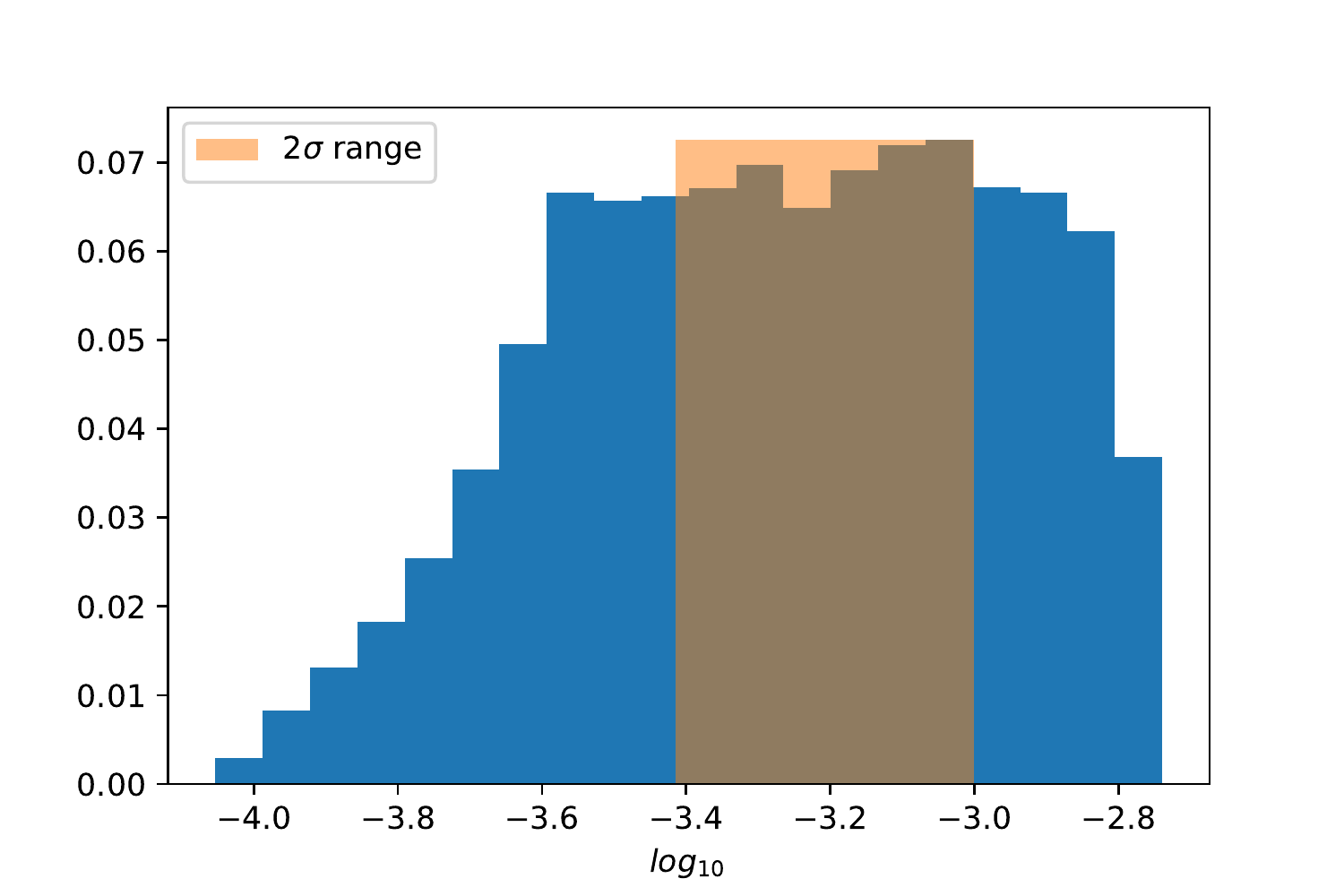}
		\subcaption{Distribution of $y_d/y_b$.}
		\label{Fig:mass_ydyb}
	\end{minipage}
	\begin{minipage}[b]{0.45\linewidth}
		\centering
		\includegraphics[keepaspectratio, scale = 0.4]{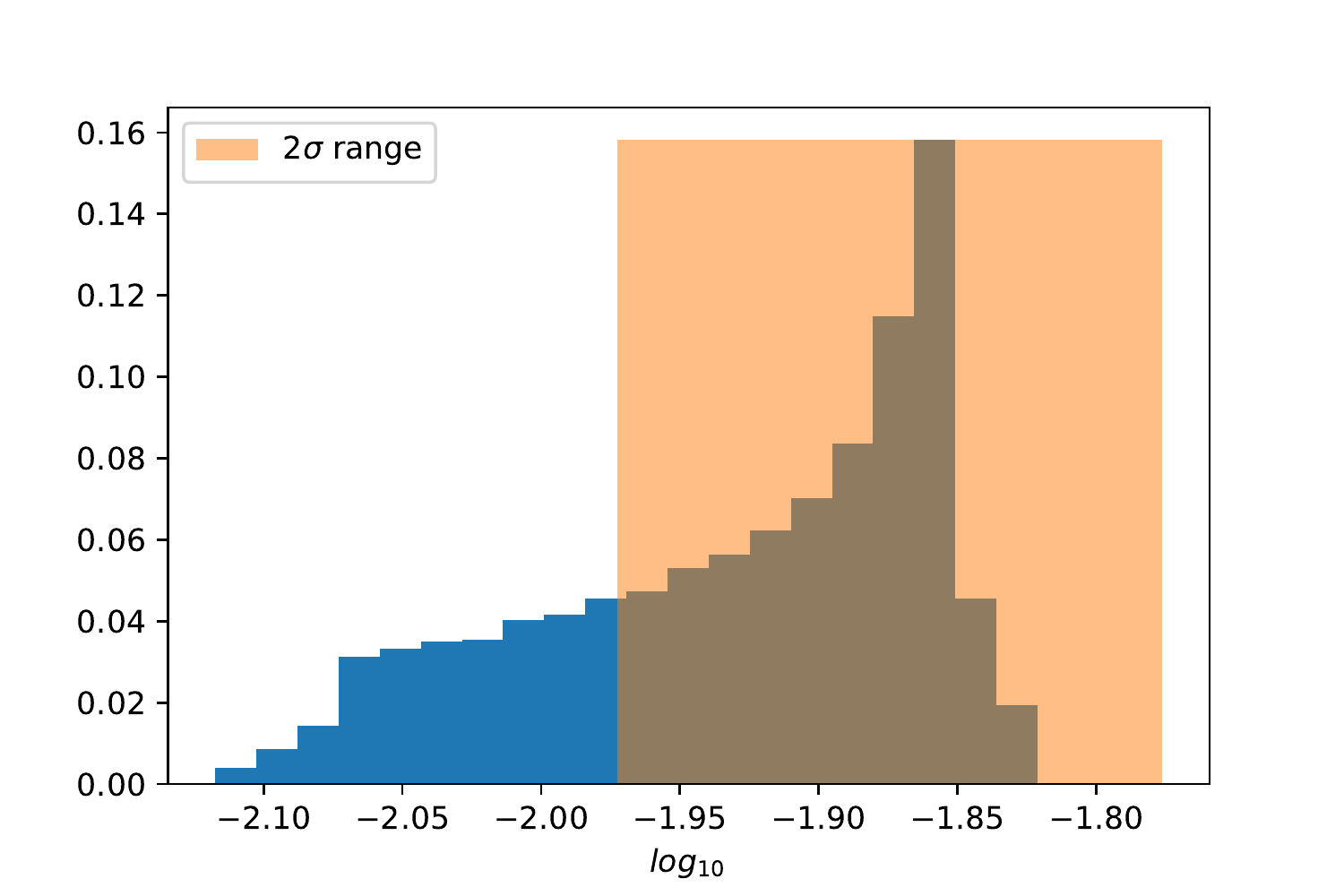}
		\subcaption{Distribution of $y_s/y_b$.}
		\label{Fig:mass_ysyb}
	\end{minipage}
	\\
\caption{Distribution of the mass ratios.
The blue bars denote the distributions of the mass ratios.
The distributions have been normalized already.
The orange region denotes the $2\sigma $ region of the corresponding value.}
\label{Fig:mass}
\end{figure}

To study the stability, 
we see the distributions of the results, 
where we randomly generate four free parameters of 
$(\alpha_u/\gamma_u, \beta_u/\gamma_u, \alpha_d/\gamma_d, \beta_d/\gamma_d)$.
Each parameter is generated as
\begin{align}
x =  \exp(p \log(10^{1/2}))x_{best},~~~~x \in \{\alpha_u/\gamma_u, \beta_u/\gamma_u, \alpha_d/\gamma_d, \beta_d/\gamma_d\},
\label{eq:uniform_log}
\end{align}
where $p$ follows the uniform distribution with minimum $-1$ and maximum $1$.
$x_{best}$ is the best fit value of the coefficients \eqref{eq:coe_best_G5}.
Therefore each of free coefficients fluctuates within the range of $\frac{x_{best}}{\sqrt{10}}<x<\sqrt{10}x_{best}$ 
to keep the same order of magnitude.
Distributions of the mixing angles and the CP-phase are shown in Figure \ref{fig:mixing}.
From the figure we find that the realistic values of the mixing angles are realized without fine-tuning of the free coefficients.
Especially $\theta_{12}$ as well as $\theta_{23}$ are localized around the observed values,  
where approximately half of the configurations reproduces the observed value of $\theta_{12}$ within $2\sigma$ range, 
and all the configurations reproduce the observed value of $\theta_{23}$.
While some configurations are out of $2 \sigma$ range for $\theta_{13}$, its order is realized without any fine-tuning 
as we expected.
$\delta_{CP}$ is also naturally realized, and more than half of the configurations can reproduce the observed value.

\begin{figure}[thb]
	\begin{minipage}[b]{0.45\linewidth}
		\centering
		\includegraphics[keepaspectratio, scale=0.4]{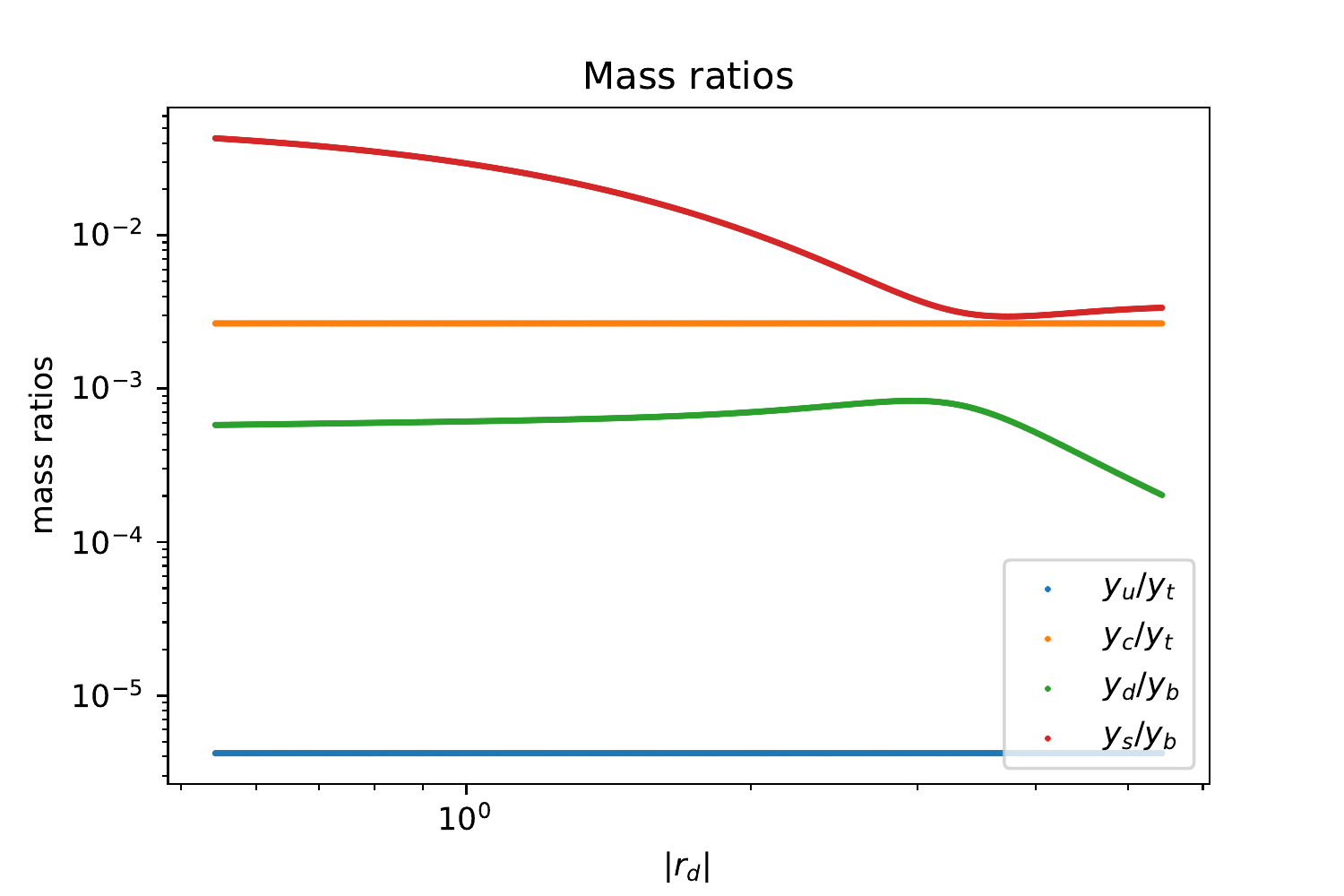}
		\subcaption{$|r_d|$ dependence of mass ratios.}
		\label{fig:mass_rd}
	\end{minipage}
	\begin{minipage}[b]{0.45\linewidth}
		\centering
		\includegraphics[keepaspectratio, scale = 0.4]{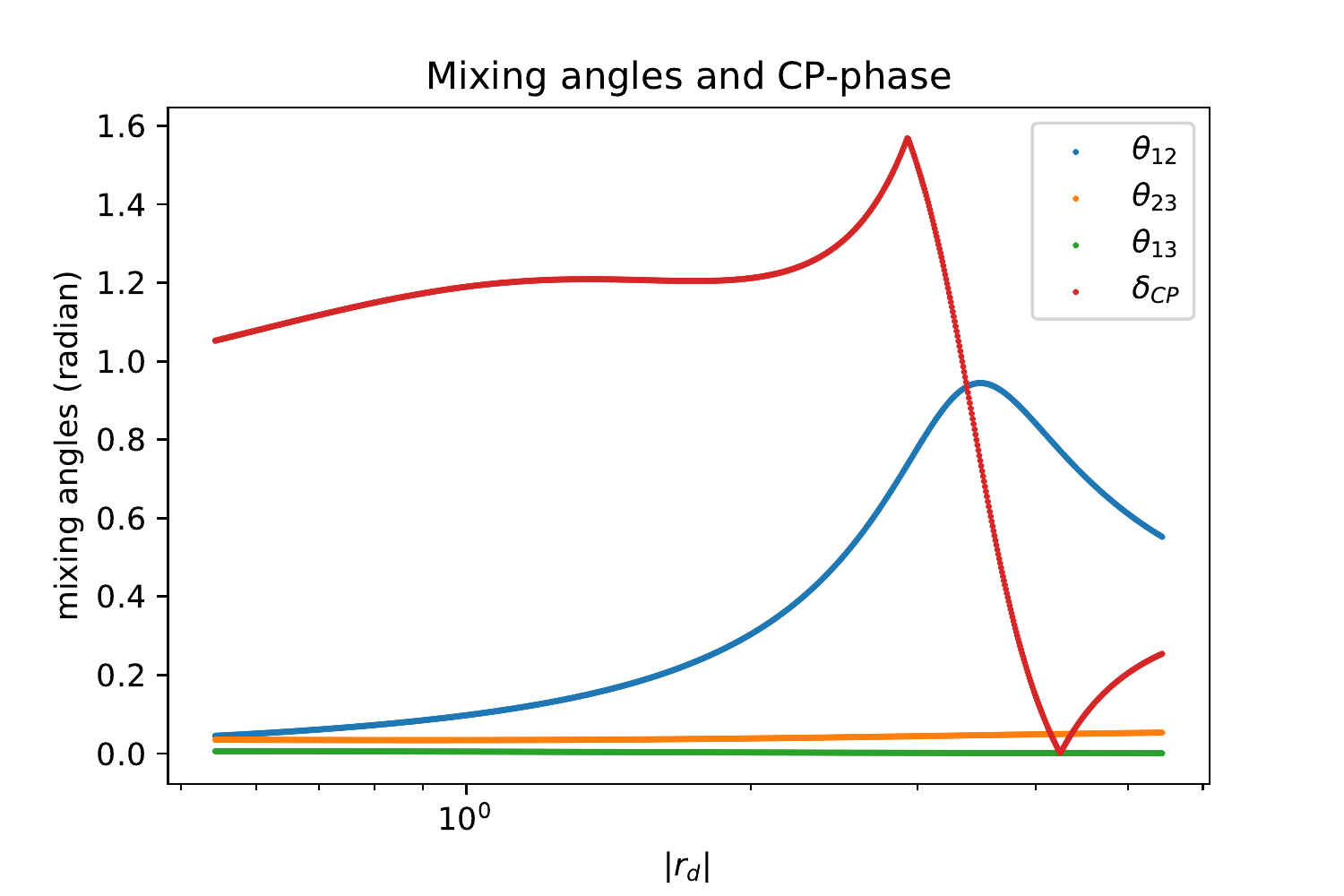}
		\subcaption{$|r_d|$ dependence of mixing angles.}
		\label{fig:mixing_rd}
	\end{minipage}
	\\
	\begin{minipage}[b]{0.45\linewidth}
		\centering
		\includegraphics[keepaspectratio, scale=0.4]{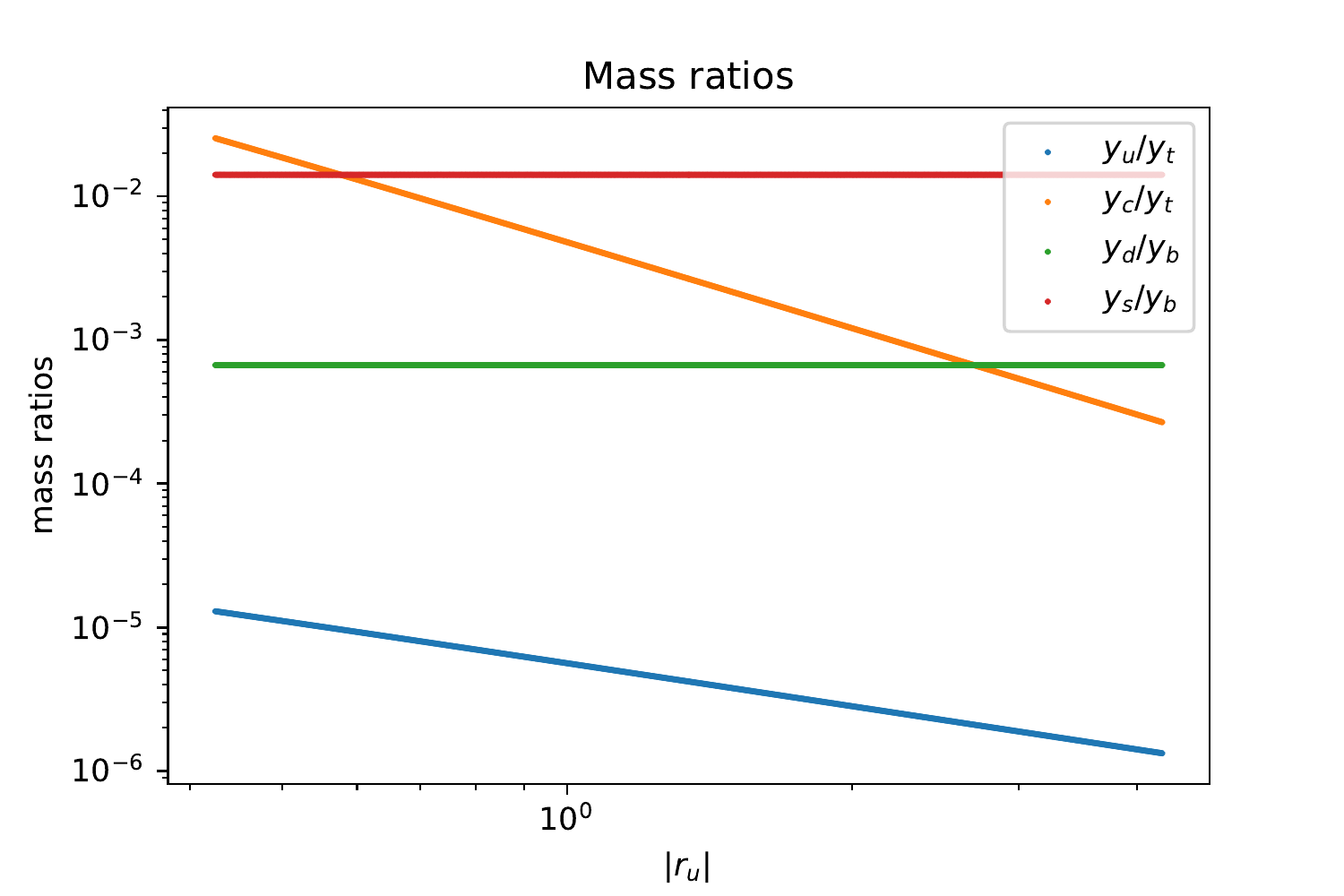}
		\subcaption{$|r_u|$ dependence of mass ratios.}
	\end{minipage}
	\begin{minipage}[b]{0.45\linewidth}
		\centering
		\includegraphics[keepaspectratio, scale = 0.4]{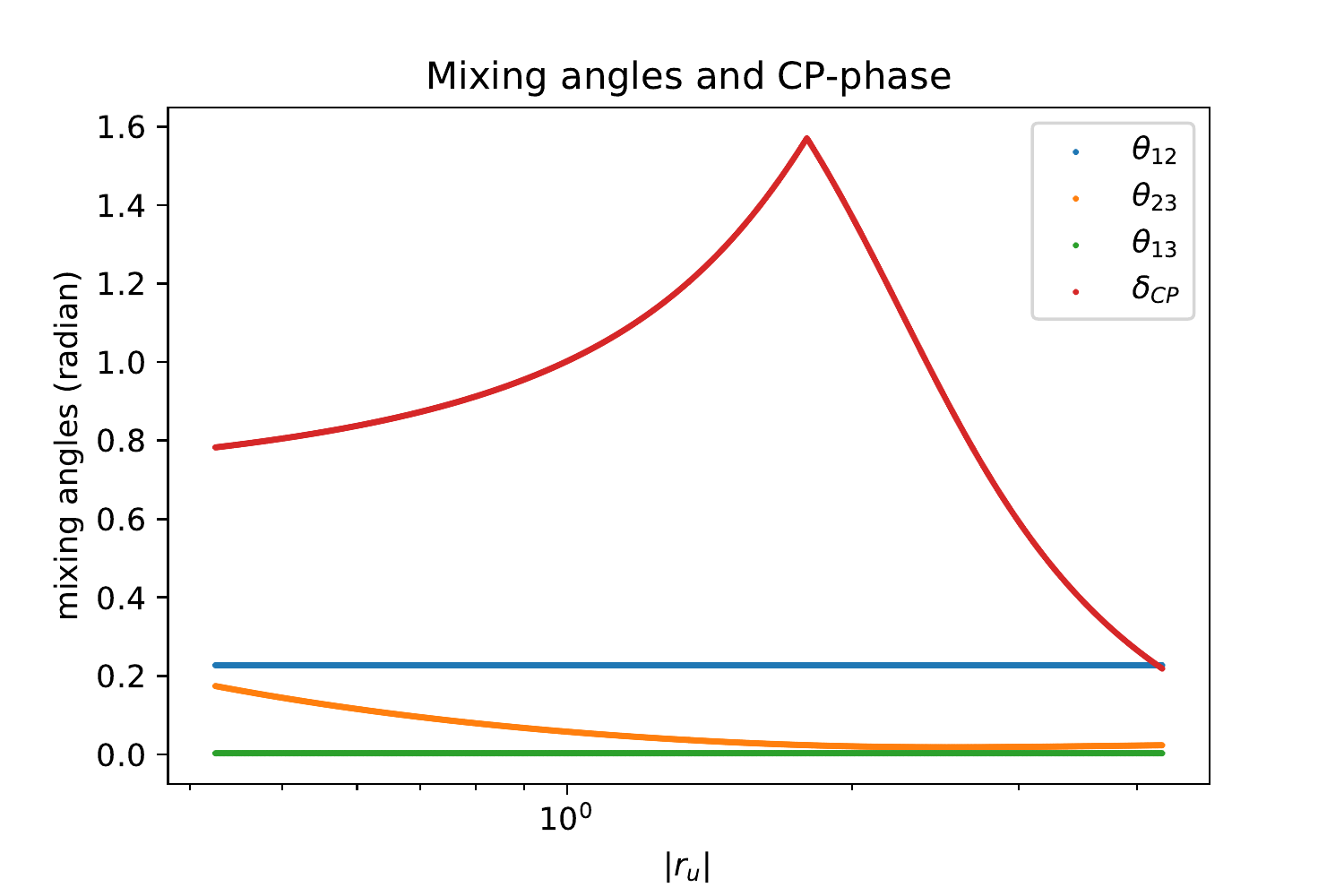}
		\subcaption{$|r_u|$ dependence of mixing angles.}
	\end{minipage}
	\\
	\begin{minipage}[b]{0.45\linewidth}
		\centering
		\includegraphics[keepaspectratio, scale=0.4]{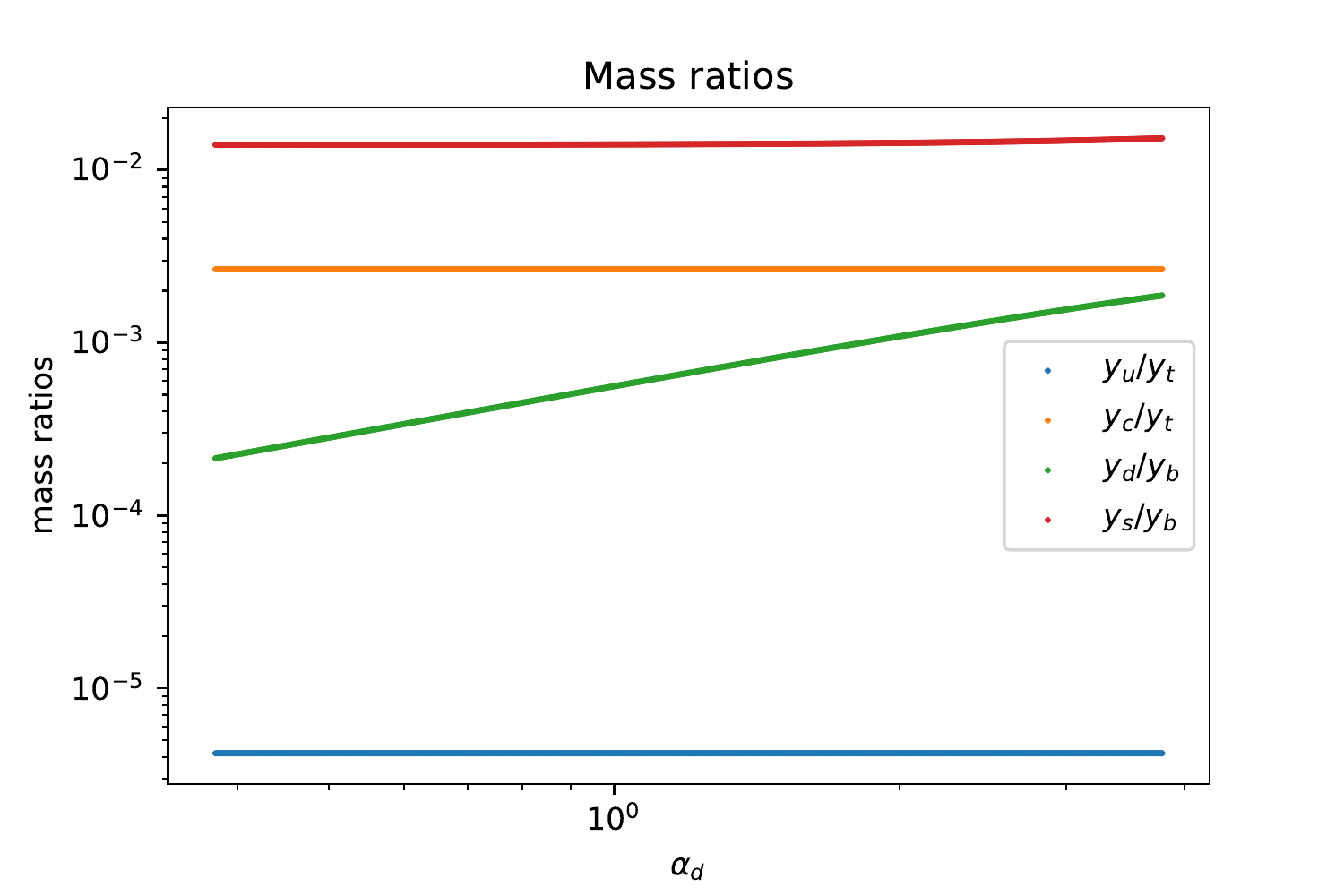}
		\subcaption{$\alpha_d$ dependence of mass ratios.}
		\label{}
	\end{minipage}
	\begin{minipage}[b]{0.45\linewidth}
		\centering
		\includegraphics[keepaspectratio, scale=0.4]{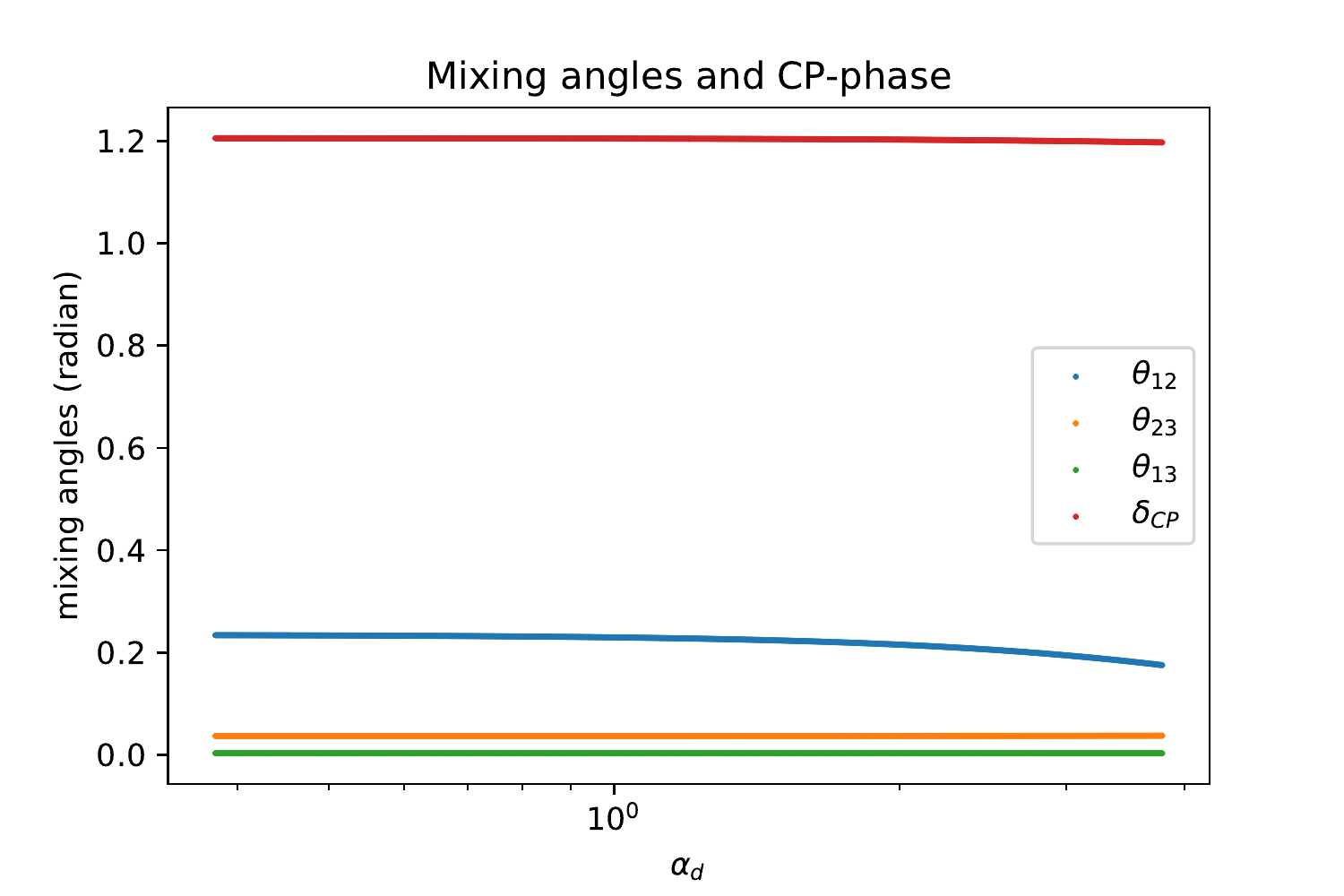}
		\subcaption{$\alpha_d$ dependence of mixing angles.}
	\end{minipage}
	\\
	\begin{minipage}[b]{0.45\linewidth}
		\centering
		\includegraphics[keepaspectratio, scale=0.4]{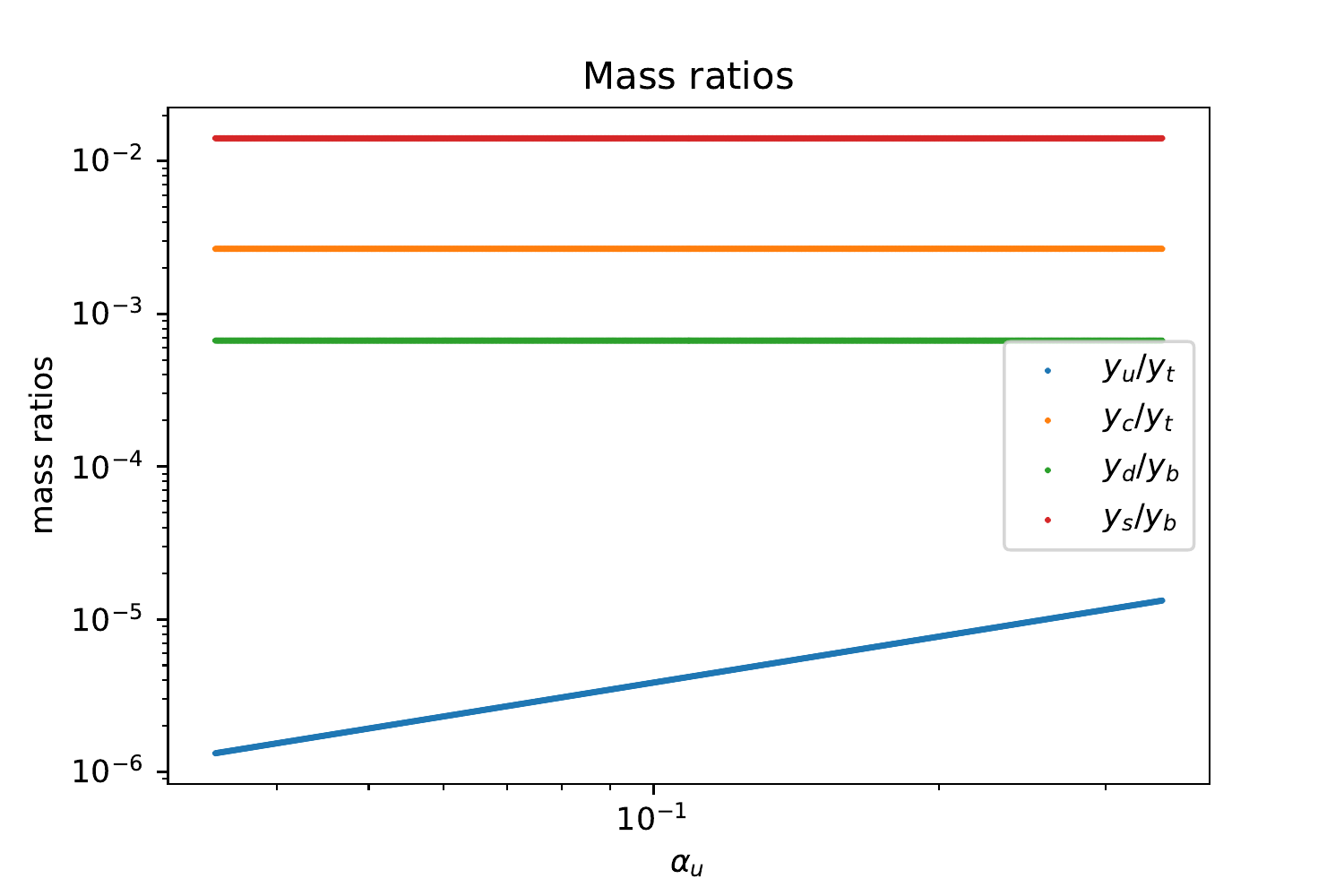}
		\subcaption{$\alpha_u$ dependence of mass ratios.}
	\end{minipage}
	\begin{minipage}[b]{0.45\linewidth}
		\centering
		\includegraphics[keepaspectratio, scale=0.4]{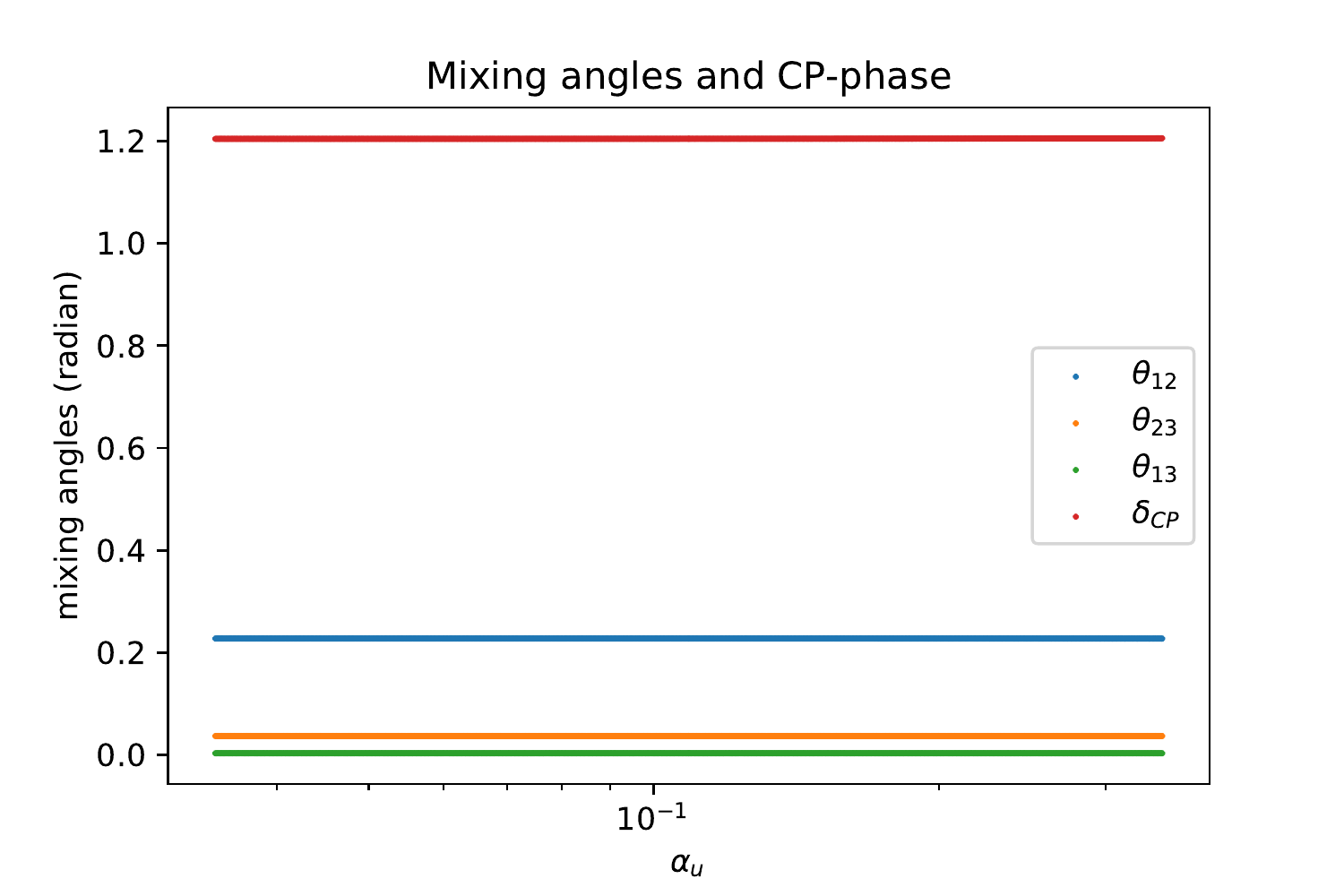}
		\subcaption{$\alpha_u$ dependence of mixing angles.}
	\end{minipage}
	\\
\caption{Parameter dependence of the mass ratios, mixing angles and CP-phase.
We show $|r_{u,d}|$ dependence since the mixing angles are explicitly depend on them at the leading order. 
We also show $\alpha_d$ and $\alpha_u$ dependence of the parameters for comparison purpose.
The observables (except for the corresponding mass eigenvalues) are almost independent of $\alpha_u$ and $\alpha_d$.}
\label{fig:r_aubu}
\end{figure}

Figure \ref{Fig:mass} shows the distributions of the mass ratios.
Figures \ref{Fig:mass_yuyt} and \ref{Fig:mass_ycyt} show that $y_u/y_t$ and $y_c/y_t$ are uniformly distributed as we expected.
On the other hand, the mass ratios of the down sectors in Figures \ref{Fig:mass_ydyb} and \ref{Fig:mass_ysyb}  do not follow the uniform distribution.
It is consistent with the fact that $M_d$ is less hierarchical than $M_u$ and the largest components of the mass matrix is exchangeable 
if the coefficients are significantly apart from the best fit point (See also Figure~\ref{fig:mass_rd} for singular behavior).  
However, the mass ratios are localized around the observed values even for $M_d$.
Thus the physical parameters are stable under perturbations of the coefficients.

The parameter dependence of the mixing angles and the mass eigenvalues are shown in Figure \ref{fig:r_aubu}.
The mass ratios in up sector and down sector depend on $|r_u|$ and $|r_d|$, respectively. 
We also find that $\theta_{12}$ drastically changes, although it is independent of $|r_{u,d}|$ in our approximation \eqref{eq:mixing_G5}.
This is because large $|r_d|$ can exchange the largest component of $M_d$ as well and 
our approximation is no longer valid in such regions.
In fact, Figure \ref{fig:mixing_rd} shows that $\theta_{12}$ have a peak around 
$|r_d| \sim 3$, and Figure \ref{fig:mass_rd} shows that $y_s$ and $y_d$ get closer at the same point, 
where our approximation becomes invalid and the mixing angles are unstable. 
Apart from such region, the order of all the mixing angles are correctly reproduced.
We also show the $\alpha_{u,d}$ dependence in comparison. 
In contrast to $r_{u,d}$, the mixing angles are almost independent of $\alpha_u$ and $\alpha_d$, 
which is consistent with the approximation in \eqref{eq:mixing_G5}.

Similar results can be obtained in the models with $\Gamma_3'$ and $\Gamma_4'$.
Therefore, in any cases of our FN-like models the approximate estimation is valid and useful 
to analyze the relation between the mass ratios and the mixing angles, 
especially for the up-type mass matrix due to its large hierarchy.
On the other hand, the down-type mass matrix is less hierarchical and 
the mixing angles may fluctuate depending on the coefficients in the down sectors. 
Nevertheless all of our models can reproduce the correct orders of the mixing angles and the mass ratios 
under perturbations of the $\mathcal{O}(1)$ coefficients.

\clearpage

\section{Conclusion}

We have studied the mass hierarchy and the mixing angles in quark sector 
based on the FN-like mechanism in the framework of the modular symmetry. 
We have assumed that each of singlet quarks have a common representation under the modular symmetry
but with a different modular weight, 
which is important to obtain the CP-phase as well as a full rank mass matrix.  
We have introduced a scalar $\phi$ with a negative modular weight.
The allowed Yukawa couplings are then suppressed by powers of the scalar vev ($\langle \phi \rangle/\Lambda$) 
due to the FN-like mechanism, and the mass hierarchy is originated from 
powers of $\langle \phi \rangle/\Lambda$ and the hierarchical modular forms. 
Using $q$-expansions of the modular forms we have analyzed an approximate expression of the Yukawa matrices, 
where the same order of suppression factors appears in each column or each row of the Yukawa matrices, 
so that our models can simply realize the FN-like matrices for the up sector and down sector simultaneously. 
We have illustrated this mechanism in models with different finite modular groups of 
$\Gamma_3', \Gamma_4'$ and $\Gamma_5'$ in detail.
As the result, all of our models can reproduce the correct orders of observed mass ratios and mixing angles 
by choosing the modular parameters and the singlet vev. 
The best model is constructed in the model with $\Gamma_4'$, 
where the Yukawa couplings and the left-handed quarks $Q$ have to be the same representation 
to obtain small mixing angles.
In this model, we have approximately reproduced all the 9 observables by tuning 8 parameters, 
for which we do not require any unnatural hierarchical coefficients.  
A statistical investigation has also been carried out to study the stability of our results.  
We have shown that the approximate estimation is valid 
and our results are stable against perturbations of the $\mathcal{O}(1)$ coefficients, 
especially for the up sector due to its large hierarchical structure.

Throughout this paper, we have not studied the lepton sector, 
while it is interesting to see if the lepton mass spectra and the neutrino mixing angles can be realized 
based on this models. 
We assume that $\phi$ is stabilized and is safely decoupled.
To stabilize the weighton, we require superpotetial in terms of $\phi$.
The superpotential of $\phi$ is modular invariant as well.
Thus it is also restricted by the weight and representations.
We should investigate its vacuum structure to see if $\phi$ can develop our desired value of vev. 
The potential of $\phi$ is also interesting from the phenomenological point of view.
$\phi$ may be related to beyond the SM physics, such as SUSY breaking, 
soft-terms, and $\mu$-terms.
The deviation between Majonara mass scale and GUT scale may be originated from the VEV of $\phi$, too.
We assume $\phi$ is the trivial singlet of the finite modular group.
$\phi$ can be non-trivial singlet such as ${\bf \hat 1}$ in $S_4'$, and such $\phi$ possibly change the phenomenology.
In this case $\phi^4$ is the trivial singlet, and it looks like a flavon of the usual $\mathbb{Z}_4$ flavor symmetric models.
Such a flavon may be important for physics around the standard model \cite{Higaki:2019ojq}.
The stringy origin of our model is also unclear and it should be investigated.
However, these topics are beyond our scope in this paper,
and we will study them elsewhere.

\section*{Acknowledgments}

H.~O. is supported in part by JSPS KAKENHI Grants 
No.\,17K14309, No.\,18H03710, and No.\,21K03554.
The authors also thank the Yukawa Institute for Theoretical Physics at Kyoto University, where this work was initiated during the YITP-W-20-08 on ``Progress in Particle Physics 2020".

\appendix
\section{The modular group of level 3}
\label{app:level_3}

In this appendix, we briefly review the modular forms of level 3 and develop our notation.
Complete explanation is not our purpose.
We would provide a minimal toolkit necessary for our analysis.

$\Gamma_3'$ is isomorphic to $T'$.
$T'$ has seven irreducible representations,
\begin{align}
{\bf 1, 1',  1'', 2,  2', 2'',  3}.
\end{align}
$T'$ is generated by $S$, $T$ and $R$.
Thus it is sufficient to study the matrix representations of these two elements.
The matrix representations of $T'$ have ambiguities.
In this paper, we follow the notation of \cite{Liu:2019khw}.
The matrix representations are summarized in Table \ref{tab:irrep_T'}.
\begin{table}[th]
\begin{center}
\begin{tabular}{l|ccc} \hline
{\bf r} & $S$ & $T$ & $R$\\
\hline\hline
${\bf 1}$ & 1 & 1  & 1
\\
${\bf 1}'$ & 1 & $\omega$ & 1
\\
${\bf 1}''$ & 1 & $\omega^2$ & 1 
\\
${\bf 2}$ &  
$-\frac 1 {\sqrt 3}
\begin{pmatrix}
i & \sqrt 2 e^{i\pi/12}\\
-\sqrt 2 e^{-i\pi/12} & i
\end{pmatrix}$
&
$\begin{pmatrix}
\omega & 0\\
0 & 1
\end{pmatrix}$
& 
$-I_{2\times2} $
\\[15pt]
${\bf 2}'$ &
$-\frac 1 {\sqrt 3}
\begin{pmatrix}
i & \sqrt 2 e^{i\pi/12}\\
-\sqrt 2 e^{-i\pi/12} & i
\end{pmatrix}$
&
$\omega
\begin{pmatrix}
\omega & 0\\
0 & 1
\end{pmatrix}$
& 
$-I_{2\times2} $
\\[15pt]
${\bf 2}''$ &
$-\frac 1 {\sqrt 3}
\begin{pmatrix}
i & \sqrt 2 e^{i\pi/12}\\
-\sqrt 2 e^{-i\pi/12} & i
\end{pmatrix}$
&
$\omega^2
\begin{pmatrix}
\omega & 0\\
0 & 1
\end{pmatrix}$
&
$-I_{2\times2} $
\\[15pt]
${\bf 3}$ 
&
$-\frac 1 3
\begin{pmatrix}
-1 & 2 & 2 \\
2 & -1 & 2 \\
2 & 2 & -1 
\end{pmatrix}$
&
$\begin{pmatrix}
1 & 0 & 0 \\
0 & \omega & 0 \\
0 & 0 & \omega^2 
\end{pmatrix}$
& $I_{3\times3}$
\\[15pt]
\hline
\end{tabular}
\caption{The matrix representations of $S$ and $T$. 
$\omega = e^{2\pi i/3}$.
Our notation is base on \cite{Liu:2019khw}.}
\label{tab:irrep_T'}
\end{center}
\end{table}

The irreducible decomposition of the tensor product of the singlets are trivial:
\begin{align}
&{\bf 1} \otimes {\bf 1} = {\bf 1}' \otimes {\bf 1}'' = {\bf 1},
\nonumber
\\
&{\bf 1} \otimes {\bf 1}' ={\bf 1}'' \otimes {\bf 1}'' = {\bf 1}',
\nonumber
\\
&{\bf 1} \otimes {\bf 1}'' =  {\bf 1}' \otimes {\bf 1}' = {\bf 1}''.
\nonumber
\end{align}
We also study the irreducible decomposition of the tensor product including ${\bf 3}$, i.e., Clebsh-Gordon coefficients.
It is given by
\begin{align}
\begin{pmatrix}
\alpha_1\\
\alpha_2\\
\alpha_3\\
\end{pmatrix}_{\bf 3}
\otimes
\begin{pmatrix}
\beta_1\\
\beta_2\\
\beta_3\\
\end{pmatrix}_{\bf 3}
=
&
\left(\alpha_1 \beta_1 +\alpha_2 \beta_3 +\alpha_3 \beta_2\right)_{\bf 1} 
+ \left(\alpha_1 \beta_2 +\alpha_2 \beta_1 +\alpha_3 \beta_3\right)_{\bf 1'} 
+ \left(\alpha_1 \beta_3 +\alpha_2 \beta_2 +\alpha_3 \beta_1\right)_{\bf 1''} 
\nonumber
\\
&
+ 
\begin{pmatrix}
2\alpha_1 \beta_1 -\alpha_2 \beta_3 - \alpha_3 \beta_2\\
2\alpha_3 \beta_3 -\alpha_1 \beta_2 - \alpha_2 \beta_1\\
2\alpha_2 \beta_2 -\alpha_1 \beta_3 - \alpha_3 \beta_1
\end{pmatrix}_{{\bf 3}S}
+ 
\begin{pmatrix}
\alpha_2 \beta_3 -\alpha_3 \beta_2 \\
\alpha_1 \beta_2 -\alpha_2 \beta_1 \\
\alpha_3 \beta_1 -\alpha_1 \beta_3 \\
\end{pmatrix}_{{\bf 3}A}.
\label{eq:CG_level_3}
\end{align}

\subsection*{Modular forms}

The modular forms of level 3 and weight $k$ are given by tensor products of the modular forms of level 3 and weight 1 \eqref{eq:level3_lowest}.
In this appendix, we consider modular forms which are ${\bf 3}$ of $T'$.
The complete set of the modular forms whose weights are lower than 6 can be found in \cite{Liu:2019khw}.
The modular forms of weight 2 are given by the tensor product,
\begin{align}
Y_{\bf 3}^{(2)} = (Y_{\bf 2}^{(1)} \otimes Y_{\bf 2}^{(1)})_{{\bf 3}S} = 
\begin{pmatrix}
e^{i\pi/6} Y_2^2\\
\sqrt{2} e^{i\pi7/12} Y_1 Y_2\\
Y_1^2\\
\end{pmatrix}
\sim
\begin{pmatrix}
\frac 1 9 e^{\pi i/6}\\
-\frac 2 3 e^{\pi i/6}q^{1/3}\\
-2 e^{\pi i/6} q^{2/3}
\end{pmatrix}.
\end{align}
We can obtain the modular forms of higher weights as
\begin{align}
Y_{\bf 3}^{(4)} = 
\begin{pmatrix}
\sqrt 2 e^{i\pi 7/12}Y_1^3 Y_2 -e^{i\pi /3} Y_2^4\\
-Y_1^4 -(1-i) Y_1 Y_2^3\\
3e^{i\pi/6} Y_1^2 Y_2^2
\end{pmatrix}
\sim 
\begin{pmatrix}
\frac 1 {81} e^{4\pi i/3}\\
\frac {2}{27} e^{4\pi i/3} q^{1/3}\\
\frac {2}{3} e^{4\pi i/3} q^{2/3}
\end{pmatrix}
\end{align}
and 
\begin{align}
Y_{{\bf 3},I}^{(6)} &= 
\begin{pmatrix}
-2(1-i)Y_1^3 Y_2^3 + i Y_2^6\\
-4e^{i\pi/6}Y_1^4 Y_2^2 - (1-i) e^{i\pi/6} Y_1 Y_2^5\\
2\sqrt2 e^{i\pi 7/12}Y_1^5 Y_2 + e^{i\pi/3} Y_1^2 Y_2^4
\end{pmatrix}
\sim
\frac i {81}
\begin{pmatrix}
\frac 1 {9} \\
-\frac {2}{3}  q^{1/3}\\
-2 q^{2/3}
\end{pmatrix}
,
\nonumber
\\
Y_{{\bf 3},II}^{(6)} &= 
\begin{pmatrix}
Y_1^6 - 2(1-i) Y_1^3 Y_2^3\\
e^{i\pi/6}Y_1^4 Y_2^2 - 2 (1-i) e^{i\pi/6} Y_1 Y_2^5\\
-4  e^{i\pi /3}Y_1^2 Y_2^4 + (1+i) e^{i\pi/3} Y_1^5 Y_2
\end{pmatrix}
\sim
\frac i {81}
\begin{pmatrix}
-24 q \\
- \frac {4}{3}  q^{1/3}\\
8  q^{2/3}
\end{pmatrix}.
\label{eq:mf_lev3_wei6}
\end{align}
These expansions are consistent with \eqref{eq:app_Y3k}.

\section{The modular forms of level 4}
\label{app:level_4}

$\Gamma_4'$ is isomorphic to $S_4' \simeq SL(2,\mathbb{Z}_4)$, which is a double covering group of $S_4$.
$S_4'$ has the following irreducible representations,
\begin{align}
{\bf 1, \hat 1, 1', \hat 1', 2, \hat 2, 3, \hat 3, 3', \hat 3'}.
\end{align}
Our notation follows \cite{Novichkov:2020eep}.
The matrix representations are summarized in Table \ref{tab:irrep_S4'}.
\begin{table}[t]
\begin{center}
\begin{tabular}{l|ccc} \hline
{\bf r} & $S$ & $T$ & $R$\\
\hline\hline
${\bf 1}$ & 1 & 1  & 1
\\
$\hat{\bf 1}$ & i & $-i$ & $-1$
\\
${\bf 1}'$ & $-1$ & $-1$ & $1$ 
\\
$\hat{\bf 1}'$ & $-i$ & $i$ & $-1$
\\
${\bf 2}$ &  
$\frac {1}{2}
\begin{pmatrix}
-1 & \sqrt 3\\
-\sqrt 3 & 1
\end{pmatrix}$
&
$\begin{pmatrix}
1 & 0\\
0 & -1
\end{pmatrix}$
& 
$I_{2\times2} $
\\[15pt]
$\hat{\bf 2}$ &  
$\frac {i}{2}
\begin{pmatrix}
-1 & \sqrt 3\\
-\sqrt 3 & 1
\end{pmatrix}$
&
$\begin{pmatrix}
-i & 0\\
0 & i
\end{pmatrix}$
& 
$-I_{2\times2} $
\\[15pt]
${\bf 3}$ 
&
$-\frac 1 2
\begin{pmatrix}
0 & \sqrt 2 & \sqrt 2 \\
\sqrt 2 & -1 & 1 \\
\sqrt 2 & 1 & -1 
\end{pmatrix}$
&
$\begin{pmatrix}
-1 & 0 & 0 \\
0 & -i & 0 \\
0 & 0 & i
\end{pmatrix}$
& $I_{3\times3}$
\\[20pt]
$\hat{\bf 3}$ 
&
$-\frac i 2
\begin{pmatrix}
0 & \sqrt 2 & \sqrt 2 \\
\sqrt 2 & -1 & 1 \\
\sqrt 2 & 1 & -1 
\end{pmatrix}$
&
$\begin{pmatrix}
i & 0 & 0 \\
0 & -1 & 0 \\
0 & 0 & 1
\end{pmatrix}$
& $- I_{3\times3}$
\\[20pt]
${\bf 3}'$ 
&
$ \frac 1 2
\begin{pmatrix}
0 & \sqrt 2 & \sqrt 2 \\
\sqrt 2 & -1 & 1 \\
\sqrt 2 & 1 & -1 
\end{pmatrix}$
&
$\begin{pmatrix}
1 & 0 & 0 \\
0 & i & 0 \\
0 & 0 & -i
\end{pmatrix}$
& $I_{3\times3}$
\\[20pt]
$\hat{\bf 3}'$ 
&
$\frac i 2
\begin{pmatrix}
0 & \sqrt 2 & \sqrt 2 \\
\sqrt 2 & -1 & 1 \\
\sqrt 2 & 1 & -1 
\end{pmatrix}$
&
$\begin{pmatrix}
-i & 0 & 0 \\
0 & 1 & 0 \\
0 & 0 & -1
\end{pmatrix}$
& $- I_{3\times3}$
\\[15pt]
\hline
\end{tabular}
\caption{The matrix representations of the generators of $\Gamma_4'$. 
Our notation is base on \cite{Novichkov:2020eep}.}
\label{tab:irrep_S4'}
\end{center}
\end{table}
The complete table of the irreducible decomposition of the tensor products, and their CG coefficients also be found in \cite{Novichkov:2020eep}.
We summarize necessary part here.
$\hat{\bf 1}$ corresponds to $e^{i \pi/2}$.
$\hat{\bf 1}^2 ={\bf 1}'$, and $({\bf 1}')^2 = {\bf  1}$.
Thus the irreducible decomposition of the tensor product of the singlets are trivial:
\begin{align}
&{\bf 1} \otimes {\bf 1} = \hat{\bf 1} \otimes \hat{\bf 1}' = {\bf 1}' \otimes {\bf 1}' = {\bf 1},~~
{\bf 1} \otimes {\bf 1}' = \hat{\bf 1} \otimes \hat{\bf 1} = \hat{\bf 1}' \otimes \hat{\bf 1}' = {\bf 1}',
\nonumber
\\
&{\bf 1} \otimes \hat{\bf 1} = {\bf 1}' \otimes \hat{l\bf 1}' =  \hat{\bf 1},~~
{\bf 1} \otimes \hat{\bf 1}' = {\bf 1}' \otimes \hat{\bf 1} =  \hat{\bf 1}',
\nonumber
\end{align}
We also require the irreducible decomposition of the tensor products including the triplets, and their Clebsh-Gordon coefficients.
They are classified to two cases.
For the first case, the irreducible decomposition are given by
\begin{align}
{\bf 3} \otimes {\bf 3} &= {\bf 1} \oplus {\bf 2} \oplus {\bf 3} \oplus {\bf 3}',
\nonumber
\\
{\bf 3} \otimes \hat{\bf 3} &= \hat{\bf 1} \oplus \hat{\bf 2} \oplus \hat{\bf 3} \oplus \hat{\bf 3}',
\nonumber
\\
{\bf 3}' \otimes  {\bf 3}' &= {\bf 1} \oplus  {\bf 2} \oplus {\bf 3} \oplus {\bf 3}',
\nonumber
\\
{\bf 3}' \otimes \hat{\bf 3}' &= \hat{\bf 1} \oplus \hat{\bf 2} \oplus \hat{\bf 3} \oplus \hat{\bf 3}',
\nonumber
\\
\hat{\bf 3} \otimes \hat{\bf 3} &= {\bf 1} \oplus  {\bf 2} \oplus {\bf 3} \oplus {\bf 3}',
\nonumber
\end{align}
and the CG-coefficients for these five cases are given as follows
\begin{align}
\begin{pmatrix}
\alpha_1\\
\alpha_2\\
\alpha_3\\
\end{pmatrix}
\otimes
\begin{pmatrix}
\beta_1\\
\beta_2\\
\beta_3\\
\end{pmatrix}
=
&
\frac 1 {\sqrt{3}}
\left(\alpha_1 \beta_1 +\alpha_2 \beta_3 +\alpha_3 \beta_2\right)
+
\frac 1 {\sqrt{2}}
\begin{pmatrix}
(2\alpha_1 \beta_1 -\alpha_2 \beta_3 -\alpha_3 \beta_2)/\sqrt{3}\\
\alpha_2 \beta_2 + \alpha_3 \beta_3
\end{pmatrix}
\nonumber
\\
&
+ 
\frac{1}{\sqrt 2}
\begin{pmatrix}
\alpha_3 \beta_3 - \alpha_2 \beta_2 \\
\alpha_1 \beta_3 + \alpha_3 \beta_1 \\
-\alpha_1 \beta_2 - \alpha_2 \beta_1 \\
\end{pmatrix}
+ 
\frac{1}{\sqrt 2}
\begin{pmatrix}
\alpha_3 \beta_2 - \alpha_2 \beta_3 \\
\alpha_2 \beta_1 - \alpha_1 \beta_2 \\
\alpha_1 \beta_3 - \alpha_3 \beta_1 \\
\end{pmatrix}.
\label{eq:CG_level4_1}
\end{align}
For the second case, the irreducible decomposition of the tensor products are given by
\begin{align}
{\bf 3} \otimes {\bf 3}' &= {\bf 1}' \oplus {\bf 2} \oplus {\bf 3} \oplus {\bf 3}',
\nonumber
\\
{\bf 3} \otimes \hat{\bf 3}' &= \hat{\bf 1}' \oplus \hat{\bf 2} \oplus \hat{\bf 3} \oplus \hat{\bf 3}',
\nonumber
\\
{\bf 3}' \otimes  \hat{\bf 3} &= \hat{\bf 1}' \oplus  \hat{\bf 2} \oplus \hat{\bf 3} \oplus \hat{\bf 3}',
\nonumber
\\
\hat{\bf 3} \otimes \hat{\bf 3} &= {\bf 1}' \oplus {\bf 2} \oplus {\bf 3} \oplus {\bf 3}',
\nonumber
\\
\hat{\bf 3}' \otimes \hat{\bf 3}' &= {\bf 1}' \oplus  {\bf 2} \oplus {\bf 3} \oplus {\bf 3}',
\nonumber
\end{align}
and the CG-coefficients are summarized by
\begin{align}
\begin{pmatrix}
\alpha_1\\
\alpha_2\\
\alpha_3\\
\end{pmatrix}
\otimes
\begin{pmatrix}
\beta_1\\
\beta_2\\
\beta_3\\
\end{pmatrix}
=
&
\frac 1 {\sqrt{3}}
\left(\alpha_1 \beta_1 +\alpha_2 \beta_3 +\alpha_3 \beta_2\right)
+
\frac 1 {\sqrt{2}}
\begin{pmatrix}
\alpha_2 \beta_2 + \alpha_3 \beta_3\\
(-2\alpha_1 \beta_1 + \alpha_2 \beta_3 + \alpha_3 \beta_2)/\sqrt{3}
\end{pmatrix}
\nonumber
\\
&
+ 
\frac{1}{\sqrt 2}
\begin{pmatrix}
\alpha_3 \beta_2 - \alpha_2 \beta_3 \\
\alpha_2 \beta_1 - \alpha_1 \beta_2 \\
\alpha_1 \beta_3 - \alpha_3 \beta_1 \\
\end{pmatrix}
+ 
\frac{1}{\sqrt 2}
\begin{pmatrix}
\alpha_3 \beta_3 - \alpha_2 \beta_2 \\
\alpha_1 \beta_3 + \alpha_3 \beta_1 \\
-\alpha_1 \beta_2 - \alpha_2 \beta_1 \\
\end{pmatrix}.
\end{align}

\subsection*{Modular forms}

The modular forms of level 4 and weight $k$ are given by tensor products of the modular forms of level 4 and weight 1.
The modular forms of weight 1 form $\hat{\bf 3}$ of $S_4'$, and they are given by
\begin{align}
Y_{\bf \hat 3}^{(1)}(\tau) =
\begin{pmatrix}
\sqrt{2} \varepsilon \theta\\
\varepsilon^2\\
-\theta^2
\end{pmatrix},
\nonumber
\end{align}
where $\varepsilon$ and $\theta$ are given by
\begin{align}
\varepsilon(\tau) \equiv \frac{2\eta^2(4\tau)}{\eta(2\tau)},~~
\theta(\tau) \equiv \frac{\eta^5(2\tau)}{\eta^2(\tau) \eta^2(4\tau)},
\nonumber
\end{align}
where $\eta(\tau)$ is the Dedekind eta function. 
$\varepsilon(\tau)$ and $\theta(\tau)$ are expanded by
\begin{align}
\varepsilon(\tau) &=  2 \sum_{k=1}^{\infty} q^{\frac{(2k-1)^2}{4}} = 2 q^{\frac 1 4} + 2 q^{\frac 9 4} + 2 q^{\frac{25}4}+...
\nonumber
\\
\theta(\tau) &= 1 + 2 \sum_{k=1}^{\infty} q^{\frac{(2k)^2}{4}} = 1 + 2 q^{1} + 2 q^{4} + 2 q^{9}+...
\nonumber
\end{align}
$\varepsilon(\tau) \sim 2 q^{1/4} $ and $\theta(\tau) \sim 1$ for large $\im \tau$.
The modular forms of higher weights are constructed by their tensor products.
We summarize the irreducible decompositions of $\mathcal{M}_k(\Gamma(4))$ in Table \ref{tab:irrep_G4'}.
\begin{table}[th]
\begin{center}
\begin{tabular}{|c|c|} \hline
weight $k$ & representations
\\
\hline \hline
1 & ${\bf \hat 3}$ 
\\ 
2 & ${\bf 2, 3'} $
\\
3 & ${\bf \hat 1', \hat 3, \hat 3'}$ 
\\
4 & ${\bf 1, 2, 3, 3'}$ 
\\
5 & ${\bf \hat 2}, 2\times {\bf \hat 3}, {\bf \hat 3'}$ 
\\
6 & ${\bf 1, 1', 2, 2', 3}, 2\times{\bf  3'}$ 
\\
\vdots & \vdots 
\\
\hline
\end{tabular}
\caption{Irreducible decomposition of $\mathcal{M}_{k}(\Gamma(4))$.}
\label{tab:irrep_G4'}
\end{center}
\end{table}
We concentrate on the modular forms which are triplets of $S_4'$ since our Yukawa couplings are triplets.%
\footnote{The complete set of the modular forms whose weights are lower than 8 can be found in \cite{Novichkov:2020eep}.}
The modular forms of weight 2 are given by
\begin{align}
Y_{3'}^{(2)}(\tau) =  (Y_{\hat 3}^{(1)}\otimes Y_{\hat 3}^{(1)})_{3'}=
\begin{pmatrix}
\frac{1}{\sqrt{2}} (\theta^4 - \epsilon^4)
\\
-2 \epsilon \theta^3\\
-2 \epsilon^3 \theta
\end{pmatrix}
\sim 
\begin{pmatrix}
\frac{1}{\sqrt{2}} \\
-4 q^{1/4} \\
-16q^{3/4}
\end{pmatrix}.
\nonumber
\end{align}
The modular forms of weight 3 are given by
\begin{align}
Y_{\hat{\bf 3}} ^{(3)}(\tau) =& 
\begin{pmatrix}
\varepsilon \theta^5 + \varepsilon^5 \theta
\\
\frac 1 {2\sqrt{2}} (5 \varepsilon^2 \theta^4 - \varepsilon^6)
\\
\frac 1 {2\sqrt{2}} (\theta^6 - 5 \varepsilon^4 \theta^2)
\end{pmatrix}
\sim
\begin{pmatrix}
2 q^{1/4} \\
\frac {10} {\sqrt{2}} q^{2/4}
\\
\frac 1 {2\sqrt{2}}
\end{pmatrix},
~~
Y_{\hat{\bf 3}'}^{(3)}(\tau) = 
\frac 1 2
\begin{pmatrix}
-4\sqrt{2} \varepsilon^3 \theta^3\\
\theta^6 + 3 \epsilon^4 \theta^2\\
-3 \e^2 \theta^4 -\e^6
\end{pmatrix}\sim
\begin{pmatrix}
-16 \sqrt{2} q^{3/4}\\
1/2\\
-6 q^{2/4}
\end{pmatrix}.
\nonumber
\end{align}
The modular forms of weight 4 are given by
\begin{align}
Y_{{\bf 3}}^{(4)}(\tau) =& \frac 3 {2\sqrt{2}} 
\begin{pmatrix}
\sqrt{2} (\e^2 \theta^6 - \e^6 \theta^2)
\\
\e^3 \theta^5 -\e^7 \theta
\\
-\e \theta^7 + \e^5 \theta^3
\end{pmatrix}
\sim
\frac 3 {2\sqrt{2}}
\begin{pmatrix}
4 \sqrt{2} q^{2/4}
\\
8 q^{3/4}
\\
-2q^{1/4}
\end{pmatrix}
,
\nonumber
\\
Y_{{\bf 3}'}^{(4)}(\tau) =&
\begin{pmatrix}
\frac 1 4 (\theta^8 -\e^8)
\\
\frac 1 {2\sqrt 2} (\e \theta^7 + 7 \e^5 \theta^3)
\\
\frac 1 {2\sqrt 2} (7 \e^3 \theta^5  +\e^7 \theta)
\end{pmatrix}
\sim
\begin{pmatrix}
\frac 1 4 
\\
\frac 1 {\sqrt 2} q^{1/3}
\\
\frac {28}{\sqrt 2} q^{3/4}
\end{pmatrix}.
\nonumber
\end{align}
The modular forms of weight 5 are given by 
\begin{align}
Y_{\hat{\bf 3},I}^{(5)} (\tau) =& 
	\begin{pmatrix}
	\frac {6\sqrt{2}} {\sqrt{5}} \e^5 \theta^5 \\
	\frac {3}{8 \sqrt{5}} (5\e^2 \theta^8 + 10 \e^6 \theta^4 +\e^{10})\\
	-\frac{3}{8\sqrt{5}} (\theta^{10} +10 \e^5 \theta^6 +5 \e^8 \theta^2)
	\end{pmatrix}
\sim 
	\begin{pmatrix}
	\frac {192 \sqrt{2}} {\sqrt{5}} q^{5/4} \\
	\frac {15}{2 \sqrt{5}} q^{2/4}\\
	-\frac{3}{8\sqrt{5}} 
	\end{pmatrix},
\nonumber
\\
~~
Y_{\hat{\bf 3},II}^{(5)} (\tau) =&
	\begin{pmatrix}
	\frac {3} {4} (\e \theta^9 - 2 \e^5 \theta^5 +\e^9 \theta)\\
	\frac {3}{\sqrt{2}} (-\e^2 \theta^8 + \e^6 \theta^4 )\\
	\frac{3}{\sqrt{2}} ( - \e^{4} \theta^6 + \e^8 \theta^2)
	\end{pmatrix}
\sim
	\begin{pmatrix}
	\frac {3} {2} q^{1/4} \\
	-\frac {12}{\sqrt{2}} q^{2/4} \\
	-\frac{48}{\sqrt{2}} q
	\end{pmatrix},
	\nonumber
\\
Y_{\hat{\bf 3}'}^{(5)} (\tau) =& 
	\begin{pmatrix}
	2 (\e^3 \theta^7 + \e^7 \theta^3)\\
	\frac {1}{4\sqrt{2}} (\theta^{10} - 14 \e^4 \theta^6 -3 \e^8 \theta^2 )\\
	\frac {1}{4\sqrt{2}} (3 \e^2 \theta^8 + 14 \e^{6} \theta^4 - \e^{10})
	\end{pmatrix}
\sim
	\begin{pmatrix}
	16 q^{3/4} \\
	\frac {1}{4\sqrt{2}} \\
	\frac {3}{\sqrt{2}} q^{2/4}
	\end{pmatrix}.
	\nonumber
\end{align}
The modular forms of weight 6 are given by
\begin{align}
Y_{{\bf 3}}^{(6)}(\tau) =& 
	\begin{pmatrix}
	\frac 32 (\e^{2}\theta^{10} - \e^{10} \theta^{2})\\
	\frac{3 }{4\sqrt 2} (5 \e^3\theta^{9} - 6 \e^7 \theta^5 +\e^{11} \theta)\\
	\frac{3 }{4\sqrt 2} (\e \theta^{11} - 6 \e^5 \theta^7 + 5 \e^9 \theta^3)
	\end{pmatrix}
\sim
	\begin{pmatrix}
	6 q^{2/4} \\
	\frac{30 }{\sqrt 2} q^{3/4}\\
	\frac{3 }{2\sqrt 2} q^{1/4}
	\end{pmatrix},
	\nonumber
\\
Y_{{\bf 3}',I}^{(6)}(\tau) =& 
	\begin{pmatrix}
	-\frac3 {8\sqrt{13}} (\theta^{12} - 3 \e^{4} \theta^{8} + 3 \e^8 \theta^4 -\e^{12} )\\
	\frac{3 \sqrt2}{\sqrt{13}} (3 \e^5 \theta^{7} + \e^{9} \theta^3)\\
	\frac{3 \sqrt2}{\sqrt{13}} (\e^3 \theta^{9} + 3\e^{7} \theta^5)\\
	\end{pmatrix}
\sim
	\begin{pmatrix}
	-\frac3 {8\sqrt{13}} \\
	\frac{36 \sqrt2}{\sqrt{13}} q^{5/4} \\
	\frac{24 \sqrt2}{\sqrt{13}} q^{3/4} \\
	\end{pmatrix},
\nonumber
\\
Y_{{\bf 3}',II}^{(6)}(\tau) =& 
	\begin{pmatrix}
	3 (\e^{4} \theta^{8} - \e^8 \theta^4)\\
	-\frac{3}{4\sqrt{2}} (\e^5 \theta^{11} + 2 \e^{5} \theta^7 -3 \e^9 \theta^3)\\
	\frac{3}{4\sqrt{2}} (3 \e^3 \theta^{9} - 2 \e^{7} \theta^5 - \e^{11} \theta)\\
	\end{pmatrix}
\sim
	\begin{pmatrix}
	48 q\\
	-\frac{24}{\sqrt{2}} q^{5/4} \\
	\frac{18}{\sqrt{2}} q^{3/4} \\
	\end{pmatrix}.
	\nonumber
\end{align}
For $k=7$, we have
\begin{align}
Y_{\hat{\bf 3}',I}^{(7)}(\tau) =&
	\begin{pmatrix}
	\frac {3}{4 \sqrt{37}} (7 \e^3 \theta^{9}  + 50 \e^{7} \theta^7 + 7 \e^{11} \theta^3)\\
	-\frac{3} {4\sqrt{74}} (\theta^{14} +14 \e^4 \theta^{10} + 49 \e^{8} \theta^6)\\ 
	\frac{3} {4 \sqrt{74}} (49 \e^6 \theta^{8} +14 \e^{10} \theta^{4} + \e^{14})\\ 
	\end{pmatrix}
\sim
	\begin{pmatrix}
	\frac {42}{\sqrt{37}} q^{3/4} \\
	-\frac{3} {4\sqrt{74}} \\ 
	\frac{2352} {\sqrt{74}} q^{6/4}\\ 
	\end{pmatrix},
\nonumber
\\
Y_{\hat{\bf 3}',II}^{(7)}(\tau) =&
	\begin{pmatrix}
	\frac {9}{4} (\e^3 \theta^{11}  -2 \e^{7} \theta^7 + \e^{11} \theta^3)\\
	\frac{9} {4\sqrt{2}} (\e^4 \theta^{10} - 2 \e^8 \theta^6 + \e^{12} \theta^2)\\ 
	-\frac{9} {4\sqrt{2}} (\e^2 \theta^{12} - 2 \e^6 \theta^{8} + \e^{10}\theta^4)\\ 
	\end{pmatrix}
\sim
	\begin{pmatrix}
	18 q^{3/4}\\
	\frac{36} {\sqrt{2}} q\\ 
	-\frac{9} {\sqrt{2}} q^{2/4} \\ 
	\end{pmatrix}.
	\nonumber
\end{align}
For $k=8$, we have
\begin{align}
Y_{{\bf 3},I}^{(8)}(\tau) =&
	\begin{pmatrix}
	9 \sqrt{\frac {2}{5}} (\e^6 \theta^{10} -  \e^{10} \theta^6)\\
	\frac{9} {16 \sqrt{5}} (5 \e^3 \theta^{13} +5 \e^7 \theta^{9} - 9 \e^{11} \theta^{5} - \e^{15} \theta)\\ 
	-\frac{9} {16 \sqrt{5}} (\e \theta^{15} + 9 \e^5 \theta^{11} - 5 \e^{9} \theta^{7} - 5 \e^{13} \theta^3)\\ 
	\end{pmatrix}
\sim
	\begin{pmatrix}
	576 \sqrt{\frac {2}{5}} q^{6/4} \\
	\frac{9\sqrt{5}} {2} q^{3/4} \\ 
	-\frac{9} {8 \sqrt{5}} q^{1/4}\\ 
	\end{pmatrix},
\nonumber
\\
~~
Y_{{\bf 3},II}^{(8)}(\tau) =&
	\begin{pmatrix}
	-\frac {9}{8} (\e^2 \theta^{14} - 2 \e^{6} \theta^{10} + 3 \e^{10} \theta^6 - \e^{14} \theta^2)\\
	\frac{9} {2\sqrt{2}} (\e^3 \theta^{13} - 2 \e^7 \theta^9 + \e^{11} \theta^{5})\\ 
	-\frac{9} {2\sqrt{2}} (\e^5 \theta^{11} - 2 \e^9 \theta^{7} + \e^{13}\theta^3)\\ 
	\end{pmatrix}
\sim
	\begin{pmatrix}
	-\frac {9}{2} q^{2/4} \\
	\frac{36} {\sqrt{2}} q^{3/4}\\ 
	-\frac{144} {\sqrt{2}} q^{5/4}\\ 
	\end{pmatrix}.
	\nonumber
\end{align}
These $q$-expansions are consistent with \eqref{eq:MF_order_lev4}.

\section{The modular forms of level 5}
\label{app:level_5}

$\Gamma_5'$ is isomorphic to $A_5' \simeq SL(2,\mathbb{Z}_5)$, which is a double covering group of $A_5$.
$A_5'$ has the following irreducible representations,
\begin{align}
{\bf 1,\hat{2}, \hat{2}', 3, 3', 4,\hat{4}, 5, \hat{6}}.
\nonumber
\end{align}
Our notation follows \cite{Wang:2020lxk}.
We concentrate on the triplet representations.
Their matrix representations are summarized in Table \ref{tab:irrep_S4'}.
\begin{table}[t]
\begin{center}
\begin{tabular}{l|ccc} \hline
{\bf r} & $S$ & $T$ & $R$\\
\hline\hline
${\bf 1}$ & 1 & 1  & 1
\\
${\bf 3}$ 
&
$\frac 1 {\sqrt{5}}
\begin{pmatrix}
1 & -\sqrt{2} & -\sqrt 2 \\
-\sqrt 2 & -(\sqrt 5 +1)/2 & (\sqrt 5 -1)/2 \\
-\sqrt 2 & (\sqrt 5 -1)/2 & -(\sqrt 5 +1)/2 
\end{pmatrix}$
&
$\begin{pmatrix}
1 & 0 & 0 \\
0 & \sigma & 0 \\
0 & 0 & \sigma^4
\end{pmatrix}$
& $I_{3\times3}$
\\[20pt]
${\bf 3}'$ 
&
$\frac 1 {\sqrt{5}}
\begin{pmatrix}
-1 & \sqrt{2} & \sqrt 2 \\
\sqrt 2 & (-\sqrt 5 + 1)/2 & (\sqrt 5 +1)/2 \\
\sqrt 2 & (\sqrt 5 +1)/2 & (-\sqrt 5 + 1)/2
\end{pmatrix}$
&
$\begin{pmatrix}
1 & 0 & 0 \\
0 & \sigma^2 & 0 \\
0 & 0 & \sigma^3
\end{pmatrix}$
& $I_{3\times3}$
\\[20pt]
\hline
\end{tabular}
\caption{The matrix representations of the generators of $\Gamma_5'$.
$\sigma = e^{2i\pi/5}$.
We concentrate on the modular forms of 3-dimensional representations,
and omit the other representations in this table.
Our notation is base on \cite{Wang:2020lxk}.}
\label{tab:irrep_A5'}
\end{center}
\end{table}
The complete table of the irreducible decomposition of the tensor products, and their CG coefficients also can be found in \cite{Wang:2020lxk}.
The irreducible decomposition of the tensor products of the triplets are given by
\begin{align}
{\bf 3} \otimes {\bf 3} &= {\bf 1} \oplus {\bf 3} \oplus {\bf 5},
\nonumber
\\
{\bf 3}' \otimes {\bf 3}' &= {\bf 1} \oplus {\bf 3}' \oplus {\bf 5},
\nonumber
\\
{\bf 3} \otimes {\bf 3}' &=  {\bf 4} \oplus {\bf 5},
\nonumber
\end{align}
and their CG-coefficients are given by
\begin{align}
\begin{pmatrix}
\alpha_1\\
\alpha_2\\
\alpha_3\\
\end{pmatrix}_{\bf 3}
\otimes
\begin{pmatrix}
\beta_1\\
\beta_2\\
\beta_3\\
\end{pmatrix}_{\bf 3}
=
&
\frac 1 {\sqrt{3}}
\left(\alpha_1 \beta_1 +\alpha_2 \beta_3 +\alpha_3 \beta_2\right)
+
\frac 1 {\sqrt{2}}
\begin{pmatrix}
\alpha_2 \beta_3 -\alpha_3 \beta_2\\
\alpha_1 \beta_2 -\alpha_2 \beta_1\\
\alpha_3 \beta_1 -\alpha_1 \beta_3\\
\end{pmatrix}
+ 
\frac{1}{\sqrt 6}
\begin{pmatrix}
2 \alpha_1 \beta_1 - \alpha_2 \beta_3 - \alpha_3 \beta_2 \\
-\sqrt 3 (\alpha_1 \beta_2 + \alpha_2 \beta_1 ) \\
\sqrt 6 \alpha_2 \beta _2\\
\sqrt 6 \alpha_3 \beta _3\\
\sqrt 3 (\alpha_1 \beta _3 + \alpha_3 \beta _1 )\\
\end{pmatrix},
\nonumber
\\
\begin{pmatrix}
\alpha_1\\
\alpha_2\\
\alpha_3\\
\end{pmatrix}_{\bf 3'}
\otimes
\begin{pmatrix}
\beta_1\\
\beta_2\\
\beta_3\\
\end{pmatrix}_{\bf 3'}
=
&
\frac 1 {\sqrt{3}}
\left(\alpha_1 \beta_1 +\alpha_2 \beta_3 +\alpha_3 \beta_2\right)
+
\frac 1 {\sqrt{2}}
\begin{pmatrix}
\alpha_2 \beta_3 -\alpha_3 \beta_2\\
\alpha_1 \beta_2 -\alpha_2 \beta_1\\
\alpha_3 \beta_1 -\alpha_1 \beta_3
\end{pmatrix}
+ 
\frac{1}{\sqrt 6}
\begin{pmatrix}
2 \alpha_1 \beta_1 - \alpha_2 \beta_3 - \alpha_3 \beta_2 \\
\sqrt 6 \alpha_3 \beta _3\\
-\sqrt 3 (\alpha_1 \beta_2 + \alpha_2 \beta_1 ) \\
\sqrt 3 (\alpha_1 \beta _3 + \alpha_3 \beta _1 )\\
\sqrt 6 \alpha_2 \beta _2
\end{pmatrix},
\nonumber
\\
\begin{pmatrix}
\alpha_1\\
\alpha_2\\
\alpha_3\\
\end{pmatrix}_{\bf 3}
\otimes
\begin{pmatrix}
\beta_1\\
\beta_2\\
\beta_3\\
\end{pmatrix}_{\bf 3'}
=
&
\frac 1 {\sqrt{3}}
\begin{pmatrix}
\sqrt 2 \alpha_2 \beta_1 + \alpha_3 \beta_2\\
-\sqrt 2 \alpha_1 \beta_2 -\alpha_3 \beta_3\\
-\sqrt 2 \alpha_1 \beta_3 -\alpha_2 \beta_2\\
\sqrt 2 \alpha_3 \beta_1 + \alpha_2 \beta_3
\end{pmatrix}
+ 
\frac{1}{\sqrt 3}
\begin{pmatrix}
\sqrt 3 \alpha_2 \beta_1 + \alpha_3 \beta_2 \\
\alpha_2 \beta_1 - \sqrt2 \alpha_3 \beta_2 \\
\alpha_1 \beta_2 - \sqrt2 \alpha_3 \beta_3 \\
\alpha_1 \beta_3 - \sqrt2 \alpha_2 \beta_2 \\
\alpha_3 \beta_1 - \sqrt2 \alpha_2 \beta_3 \\
\end{pmatrix}.
\label{eq:CG_lev5}
\end{align}

\subsection*{Modular forms}

The modular forms of level 5 and weight $k$ are given by tensor products of the modular forms of level 5 and weight 1.
The modular forms of weight 1 form $\hat{\bf 6}$ of $A_5'$, and they are given by
\begin{align}
Y_{\bf \hat 6}^{(1)}(\tau) \equiv
\begin{pmatrix}
Y_1\\
Y_2\\
Y_3\\
Y_4\\
Y_5\\
Y_6
\end{pmatrix}
=
\begin{pmatrix}
\hat e_1 -3 \hat e_6\\
5 \sqrt 3 \hat e_2\\
10 \hat e_3\\
10 \hat e_4\\
5\sqrt 2 \hat e_5\\
-3 \hat e_1 - \hat e_6
\end{pmatrix},~{\rm with}~
\hat e_i = \frac{\eta^{15}(5\tau)}{\eta^3(\tau)} \mathfrak{k}^{i-1}_{\frac 2 5, \frac 0 5}(5\tau) \mathfrak{k}^{6-i}_{\frac 2 5, \frac 0 5}(5\tau),
\end{align}
where $\mathfrak{k}_{r_1, r_2}(\tau)$ is the Klein form defined by
\begin{align}
\mathfrak{k}_{r_1, r_2}(\tau) = q_z^{(r_1-1)/2} (1-q_z) \prod_{n=1}^{\infty} (1-q^n q_z)(1-q^n q_z^{-1})(1-q^n)^{-2},
\end{align}
with $(r_1, r_2)$ being a pair of rational number.
$z \equiv \tau r_1 +r_2$ and $q_z \equiv e^{2\pi i z}$.
The modular forms of higher weights are constructed by its products.
\begin{align}
\hat e_i \sim q^{(i -1)/5}
\end{align}
for large $\im \tau$.
We summarize the irreducible decompositions of the $\mathcal{M}_k(\Gamma(5))$ in Table \ref{tab:irrep_G5'}.
\begin{table}[th]
\begin{center}
\begin{tabular}{|c|c|} \hline
weight $k$ & representations
\\
\hline \hline
1 & ${\bf \hat 6}$ 
\\ 
2 & ${\bf 3, 3', 5} $
\\
3 & ${\bf \hat 4}, 2\times {\bf \hat 6}$ 
\\
4 & ${\bf 1, 3, 3', 4}, 2\times {\bf 5}$ 
\\
5 & ${\bf \hat 2, \hat 2', \hat 4}, 3\times {\bf \hat 6}$ 
\\
6 & ${\bf 1}, 2\times{\bf  3}, 2\times {\bf 3'}, 2\times{\bf  4},  2\times{\bf  5} $ 
\\
\vdots & \vdots 
\\
\hline
\end{tabular}
\caption{Irreducible decomposition of $\mathcal{M}_{k}(\Gamma(5))$.
$\mathcal{M}_{k}(\Gamma(5))$ is a $5k+1$ dimensional linear space.}
\label{tab:irrep_G5'}
\end{center}
\end{table}
We concentrate on the modular forms which are triplets of $\Gamma_5'$.
The weights of the triplet modular forms always be positive and even for $\Gamma_5'$.
The complete set of the modular forms whose weights are lower than 6 can be found in  \cite{Wang:2020lxk}.
The triplet modular forms of weight 2 are given by
\begin{align}
Y_{\bf 3}^{(2)}(\tau) &=  
-3
\begin{pmatrix}
Y_1^2-3Y_1 Y_6 -Y_6^2\\
Y_1Y_2\\
-Y_5 Y_6
\end{pmatrix}
\sim -3
\begin{pmatrix}
1 \\
5\sqrt 5 q^{1/5} \\
15 \sqrt{2} q^{4/5}
\end{pmatrix},
\nonumber
\\
Y_{\bf 3'}^{(2)}(\tau) &=  
\frac 1 2
\begin{pmatrix}
\sqrt 6 (Y_1^2- 2Y_1 Y_6 -Y_6^2)\\
-\sqrt 3 Y_3 (Y_1 + Y_6)\\
\sqrt 3 Y_4 (Y_1 - Y_6)\\
\end{pmatrix}
\sim \frac 1 2
\begin{pmatrix}
-2 \sqrt 6\\
20 \sqrt 3 q^{2/5}\\
40\sqrt 3 q^{3/5}\\
\end{pmatrix}.
\nonumber
\end{align}
The triplet modular forms of weight 4 are given by
\begin{align}
Y_{{\bf 3}} ^{(4)}(\tau) =& 
\frac{\sqrt 3}{4}
\begin{pmatrix}
(Y_1^2 + Y_6^2)(7Y_1^2 - 18 Y_1 Y_6 -7 Y_2^2)\\
Y_2 (13Y_1^3 -3Y_1^2 Y_6 -29 Y_1 Y_6^2 - 9 Y_6^3 )\\
-Y_5(9 Y_1^3 -29 Y_1^2 Y_6 + 3 Y_1 Y_6^2 +13 Y_6^3)
\end{pmatrix}
\sim 
\frac{\sqrt 3}{4}
\begin{pmatrix}
-20\\
20 \sqrt 2 q^{1/5} \\
1140 \sqrt 2 q^{4/5}
\end{pmatrix},
\nonumber
\\
Y_{{\bf 3}'}^{(4)}(\tau) =& 
-\frac 1 2
\begin{pmatrix}
\sqrt 2(Y_1^2 + Y_6^2)(4 Y_1^2 -11 Y_1 Y_6 -4 Y_6^2)\\
-Y_3 (Y_1 -2 Y_6) (7 Y_1^2 -3 Y_1 Y_6 -2 Y_6^2)\\
Y_4 (2Y_1 + Y_6 ) (3Y_1^2 - 3Y_1 Y_6 -7Y_6^2)
\end{pmatrix}
\sim
-\frac 1 2
\begin{pmatrix}
10 \sqrt 2 \\
140 q^{2/5} \\
510 q^{3/5}
\end{pmatrix}.
\nonumber
\end{align}
The triplet modular forms of weight 6 are given by
\begin{align}
Y_{{\bf 3},I}^{(6)}(\tau)=&
\frac {9\sqrt 2}{16} 
(Y_1^2 - 4 Y_1 Y_6 -Y_6^2)
\begin{pmatrix}
(Y_1 -3 Y_6)(3 Y_1 +Y_6)(3Y_1^2 -2 Y_1 Y_6 -3 Y_6^2)
\\
2Y_2(2Y_1^3 - 9Y_1 Y_6^2 - 3Y_6^3)
\\
2Y_5(3Y_1^3 - 9Y_1^2 Y_6 + 2Y_6^3)
\end{pmatrix}\sim
\frac {9\sqrt 2}{4} 
\begin{pmatrix}
-1800 q
\\
20\sqrt 2 q^{1/5} 
\\
-240 \sqrt 2 q^{4/5}
\end{pmatrix},
\nonumber
\\
Y_{{\bf 3}, II}^{(6)}(\tau) =&
3\sqrt 2 (Y_1^4 - 3 Y_1^3 Y_6 - Y_1^2 Y_6^2 + 3Y_1 Y_6^3 +Y_6^4)
\begin{pmatrix}
Y_1^2-3Y_1 Y_6 -Y_6^2\\
Y_1Y_2\\
-Y_5 Y_6
\end{pmatrix}
\sim
3\sqrt 2 
\begin{pmatrix}
1 \\
5 \sqrt 2 q^{1/5} \\
15\sqrt 2 r^{4/5}
\end{pmatrix},
\nonumber
\\
Y_{{\bf 3}',I}^{(6)}(\tau)=&
-\frac {\sqrt 3}{2} 
(Y_1^2 - 4 Y_1 Y_6 -Y_6^2)^2
\begin{pmatrix}
(3Y_1+Y_6)(Y_1-3 Y_6)
\\
\sqrt 2 Y_1 Y_3
\\
\sqrt 2 Y_4 Y_6
\end{pmatrix}
\sim
-8 \sqrt 3 
\begin{pmatrix}
100 q\\
10 \sqrt 2 q^{2/5}
\\
-30 \sqrt 2 q^{3/5}
\end{pmatrix},
\nonumber
\\
Y_{{\bf 3}', II}^{(6)}(\tau) =&
-\frac{\sqrt{6}}{2} (Y_1^4 -3 Y_1^3 Y_6 - Y_1^2 Y_6^2 + 3Y_1 Y_6^3 +Y_6^4)
\begin{pmatrix}
\sqrt 2 (Y_1^2- 2Y_1 Y_6 -Y_6^2)\\
- Y_3 (Y_1 + Y_6)\\
 Y_4 (Y_1 - Y_6)\\
\end{pmatrix}
\sim
-\frac{\sqrt{6}}{2} 
\begin{pmatrix}
-2 \sqrt 2 \\
20 q^{2/5} \\
40 q^{3/5}\\
\end{pmatrix}.
\nonumber
\end{align}
These $q$-expansions are consistent with $\rho(T)$.

\end{document}